\documentclass[12pt]{aastex}
\usepackage{eqsection}
\usepackage{epsf}

\def\beq{\begin{equation}}
\def\eeq{\end{equation}}
\def\bea{\begin{eqnarray}}
\def\eea{\end{eqnarray}}
\def\nn{\nonumber}

\def\analcont{\longrightarrow}
\relax
\begin{document}

\begin{titlepage}
\begin{center}
{\large\bf Parafermionic theory with the symmetry $Z_{5}$.}\\[.5in] 
{\bf Vladimir S.~Dotsenko\bf${}^{(1)}$,
 \bf Jesper Lykke Jacobsen\bf${}^{(2)}$
 \bf and Raoul Santachiara\bf${}^{(1)}$}\\[.2in]
{\bf (1)} {\it LPTHE\/}\footnote{Laboratoire associ\'e No. 280 au CNRS},
         {\it Universit{\'e} Pierre et Marie Curie, Paris VI\\
               Bo\^{\i}te 126, Tour 16, 1$^{\it er}$ {\'e}tage,
               4 place Jussieu, F-75252 Paris Cedex 05, France.}\\
{\bf (2)} {\it Laboratoire de Physique Th\'eorique et Mod\`eles
               Statistiques, \\
               Universit\'e Paris-Sud, B\^atiment 100, F-91405 Orsay,
               France.}\\[.2in]
dotsenko@lpthe.jussieu.fr, jacobsen@ipno.in2p3.fr,
santachiara@lpthe.jussieu.fr 
\end{center}

\underline{Abstract.}

A parafermionic conformal theory with the symmetry $Z_{5}$ is constructed,
based on the second solution of Fateev-Zamolodchikov for the
corresponding parafermionic chiral algebra.

The primary operators of the theory, which are the singlet,
doublet 1, doublet 2, and disorder operators, are found
to be accommodated by the weight lattice of the classical Lie
algebra $B_{2}$. The finite Kac tables for unitary theories
are defined and the formula for the conformal dimensions
of primary operators is given.

\end{titlepage}

\newpage

\section{Introduction}

Extra discrete group symmetries in two-dimensional critical phenomena are
naturally realised by parafermionic chiral algebras. The most well-known 
and
widely used parafermionic conformal theory is due to Fateev and 
Zamolodchikov
[1]. It describes, in particular, the self-dual fixed points of
a particular lattice model with $Z_N$ symmetry [1,2].

In terms of the associativity constraint for physically consistent chiral 
algebras,
the parafermions of the theory [1] represent the first solution, 
with the
minimal possible values of the conformal dimensions (or spins) of the
parafermionic
currents. In this solution the central charge of the corresponding 
Virasoro
algebra is a function of $N$ only, i.e., for each $N$ of the group $Z_N$ 
there
is just one conformal theory, with a given, fixed value $c_N$ of the 
central
charge.

These same authors gave in the Appendix of Ref.~[1] a second
solution of the associative parafermionic chiral algebra, with
the next allowed values of the spins of the parafermions. In the second
$Z_N$ symmetric solution the central charge remains a free parameter,
for each $N$.

The case $N=2$ of this second solution corresponds to the superconformal
theory. The conformal theory corresponding to the 
second solution with $Z_3$ symmetry has been fully constructed
by Fateev and Zamolodchikov in a subsequent work, Ref.~[3]. It
produces an infinite set of conformal theories which are selected by
applying the degeneracy condition to the representations generated by
physical fields, similar to the case of minimal models of purely conformal
(non-extended) conformal field theory. This infinite set of conformal 
theories
was supposed [3] to correspond to higher, multicritical fixed 
points
of physical statistical systems with the symmetry $Z_3$. And in fact, the 
first
theory in this set (the one with the least value of the central charge) 
was
shown to correspond to the tricritical Potts model.

For the second associative solution of the chiral algebra [1],
the conformal theories with $N>3$ (i.e., with symmetries $Z_4$, $Z_5$,
$Z_6$ etc.) have not been constructed so far. The purpose of our present
work is to build the corresponding theories. In particular, we shall
define the Kac formula giving the conformal dimensions of physical
(primary) fields, which realise degenerate representations of the 
corresponding
parafermionic chiral algebras.

The first of these theories, the one of $N=4$, turns out to be trivial
in the sense that it factorises into two independent superconformal chiral
algebras, with fields of dimensions $2$ and $3/2$.

The first new theory is therefore that of $Z_5$. This theory in itself 
turns
out to be so rich, and also representative for the whole class (whilst 
being
the simplest at the same time), that it appeared reasonable for us to 
present
it separately, in all the details. 
The present paper is therefore devoted to the theory $Z_5$,
and generalisations will be given in a following work [4].

One extra result which we have used in our construction is that the 
central charge for the
second solution of the $Z_{N}$ parafermionic
algebra agrees with that of the coset [5]
\beq
 \frac{SO_{k}(N)\times SO_{2}(N)}{SO_{k+2}(N)},
 \label{SOcoset}
\eeq
which is
\bea
 c &=& (N-1) \left( 1-\frac{N(N-2)}{p(p+2)} \right), \label{cparamp} \\
 p &=& N-2+k. \label{paramp}
\eea
In the construction of the representation theory of the parafermionic
algebra this observation turns out to be extremely useful.

Our paper is organised as follows. In Section~\ref{sec2} we review some
technical details on the second associative solution of the chiral algebra
as found in Ref.~[1]. We then specialise to the case of $Z_5$.
The structure of the representation modules of the various physical operators
(singlets, doublets, disorder operators) are fixed. We present explicit
calculations of the first degeneracies in these modules. The resulting
conformal dimensions, and the levels on which singular states are found,
serve as initial conditions for dressing the entire theory.
In Section~\ref{sec3} we give the Kac formula of the theory.
We give the boundary
terms corresponding to the various sectors of the theory.

The general question
of determining the appropriate sector label for each operator in the theory
is addressed in Section~\ref{sec4}. We give a first argument, based on
fusion rules and an assumed Coulomb gas construction. Some aspects of
this argument are then verified by considering differential and
characteristic equations for three-point functions. We also give the general
formula for the eigenvalues of the 
parafermionic zero modes for the disorder operators. The conclusions
are given in Section~\ref{sec5}.

Four appendices deal with technical points of the calculations.

\section{Parafermionic algebras in different sectors. First degeneracies}
\label{sec2}

The operator product expansions of chiral fields within the second 
solution of
the $Z_N$ symmetric parafermionic algebra is given in Appendix~\ref{appA} of
Ref.~[1]. They take the form
\bea
 \Psi^{q}(z)\Psi^{q'}(z') &=& 
\frac{\lambda^{q,q'}_{q+q'}}{(z-z')^{\Delta_{q}
 +\Delta_{q'}-\Delta_{q+q'}}} \label{ope1} \\
 &\times& \left \{\Psi^{q+q'}(z')+(z-z')
 \frac{\Delta_{q+q'}+\Delta_{q}-\Delta_{q'}}{2\Delta_{q+q'}}
 \partial_{z'}\Psi^{q+q'}(z')+\ldots \right \},\quad q+q'\neq 0 \nn \\
 \Psi^{q}(z)\Psi^{-q}(z') &=& \frac{1}{(z-z')^{2\Delta_{q}}}
 \left \{1+(z-z')^{2}
 \frac{2\Delta_{q}}{c}T(z')+\ldots \right \}
 \label{ope2}
\eea
Here the structure constants $\lambda_{q+q'}^{q,q'}$ and the central 
charge
$c$ (of the Virasoro algebra) are given by the expressions:
\bea
 (\lambda^{q,q'}_{q+q'})^{2} &=& \frac{\Gamma(q+q'+1)\Gamma(N-q+1)
 \Gamma(N-q'+1)}{\Gamma(q+1)\Gamma(q'+1)\Gamma(N-q-q'+1)\Gamma(N+1)}\nn\\
 &\times& \frac{\Gamma(q+q'+v)\Gamma(N+v-q)\Gamma(N+v-q')\Gamma(v)}
 {\Gamma(N+v-q-q')\Gamma(q+v)\Gamma(q'+v)\Gamma(N+v)}, \label{struct} \\
 c &=& \frac{4(N-1)(N+v-1)v}{(N+2v)(N+2v-2)}. \label{cparamv}
\eea
The dimensions of the fields $\{\Psi^{q}(z)\}$ have the
following form:
\beq
 \Delta_{q}=\Delta_{N-q}=\frac{2q(N-q)}{N}.
 \label{chidim2}
\eeq
Note in particular that the field $\Psi^{-q}$ in (\ref{ope2}) is assumed
to have dimension $\Delta_{N-q}$, in the sense that the indices $q$
referring to the $Z_N$ charge are always defined modulo $N$. Thus,
\beq
 \Psi^{N-q}\equiv\Psi^{-q}\equiv(\Psi^{q})^{+},\quad \Delta_{N-q}
 \equiv\Delta_{-q}.
\eeq

In the expressions (\ref{struct})--(\ref{cparamv}) $v$ is a free 
parameter,
which was chosen in [1] to parametrise this solution. If one 
defines
\beq
 v=2k
 \label{vkparam}
\eeq
then the expression (\ref{cparamv}) for the central charge $c$ could also
be represented as in
(\ref{cparamp}), making connection to the coset (\ref{SOcoset}) 
[5].

{}From now on we specialise in this paper to the case $N=5$, for the
reasons given in the Introduction. In this case the $Z_{5}$ charge $q$
takes the values $0,\pm 1,\pm 2,$ and we have the following
parafermionic fields:
\beq
 \Psi^{\pm 1}(z),\quad \Psi^{\pm 2}(z).
\eeq
Together with the identity and the stress-energy operator $T(z)$ these 
fields
make a closed operator algebra, given by Eqs.~(\ref{ope1})--(\ref{ope2}).
Their dimensions are
\beq
 \Delta_{1}=\frac{8}{5},\quad \Delta_{2}=\frac{12}{5}
\eeq
according to Eq.~(\ref{chidim2}).

On the side of the representation fields, primary fields of the algebra
(\ref{ope1})--(\ref{ope2}), we
should expect the singlet
operators:
\beq
 \Phi^{0}(z,\bar{z})\quad (q=0)
 \label{singlet}
\eeq
the doublet 1 operators:
\beq
 \Phi^{\pm 1}(z,\bar{z})\quad (q=\pm 1)
\eeq
and the doublet 2 operators:
\beq
 \Phi^{\pm 2}(z,\bar{z})\quad (q=\pm 2).
\eeq

\begin{figure}
\begin{center}
 \leavevmode
 \epsfysize=150pt{\epsffile{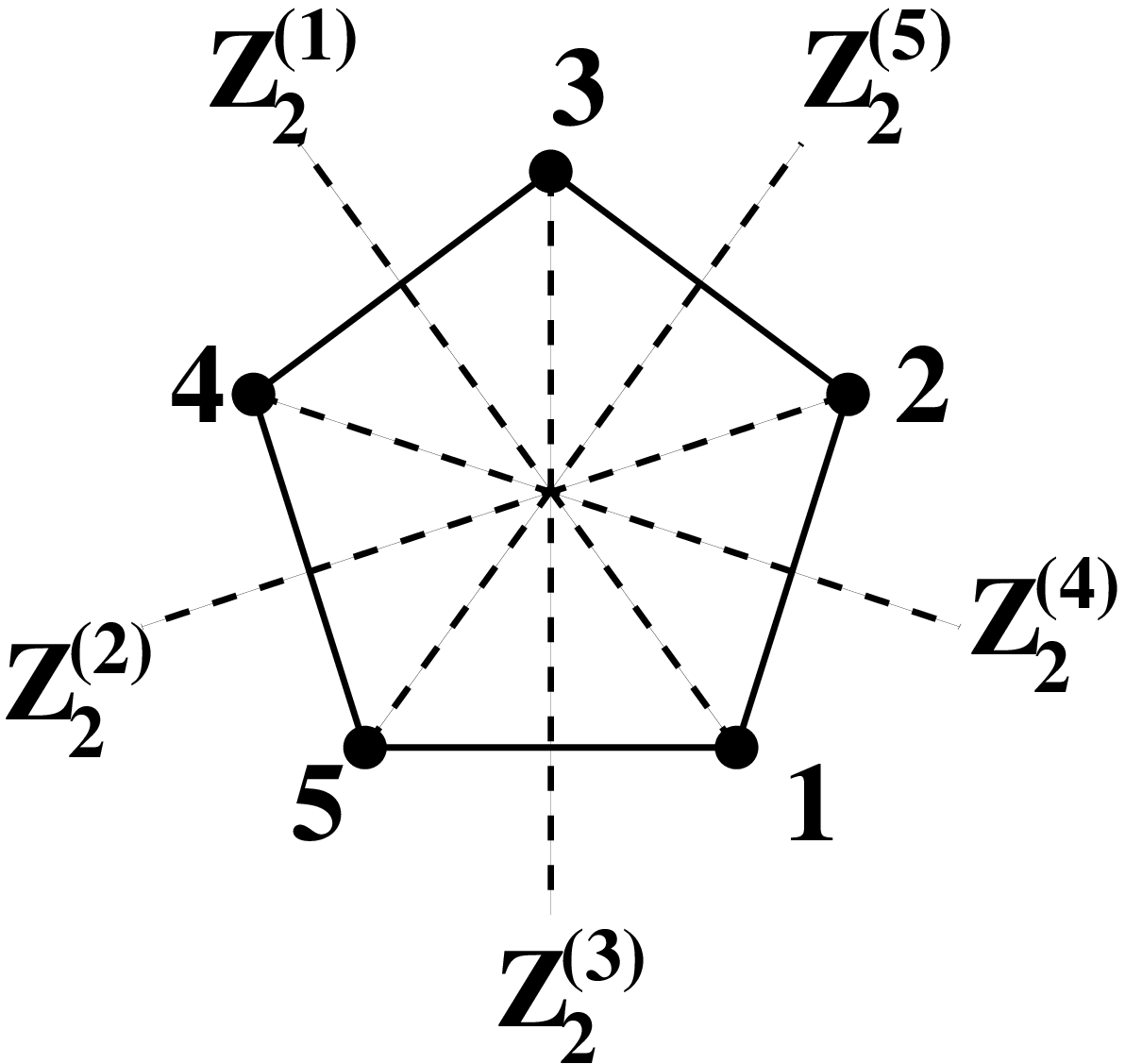}}
 \end{center}
 \protect\caption[3]{\label{fig1}The action of the parafermionic current
 $\Psi^k$ can be represented pictorially as a $2\pi k/5$ rotation of a
 pentagon. A quintuplet of disorder operators augments the cyclic symmetry
 to a dihedral one, each operator corresponding to the $Z_2$ reflection 
with
 respect to an axis passing through one of the vertices of the pentagon.
 The labeling of the five axes, as shown on the figure, is also used to
 label the corresponding disorder operators.}
\end{figure}

In addition to these usual representations of the $Z_5$ group we
expect the spectrum to include a quintuplet (5-plet) of 
$Z_2$ disorder operators:
\beq
 \{R^a(z,\bar{z}),\quad a=1,2,3,4,5\}.
 \label{disorder}
\eeq
The presence of disorder operators reflects the fact that the theory is,
in fact, invariant under a larger group than $Z_5$, viz.~the dihedral
group $D_{5}$. This group includes, in addition to $Z_5$, also the
$Z_{2}$ reflections with respect to the five axes shown in 
Fig.~\ref{fig1}.

This higher symmetry stems from the fact that in the solution
(\ref{ope1})--(\ref{chidim2}) everything is symmetric with respect to
the $Z_N$ charge conjugation:
\beq
 q\rightarrow N-q.
\eeq
(For the structure constants $(\lambda^{q,q'}_{q+q'})^{2}$ in
Eq.~(\ref{struct}) the demonstration requires a little algebra with products.)

The disorder fields have previously been shown to be in the spectrum of 
the
$Z_3$ theory, second solution, as constructed in Ref.~[3]. They 
appear
as triplet representations in this case. The disorder fields also appear 
in
the first solution [1] (for the case of general $Z_{N}$), and have
been defined and fully treated in Ref.~[6].

One could object that the disorder fields $\{R_{a}(z,\bar{z})\}$
in Eq.~(\ref{disorder}) are on a somewhat different footing than the
standard (singlet and doublet) representation fields. Namely, the
disorder fields complete the cyclic group $Z_{5}$, as generated by
$1=\Psi^0(z)$, $\Psi^{\pm 1}(z)$ and $\Psi^{\pm 2}(z)$, to the dihedral
group $D_{5}$. However, there is a crucial difference between the $\Psi$
and the $R$ fields. The first are chiral, or holomorphic, whilst the
latter are non-chiral, like the rest of the representation operators.
So despite of the fact that the symmetry of the theory is $D_{5}$,
not all of the elements in the dihedral group are represented in the
chiral algebra but only those of the biggest abelian subalgebra, $Z_{5}$.
The rest of the group elements find
themselves in the representation space, as disorder operators.

This is at least the structure of the present theory, and of the theory 
$Z_{3}$ in Ref.~[3].

Summarising, we expect that the representation space of the theory $Z_5$
can be divided into singlet, doublet 1, doublet 2, and disorder operators,
cf.~Eqs.~(\ref{singlet})--(\ref{disorder}).

It is already a standard point that the spectrum, for a given chiral
algebra with a free parameter (the central charge (\ref{cparamv}), or
the parameter $v$), is to be defined by the degeneracy condition of
the representations. The reasoning why this is so can be given
in various ways, for example by demanding the closure of the operator
algebra of primary (physical) fields.

To define the spectrum of dimensions of singlets, doublets, and
disorder operators, we have to define the representations of the
chiral algebra in the corresponding sectors. One then constructs the
modules induced by the various primary fields, and demands their 
degeneracy.

As usual, it will be sufficient to analyse directly the degeneracies on 
the
first few levels in the modules. The physical operators defined in this
way become the basic ones in the Kac table. Using these basic fields 
we shall be able, in the following Sections, to define the whole set, the
Kac table and the formula for the dimensions.

\subsection{Structure of the representation modules}

\begin{figure}
\begin{center}
 \leavevmode
 \epsfysize=150pt{\epsffile{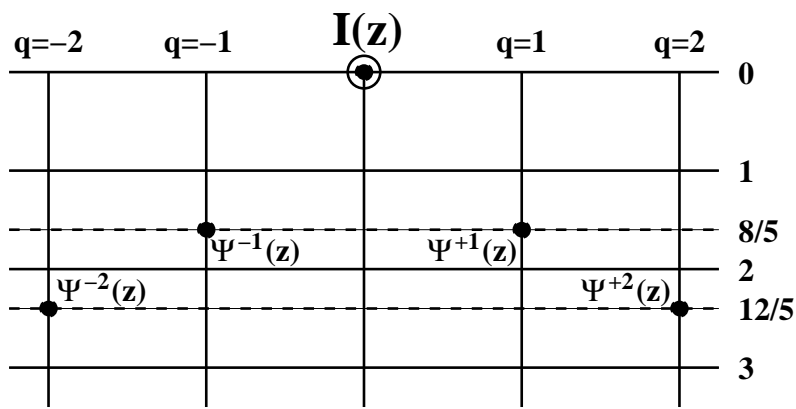}}
 \end{center}
 \protect\caption[3]{\label{fig2}The position of the parafermionic 
currents
 in the identity module. The horizontal axis shows the $Z_5$ charge $q$.
 The vertical axis gives the level of a given state, here with respect to
 the identity operator placed at the summit of the diagram.}
\end{figure}

The level structure of the modules induced by the singlets and by the two
types of doublets can be defined as follows. We first place the chiral 
fields,
i.e., the parafermions $\Psi^{\pm 1}(z)$, $\Psi^{\pm 2}(z)$ as well as the
stress-energy operator $T(z)$, in the module of identity.
The levels of
the various operators in this module correspond to their conformal 
dimensions:
$\Delta_{\pm 1}=8/5$ and $\Delta_{\pm 2}=12/5$. This partially filled 
identity
module is shown in Fig.~\ref{fig2}. To make apparent the crucial
role played by the $Z_5$ charge $q$, we depict the modules by separating
the various charge sectors horizontally.

\begin{figure}
\begin{center}
 \leavevmode
 \epsfysize=150pt{\epsffile{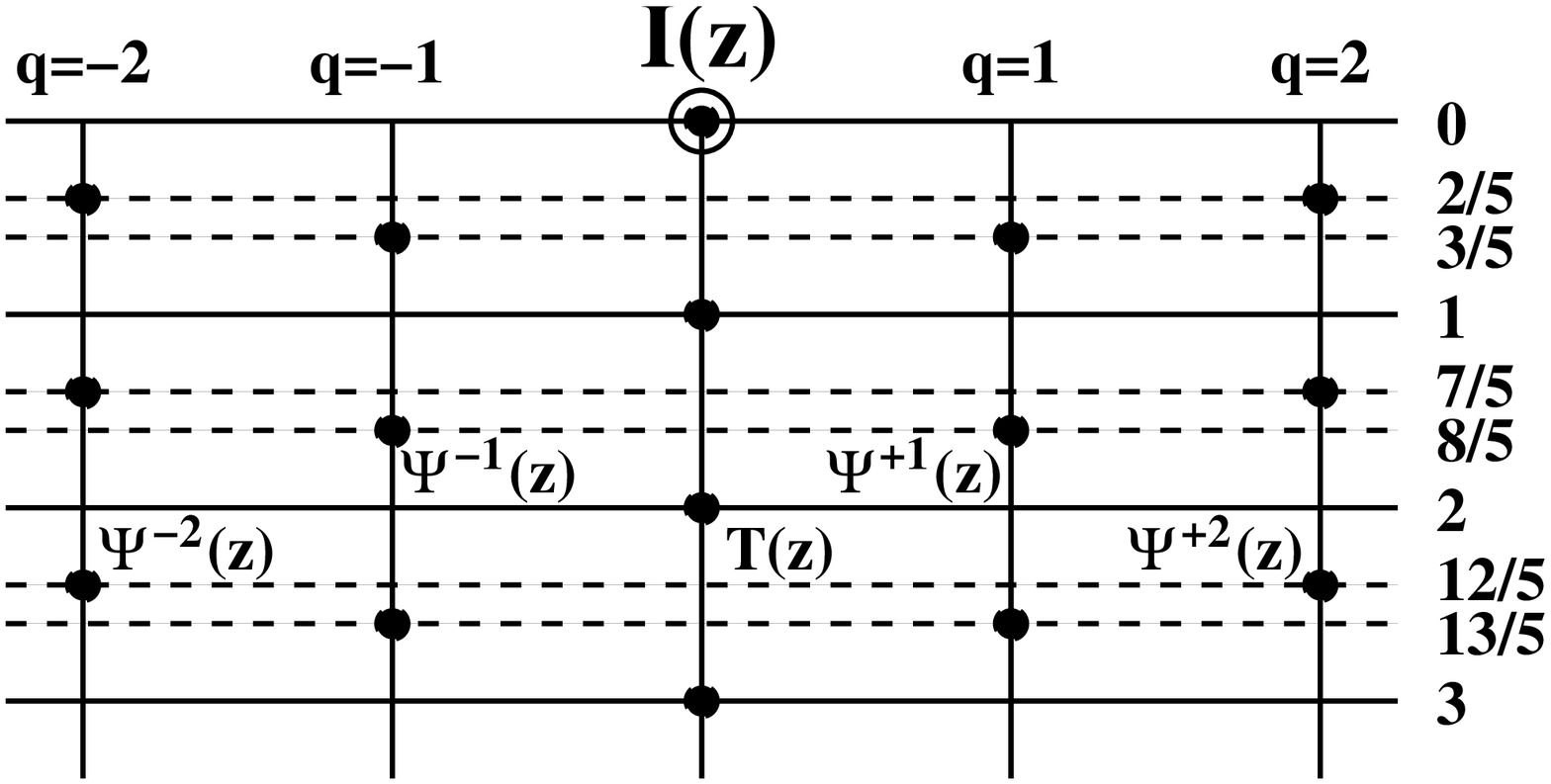}}
 \end{center}
 \protect\caption[3]{\label{fig3}Completion of the identity module. Each
 state is shown as a filled circle.}
\end{figure}

Next, the level spacing within each charge sector (along a given vertical 
line
on the figure) should be equal to 1, as the monodromy of the $\Psi$ fields 
is
abelian. This is so with respect to the identity operator, and, more
generally, with respect to the singlet and doublet operators forming the 
usual
representations of the abelian group $Z_5$. So we can complete the levels 
in
the identity module as shown in Fig.~\ref{fig3}. The states below $I$ and
above $\Psi^{\pm 1}$, on levels 2/5, 3/5, 1, and 7/5 are actually empty.
They will however become occupied when the identity operator $I$, placed
at the summit of the module, is replaced
by a more general singlet state, $\Phi^{0}$.

\begin{figure}
\begin{center}
 \leavevmode
 \epsfysize=150pt{\epsffile{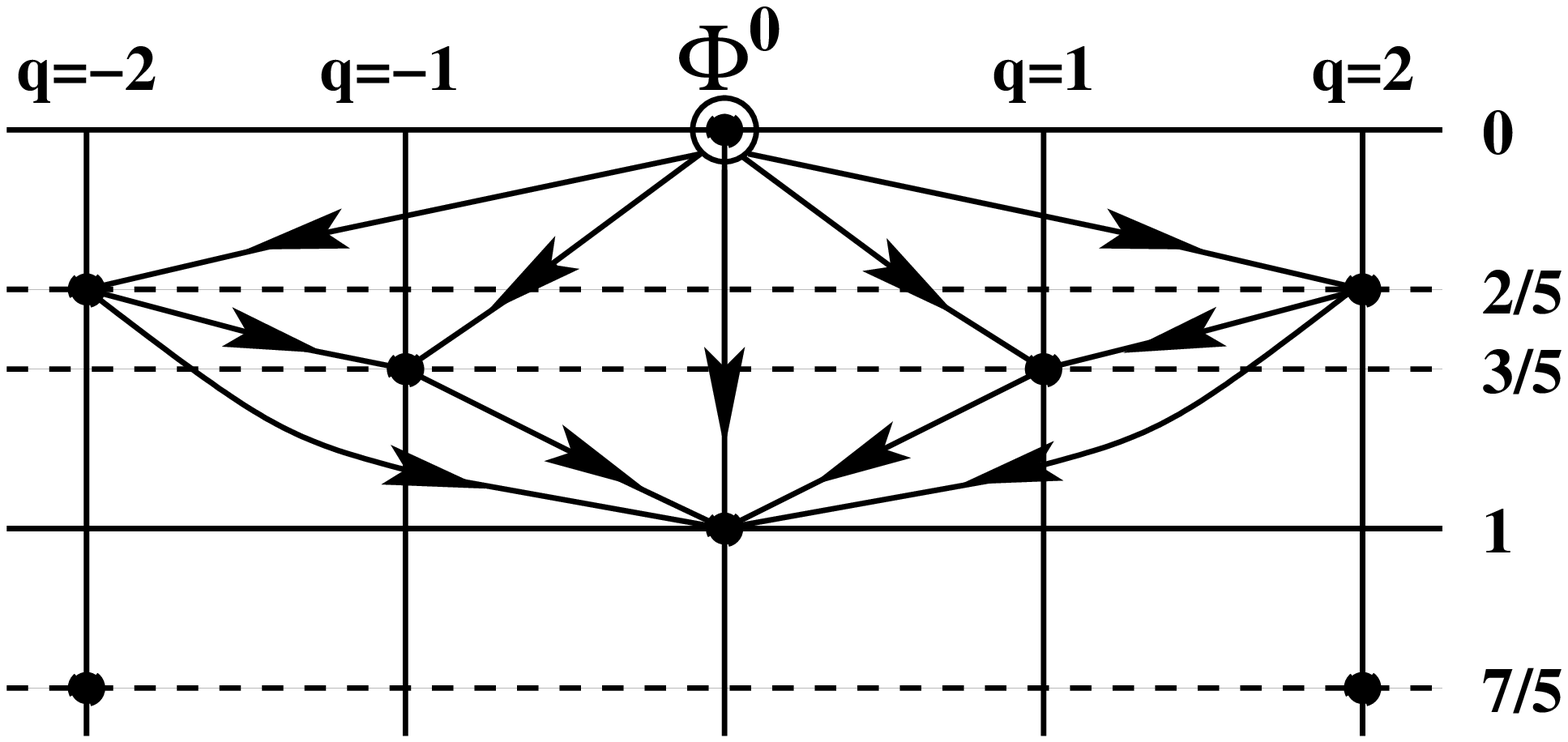}}
 \end{center}
 \protect\caption[3]{\label{fig4}Representation module of a singlet 
operator.
 Arrows depict (some of) the possible actions by the parafermionic mode
 operators (with an appropriate change of the $Z_5$ charge $q$) and by
 the Virasoro generators (with no change in $q$).}
\end{figure}

\begin{figure}
\begin{center}
 \leavevmode
 \epsfysize=150pt{\epsffile{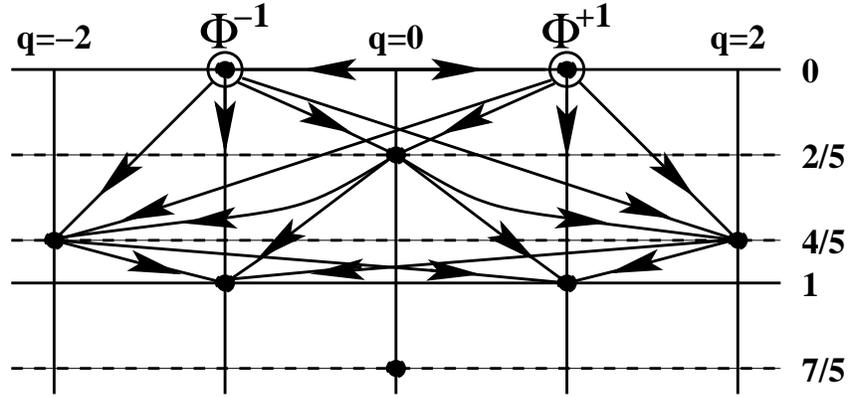}}
 \end{center}
 \protect\caption[3]{\label{fig5}Representation module of a doublet of
 operators of charge $q=\pm 1$, henceforth referred to as doublet 1.
 Note the existence of a zero mode linking the two states at the summit of
 the diagram.}
\end{figure}

\begin{figure}
\begin{center}
 \leavevmode
 \epsfysize=135pt{\epsffile{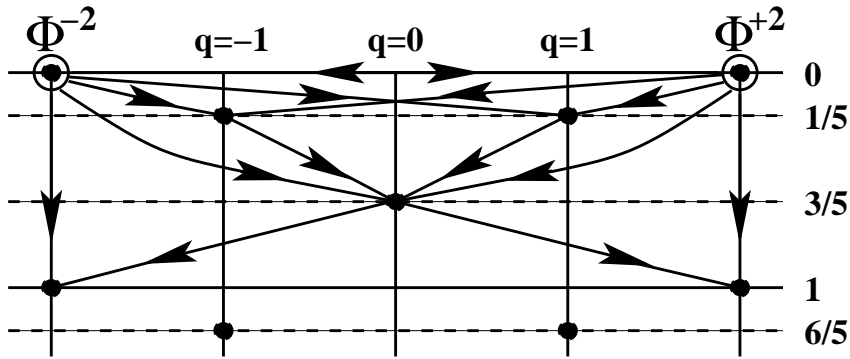}}
 \end{center}
 \protect\caption[3]{\label{fig6}Representation module of a doublet 2
 operator. The two states at the summit are linked by a zero mode.}
\end{figure}

We can now read off from Fig.~\ref{fig3} the level structure of the
doublet 1 (with $\Phi^{\pm 1}$ at the summit) and the doublet 2
(with $\Phi^{\pm 2}$ at the summit) modules by inspecting their 
corresponding
submodules in Fig.~\ref{fig3}. For instance, for the doublet 1 we should
consider the two states in Fig.~\ref{fig3} on level 3/5 as being at the 
summit,
i.e., by imposing the levels above these states to be empty. Then,
everything has to be shifted upwards by 3/5, putting these states at
level 0. Proceeding in this way one obtains the level structure of the 
modules
for a singlet, a doublet 1, and a doublet 2 operator, as shown in
Figs.~\ref{fig4}, \ref{fig5} and \ref{fig6}.
The consistency of these manipulations is ensured by the associativity
of the algebra of $\Psi$ fields.

In accordance with the structure of the modules, as depicted in
Figs.~\ref{fig4}--\ref{fig6}, the local developments of the chiral fields
$\Psi$ in the basis of the representation fields, $\Phi^{0}$, $\Phi^{\pm 
1}$
and $\Phi^{\pm 2}$, should be of the forms given below.


In the singlet basis one has:
\bea
 \Psi^{\pm 1}(z)\Phi^{0}(0) &=& \sum_{n}\frac{1}{(z)^{\Delta_{1}-\frac
 {3}{5}+n}}A^{\pm 1}_{-\frac{3}{5}+n}\Phi^{0}(0),\nn\\
 A^{\pm 1}_{-\frac{3}{5}+n}\Phi^{0}(0) &=& 0,\quad n>0 \\
 \Psi^{\pm 2}(z)\Phi^{0}(0) &=& \sum_{n}\frac{1}{(z)^{\Delta_{2}-\frac
 {2}{5}+n}}A^{\pm 2}_{-\frac{2}{5}+n}\Phi^{0}(0), \nn\\
 A^{\pm 2}_{-\frac{2}{5}+n}\Phi^{0}(0) &=& 0,\quad n>0.
\eea
The fact that positive-index mode operators annihilate $\Psi^0(0)$ is
the usual highest-weight condition, expressing the fact that we are
trying to make the representation fields primary with respect to the
chiral algebra. Conversely, one gets:
\bea
 A^{\pm 1}_{-\frac{3}{5}+n}\Phi^{0}(0) &=& \frac{1}{2\pi i}\oint_{C_{0}}
 {\rm d}z \, (z)^{\Delta_{1}-\frac{3}{5}+n-1}\Psi^{\pm 1}(z)\Phi^{0}(0),
 \label{mode10} \\
 A^{\pm 2}_{-\frac{2}{5}+n}\Phi^{0}(0) &=& \frac{1}{2\pi i}\oint_{C_{0}}
 {\rm d}z \, (z)^{\Delta_{2}-\frac{2}{5}+n-1}\Psi^{\pm 2}(z)\Phi^{0}(0).
 \label{mode20}
\eea
The arrows in Figs.~\ref{fig4}--\ref{fig6} represent the actions of the
mode operators $A^{\pm 1}_{\mu}$ and $A^{\pm 2}_{\mu}$.
For the sake of clarity not all possible arrows are shown.
In particular, we do not show the upwards arrows which would represent the
action on a descendent state. The ``gaps'' in the modules, i.e., the
level of the first descendents, show up in the fractional shifts
of the indices of the operators $A^{\pm 1}_{\mu}$ and $A^{\pm 2}_{\mu}$
(e.g., $\mu=-3/5+n$ in Eq.~(\ref{mode10}), etc.).


In a similar way, for the doublet 1 one gets:
\bea
 \Psi^{\pm 1}(z)\Phi^{\mp 1}(0) &=& \sum_{n}\frac{1}{(z)^{\Delta_{1}-
 \frac{2}{5}+n}}A^{\pm 1}_{-\frac{2}{5}+n}\Phi^{\mp 1}(0), \label{mode11-} 
\\
 \Psi^{\pm 1}(z)\Phi^{\pm 1}(0) &=& \sum_{n}\frac{1}{(z)^{\Delta_{1}-\frac
 {4}{5}+n}}A^{\pm 1}_{-\frac{4}{5}+n}\Phi^{\pm 1}(0), \label{mode11+} \\
 \Psi^{\pm 2}(z)\Phi^{\mp 1}(0) &=& \sum_{n}\frac{1}{(z)^{\Delta_{2}+n}}
 A^{\pm 2}_{n}\Phi^{\mp 1}(0), \label{mode21-} \\
 \Psi^{\pm 2}(z)\Phi^{\pm 1}(0) &=& \sum_{n}\frac{1}{(z)^{\Delta_{2}-
 \frac{4}{5}+n}}A^{\pm 2}_{-\frac{4}{5}+n}\Phi^{\pm 1}(0). \label{mode21+}
\eea

We remark that the action of $\Psi^{\pm 2}(z)$ on $\Phi^{\mp 1}(0)$
in Eq.~(\ref{mode21-}) involves a zero mode $A^{\pm 2}_{0}$, which acts 
between the states $\Phi^{\pm 1}(0)$ at the summit:
\beq
 A^{\mp 2}_{0}\Phi^{\pm 1}(0)=h_{2,1}\Phi^{\mp 1}(0).
 \label{eig1}
\eeq
The index of the eigenvalue $h_{2,1}$ refers to the indices of
$A$ and $\Phi$. Another zero mode appears in the action
of $\Psi^{\pm 1}(z)$ on $\Phi^{\pm 2}(0)$:
\beq
 A^{\pm 1}_{0}\Phi^{\pm 2}(0)=h_{1,2}\Phi^{\mp 2}(0).
 \label{eig2}
\eeq
(The corresponding expansion, in the basis of $\Phi^{\pm 2}(0)$, 
is given below.)
The eigenvalues $h$ in Eqs.~(\ref{eig1})--(\ref{eig2}) characterise the 
representations, in addition to the values of the conformal dimensions
of the fields $\Phi^{\pm 1}$, $\Phi^{\pm 2}$ (which are the eigenvalues
of the Virasoro zero mode $L_0$). 

The relations which are the inverse of Eqs.~(\ref{mode11-})--(\ref{mode21+})
read:
\bea
 A^{\pm 1}_{-\frac{2}{5}+n}\Phi^{\mp 1}(0) &=& \frac{1}{2\pi i}\oint_{C_0}
 {\rm d}z \, (z)^{\Delta_{1}-\frac{3}{5}+n-1}\Psi^{\pm 1}(z)\Phi^{\mp 
1}(0),
 \label{mode1-1} \\
 A^{\pm 1}_{-\frac{4}{5}+n}\Phi^{\pm 1}(0) &=& \frac{1}{2\pi i}\oint_{C_0}
 {\rm d}z \, (z)^{\Delta_{1}-\frac{4}{5}+n-1}\Psi^{\pm 1}(z)\Phi^{\pm 
1}(0), \\
 A^{\pm 2}_{n}\Phi^{\mp 1}(0) &=& \frac{1}{2\pi i}\oint_{C_0}
 {\rm d}z \, (z)^{\Delta_{2}+n-1}\Psi^{\pm 2}(z)\Phi^{\mp 1}(0), \\
 A^{\pm 2}_{-\frac{4}{5}+n}\Phi^{\pm 1}(0) &=& \frac{1}{2\pi i}\oint_{C_0}
 {\rm d}z \, (z)^{\Delta_{2}-\frac{4}{5}+n-1}\Psi^{\pm 2}(z)\Phi^{\pm 
1}(0).
 \label{mode21}
\eea


Finally, for the doublet 2 one gets the expansions:
\bea
 \Psi^{\pm 1}(z)\Phi^{\mp 2}(0) &=& \sum_{n}\frac{1}{(z)^{\Delta_{1}-\frac
 {1}{5}+n}}A^{\pm 1}_{-\frac{1}{5}+n}\Phi^{\mp 2}(0), \\
 \Psi^{\pm 1}(z)\Phi^{\pm 2}(0) &=& \sum_{n}\frac{1}{(z)^{\Delta_{1}
 +n}}A^{\pm 1}_{n}\Phi^{\pm 2}(0), \\
 \Psi^{\pm 2}(z)\Phi^{\mp 2}(0) &=& \sum_{n}\frac{1}{(z)^{\Delta_{2}-\frac
 {3}{5}+n}}A^{\pm 2}_{-\frac{3}{5}+n}\Phi^{\mp 2}(0), \\
 \Psi^{\pm 2}(z)\Phi^{\pm 2}(0) &=& \sum_{n}\frac{1}{(z)^{\Delta_{2}-\frac
 {1}{5}+n}}A^{\pm 2}_{-\frac{1}{5}+n}\Phi^{\pm 2}(0),
\eea
the inverse relations being:
\bea
 A^{\pm 1}_{-\frac{1}{5}+n}\Phi^{\mp 2}(0) &=& \frac{1}{2\pi i}\oint_{C_0}
 {\rm d}z \, (z)^{\Delta_{1}-\frac{1}{5}+n-1}\Psi^{\pm 1}(z)\Phi^{\mp 
2}(0),
 \label{mode1-2} \\
 A^{\pm 1}_{n}\Phi^{\pm 2}(0) &=& \frac{1}{2\pi i}\oint_{C_0}
 {\rm d}z \, (z)^{\Delta_{1}+n-1}\Psi^{\pm 1}(z)\Phi^{\pm 2}(0), \\
 A^{\pm 2}_{-\frac{3}{5}+n}\Phi^{\mp 2}(0) &=& \frac{1}{2\pi i}\oint_{C_0}
 {\rm d}z \, (z)^{\Delta_{2}-\frac{3}{5}+n-1}\Psi^{\pm 2}(z)\Phi^{\mp 
2}(0), \\
 A^{\pm 2}_{-\frac{1}{5}+n}\Phi^{\pm 2}(0) &=& \frac{1}{2\pi i}\oint_{C_0}
 {\rm d}z \, (z)^{\Delta_{2}-\frac{1}{5}+n-1}\Psi^{\pm 2}(z)\Phi^{\pm 
2}(0).
 \label{mode22}
\eea

\subsection{Degeneracy of the modules}

Having defined the mode operators, $A^{\pm 1}_{\mu}$ and $A^{\pm 
2}_{\mu}$,
we now turn to the problem of degeneracies in the modules.

To analyse the degeneracies, we shall need to compute various matrix
elements of the mode operators. To this end, we shall make extensive
use of the commutation relations of the mode operators, in the
various sectors. For convenience, we
have listed the complete set of commutation relations in Appendix~\ref{appA}:
these are all obtained in a way analogous to that used in
Refs.~[1,3]. We also exemplify in Appendix~\ref{appA}
the computation of
some of the matrix elements needed in the subsequent analysis.

\subsubsection{Doublet 1}
\label{sec:D1}

For the doublet 1 module, depicted on Fig.~\ref{fig5}, we begin by 
imposing
degeneracy at level $2/5$. Forming the linear combination%
\footnote{We adopt the convention of denoting singular vectors as
$\chi^q_{-\mu}$, where $q$ denotes the $Z_5$ charge, and $\mu$ is the
level.}
\beq
 \chi^{0}_{-\frac{2}{5}}= a A^{1}_{-\frac{2}{5}}\Phi^{-1}+b A^{-1}_{-\frac
 {2}{5}}\Phi^{1},
 \label{singD1a}
\eeq
we wish to make it into a primary operator, i.e., to ensure that it is
annihilated upon action by positive index mode operators.
In this case it will be sufficient to verify that
\beq
 A^{1}_{+\frac{2}{5}}\chi^{0}_{-\frac{2}{5}}=0\quad\mbox{and}\quad
 A^{-1}_{+\frac{2}{5}}\chi^{0}_{-\frac{2}{5}}=0.
 \label{condD1a}
\eeq 
The others arrows going upwards in Fig.~\ref{fig5} will be zero for 
trivial
reasons.

The conditions (\ref{condD1a}) for the state (\ref{singD1a}) can be 
rewritten as:
\bea
 aA^{1}_{\frac{2}{5}}A^{1}_{-\frac{2}{5}}\Phi^{-1}
 +bA^{1}_{\frac{2}{5}}A^{-1}_{-\frac{2}{5}}\Phi^{1} &=& 0, \nn\\
 aA^{-1}_{\frac{2}{5}}A^{1}_{-\frac{2}{5}}\Phi^{-1}
 +bA^{-1}_{\frac{2}{5}}A^{-1}_{-\frac{2}{5}}\Phi^{1} &=&0,
 \label{condD1a1}
\eea
where, obviously, $A^{1}_{\frac{2}{5}}A^{1}_{-\frac{2}{5}}\Phi^{-1}
\propto\Phi^{1}$ etc. It is convenient to define matrix elements $\mu$
by
\bea
 A^{1}_{\frac{2}{5}}A^{1}_{-\frac{2}{5}}\Phi^{-1} &=& 
\mu^{(1,1;-1)}_{\frac
 {2}{5},-\frac{2}{5}}\Phi^{1}, \\
 A^{1}_{\frac{2}{5}}A^{-1}_{-\frac{2}{5}}\Phi^{1} &=& 
\mu^{(1,-1;1)}_{\frac
 {2}{5},-\frac{2}{5}}\Phi^{1}, \\
 A^{-1}_{\frac{2}{5}}A^{1}_{-\frac{2}{5}}\Phi^{-1} &=& 
\mu^{(-1,1;-1)}_{\frac
 {2}{5},-\frac{2}{5}}\Phi^{-1}, \\
 A^{-1}_{\frac{2}{5}}A^{-1}_{-\frac{2}{5}}\Phi^{1} &=& 
\mu^{(-1,-1;1)}_{\frac
 {2}{5},-\frac{2}{5}}\Phi^{-1}.
\eea
By the $Z_2$ reflection symmetry $(q \leftrightarrow -q)$ we have
\bea
 \mu^{(1,1;-1)} &=& \mu^{(-1,-1;1)}, \\
 \mu^{(1,-1;1)} &=& \mu^{(-1,1;-1)}.
\eea

To compute the remaining two matrix elements in Eq.~(\ref{condD1a1}) we
need the commutation relations of the mode operators $A^{\pm 1}_{\mu}$.
In Appendix~\ref{appA} it is shown that
\bea
 \mu^{(1,1;-1)}_{\frac{2}{5},-\frac{2}{5}} &=& \lambda^{1,1}_{2}h_{2,1}, 
\\
 \mu^{(1,-1;1)}_{\frac{2}{5},-\frac{2}{5}} &=& -\frac{2}{25}+\frac
 {2\Delta_{1}}{c}\Delta_{\Phi^{1}}.
\eea
Here, $\lambda^{1,1}_{2}$ is one of the structure constants (\ref{struct}) 
of
the chiral algebra (\ref{ope1}) and $h_{2,1}$ is the eigenvalue of
$\Phi^{\mp 1}$ with respect to the zero mode $A^{\pm 2}$, as defined in
Eq.~(\ref{eig1}). Finally, $\Delta_{1}=8/5$ is the dimension of $\Psi^{\pm 
1}$,
and $\Delta_{\Phi^{1}}$ is the as yet unknown dimension of the operator
$\Phi^{\pm 1}$.

A non-trivial solution of the system (\ref{condD1a1}) exists provided that
\beq
 \left| \begin{array}{cc} \mu^{(1,1;-1)} & \mu^{(1,-1;1)}\\
 \mu^{(-1,1;-1)} & \mu^{(-1,-1,1)}\end{array}\right|=0,
 \label{detD1}
\eeq
from which one obtains:
\beq
 \left( \lambda^{1,1}_{2}h_{2,1} \right)^2 =
 \left( -\frac{2}{25}+\frac{2}{5c} \Delta_{\Phi^{1}} \right)^2.
 \label{fixeig1}
\eeq
We recall that, for a given operator $\Phi^{1}$ or $\Phi^{-1}$, we need
to determine both the eigenvalue $h_{2,1}$ and the dimension
$\Delta_{\Phi^{1}}$. By imposing degeneracy at level 2/5, we have obtained
Eq.~(\ref{fixeig1}), which defines $h_{2,1}$ as a function of
$\Delta_{\Phi^{1}}$ and $c$. (Note also that the structure constant
$\lambda^{1,1}_{2}$  is a function of $c$,
by Eqs.~(\ref{struct})--(\ref{cparamv}).)

In conclusion, degeneracy at level 2/5 is not sufficient to define
also $\Delta_{\Phi^{1}}$ (as a function of $c$). We have two unknowns,
$h_{2,1}$ and $\Delta_{\Phi^{1}}$. So we need one more constraint, in
addition to that in Eq.~(\ref{fixeig1}).

We shall therefore require that the doublet 1 module be degenerate also
at level 4/5. This is obtained by demanding the degeneracy of the state
\beq
 \chi^{2}_{-\frac{4}{5}}=\tilde{a}A^{1}_{-\frac{4}{5}}\Phi^{1}+\tilde{b}
 A^{-2}_{-\frac{4}{5}}\Phi^{-1}.
 \label{singD1b}
\eeq
The reason for considering a linear combination of two states only is
that the ``indirect'' descendent at level 4/5, obtained by descending
through the state which remains at level 2/5 after imposing the
degeneracy constraint at that level, is linearly dependent on the
direct descendents. More precisely, by using the mode operator algebra
given in Appendix~\ref{appA}, one finds (in Appendix~\ref{appA}) that
\beq
 A^{2}_{-\frac{2}{5}}A^{1}_{-\frac{2}{5}}\Phi^{-1}=
 A^{2}_{-\frac{2}{5}}A^{-1}_{-\frac{2}{5}}\Phi^{1}
 \label{lindep1}
\eeq
 is linearly dependent on the states
\beq
 A^{1}_{-\frac{4}{5}}\Phi^{1}\quad\mbox{and}\quad
 A^{-2}_{\frac{4}{5}}\Phi^{-1}.
\eeq
(In Eq.~(\ref{lindep1}) we have used the linear dependence resulting from
the degeneracy at level 2/5.)

The degeneracy of the state (\ref{singD1b}) takes the form
\beq
 \left| \begin{array}{cc} \mu^{(-1,1;1)}_{\frac{4}{5},-\frac{4}{5}} 
 & \mu^{(-1,-2;-1)}_{\frac{4}{5},-\frac{4}{5}}\\
 \mu^{(2,1;1)}_{\frac{4}{5},-\frac{4}{5}} 
 & \mu^{(2,-2,-1)}_{\frac{4}{5},-\frac{4}{5}}\end{array}\right|=0,
 \label{condD1b}
\eeq
involving the following matrix elements (see Appendix~\ref{appA} for details):
\bea
 A^{-1}_{\frac{4}{5}}A^{1}_{-\frac{4}{5}}\Phi^{1} &=& \left( \frac{3}{25}+
 \frac{2\Delta_{1}}{c}\Delta_{\Phi^{1}} \right) \Phi^{1}\equiv
 \mu^{(-1,1;1)}_{\frac{4}{5},-\frac{4}{5}}\Phi^{1}, \label{matrx1} \\
 A^{-1}_{\frac{4}{5}}A^{2}_{-\frac{4}{5}}\Phi^{1} &=& \lambda^{21}_{-2}
 h_{2,1}\Phi^{1}\equiv\mu^{(-1,-2;-1)}_{\frac{4}{5},-\frac{4}{5}}\Phi^{1}, 
\\
 A^{2}_{\frac{4}{5}}A^{1}_{-\frac{4}{5}}\Phi^{1} &=& \lambda^{21}_{-2}
 h_{2,1}\Phi^{1}\equiv\mu^{(-2,1;1)}_{\frac{4}{5},-\frac{4}{5}}\Phi^{1}, 
\\
 A^{2}_{\frac{4}{5}}A^{-2}_{-\frac{4}{5}}\Phi^{1} &=&
 \left( \frac{4}{(\lambda^{1,1}_{2})^{2}} \left[ \frac{4}{5}y^{2}-
 \frac{2\Delta_{1}}{c}(\Delta_{\Phi^{1}}+\frac{2}{5})y \right] +
 \frac{9}{5(\lambda^{1,1}_{2})^{2}}y^{2}+\frac{3}{2}y \right)
 \Phi^{-1} \nn \\
 &\equiv& \mu^{(2,-2;-1)}_{\frac{4}{5},-\frac{4}{5}}\Phi^{-1}, 
\label{matrx4}
\eea
where
\beq
 y=\lambda^{1,1}_{2}h_{2,1}= -\frac{2}{25}+\frac{2\Delta_{1}}{c}\Delta_
 {\Phi^{1}}.
\eeq

The solutions of Eq.~(\ref{condD1b}), with the ingredients
(\ref{matrx1})--(\ref{matrx4}) as well as Eq.~(\ref{fixeig1}),
have a rather simple form, in spite of its appearances.
They can be presented as:
\bea
 \Delta^{(1)}_{\Phi^{1}} &=& \frac{3}{5} \, \frac{p+5}{p},\quad
 \Delta^{(2)}_{\Phi^{1}}  =  \frac{3}{5} \, \frac{p-3}{p+2}, 
\label{soluD1} \\
 \Delta^{(3)}_{\Phi^{1}} &=& \frac{1}{10} \, \frac{p^2+2p-15}{p(p+2)},
\eea
where the parameter $p$ has been defined in Eq.~(\ref{paramp}). For $N=5$,
\beq
 p=3+k=3+\frac{v}{2}
 \label{soluD1par}
\eeq
where $v$ is the parameter in the solution (\ref{ope1})-(\ref{chidim2})
of the chiral algebra; we recall from Eq.~(\ref{vkparam}) that $v=2k$.

By introducing the parameters
\beq
 \alpha^{2}_{+}=\frac{p+2}{p},\quad \alpha^{2}_{-}=\frac{p}{p+2}
 \label{CGparam}
\eeq
which are suggestive from the formula for the central charge 
(\ref{cparamp}),
the solutions for the dimensions in Eqs.~(\ref{soluD1})--(\ref{soluD1par}) take
the form
\bea
 \Delta^{(1)}_{\Phi^{1}} &=& \frac{3}{2}\alpha^{2}_{+}-\frac{9}{10},
 \label{soluD1a} \qquad
 \Delta^{(2)}_{\Phi^{1}}  =  \frac{3}{2}\alpha^{2}_{-}-\frac{9}{10}, \\
 \Delta^{(3)}_{\Phi^{1}} &=& -\frac{3}{8}(\alpha^{2}_{+}+\alpha^{2}_{-})
 +\frac{17}{20}. \label{soluD1c}
\eea
We shall defer to the next section further discussion on the significance
of the solutions (\ref{soluD1a})--(\ref{soluD1c}).

We state once again that in the $Z_{5}$ theory the modules have to be
doubly degenerate, on two levels simultaneously, to reach the goal that
the dimension of the primary field gets fixed as a function of $c$.

\subsubsection{Doublet 2}

The module of $\Phi^{\pm 2}$, shown in Fig.~\ref{fig6}, has its first
descendents at level 1/5, with $q=\pm 1$. We could consider one side,
$q=-1$ for instance, because what is constructed for $q=-1$ will also
be confirmed by the algebra for $q=+1$, due to the $Z_{2}$ reflection
symmetry.

At level 1/5 it would appear that there are two states:
\beq
 A^{1}_{-\frac{1}{5}}\Phi^{-2}\quad\mbox{and}\quad A^{2}_{-\frac{1}{5}}
 \Phi^{2}. \label{twostat}
\eeq
However, it turns out that they are proportional to one another.
Indeed, by the commutation relation $\{\Psi^1,\Psi^1\}\Phi^2$ (see 
Appendix~\ref{appA})
with $n=m=0$, one gets:
\bea
 \lambda^{1,1}_{2}A^{2}_{-\frac{1}{5}}\Phi^{2} &=& 2A^{1}_{-\frac{1}{5}}
 A^{1}_{0}\Phi^{2}=2A^{1}_{-\frac{1}{5}}h_{1,2}\Phi^{-2}, \\
 \lambda^{1,1}_{2}A^{2}_{-\frac{1}{5}}\Phi^{2} &=& 2h_{1,2}
 A^{1}_{-\frac{1}{5}}\Phi^{-2}, \label{proprl}
\eea
where the number $h_{1,2}$ is the zero mode eigenvalue, yet to be fixed.

With this simplification, in order to have a degeneracy at level 1/5,
 it suffices to require that the state
\beq
 \chi^{-1}_{-\frac{1}{5}}=A^{1}_{-\frac{1}{5}}\Phi^{-2}
 \label{stateD2}
\eeq
be primary. Also, it is sufficient to require that it be annihilated by
$A^{-1}_{+\frac{1}{5}}$ only, and not by both $A^{-1}_{+\frac{1}{5}}$ and
$A^{2}_{+\frac{1}{5}}$. The algebraic reason for this follows from a
relation analogous to Eq.~(\ref{proprl}), but for positive index mode
operators.

We therefore require that
\beq
 A^{-1}_{\frac{1}{5}}A^{1}_{-\frac{1}{5}}\Phi^{-2}=0,
\eeq
and by means of the commutation relation $\{\Psi^1,\Psi^{-1}\}\Phi^{-2}$
with $n=m=0$
\beq
 \left( A^{1}_{0}A^{-1}_{0}+A^{-1}_{\frac{1}{5}}A^{1}_{-\frac{1}{5}} 
\right)
 \Phi^{-2}=
 \left( -\frac{3}{25}+\frac{2\Delta_{1}}{c}\Delta_{\Phi^{2}} \right) 
\Phi^{-2}
\eeq
this takes the form
\beq
 \left( h_{1,2} \right)^2 =
 -\frac{3}{25}+\frac{2\Delta_{1}}{c}\Delta_{\Phi^{2}}.
 \label{fixeig2}
\eeq
If we impose the condition (\ref{fixeig2}),
the state (\ref{stateD2}) can be put equal to
zero, thus reducing the module. As a consequence of the proportionality
(\ref{proprl}) the second state in Eq.~(\ref{twostat}) will then also vanish.
Thus, after this reduction, level 1/5 will be completely degenerate,
or empty.

Analogous to the case of doublet 1, the constraint (\ref{fixeig2}) only
defines $(h_{1,2})^{2}$ as a function of $\Delta_{\Phi^{2}}$ and $c$.
We therefore turn to level 3/5.

By the commutation relation $\{\Psi^1,\Psi^1\}\Phi^{-2}$ with $n=m=0$ we 
have
\beq
 2A^{1}_{-\frac{2}{5}}A^{1}_{-\frac{1}{5}}\Phi^{-2}=\lambda^{1,1}_{2}
 A^{2}_{-\frac{3}{5}}\Phi^{-2}.
\eeq
Since now $A^{1}_{-\frac{1}{5}}\Phi^{-2}=0$, by the previous degeneracy,
one obtains
\beq
 A^{2}_{-\frac{3}{5}}\Phi^{-2}=0.
\eeq
Similarly, because of $A^{-1}_{-\frac{1}{5}}\Phi^{2}=0$, one gets
\beq
 A^{-2}_{-\frac{3}{5}}\Phi^{2}=0.
\eeq
In conclusion, the complete degeneracy at level 1/5 implies that, after
the reduction consisting in factoring out the degenerate submodule,
level 3/5 is also empty.

To fix $\Delta_{\Phi^{2}}$ we therefore go on to examine the next 
available
level, which is level 1. It is not difficult to verify that at
level 1 we have to consider the state
\beq
 \chi^{-2}_{-1}=a L_{-1}\Phi^{-2}+bA^{1}_{-1}\Phi^{2}
\eeq
and require the following degeneracy conditions to be satisfied:
\beq
 L_{1}\chi^{-2}_{-1}=0,\quad A^{-1}_{1}\chi^{-2}_{-1}=0.
\eeq

In terms of the matrix elements $\mu_{ij}$ defined by
\bea
 L_{1}L_{-1}\Phi^{-2} &=&\mu_{11}\Phi^{-2}, \\
 L_{1}A_{-1}^{1}\Phi^{2} &=& \mu_{12}\Phi^{-2}, \\
 A_{1}^{-1}L_{-1}\Phi^{-2} &=& \mu_{21}\Phi^{2}, \\
 A_{1}^{-1}A_{-1}^{1}\Phi^{2} &=& \mu_{22}\Phi^{2},
\eea
the degeneracy criterion reads
\beq
 \mu_{11}\mu_{22}-\mu_{12}\mu_{21}=0.
\eeq
Using the Virasoro algebra and the commutation relations
$\{\Psi^1,\Psi^{-1}\}\Phi^{-2}$ and $\{T,\Psi\}\Phi$
the required matrix elements are readily computed:
\bea
 \mu_{11} &=& 2\Delta_{\Phi^{2}}, \label{matele11} \\
 \mu_{12} &=& \frac{8}{5}h_{1,2}, \label{matele12} \\
 \mu_{21} &=& \frac{8}{5}h_{1,2}, \\
 \mu_{22} &=& \frac{1}{5}(h_{1,2})^{2}+\frac{7}{25}+\frac{2\Delta_{1}}
 {c}\Delta_{\Phi^{2}},
\eea
and the resulting criterion is
\beq
 2\Delta_{\Phi^{2}} \left( \frac{1}{5}(h_{1,2})^{2}+\frac{7}{25}+
 {2\Delta_{1}}{c}\Delta_{\Phi^{2}} \right) -
 \frac{64}{25} \left( h_{1,2} \right)^2=0.
 \label{soluD2}
\eeq

Solving Eqs.~(\ref{fixeig2}) and (\ref{soluD2}) one gets the following
solutions:
\beq
 \Delta^{(1)}_{\Phi^{2}}=\alpha^{2}_{+}-\frac{3}{5},\quad
 \Delta^{(2)}_{\Phi^{2}}=\alpha^{2}_{-}-\frac{3}{5}
 \label{soluD2a}
\eeq
with $\alpha^{2}_{\pm}$ defined in Eq.~(\ref{CGparam}).

\subsubsection{Singlet}

For the singlet module, depicted in Fig.~\ref{fig4}, the first descendents
are at level $2/5$ in the sector $q = \pm 2$. By charge conjugation 
symmetry
it is sufficient to consider one of them. One may thus define the 
state
\beq
 \chi_{-\frac25}^2 = A^2_{-\frac25} \Phi^0,
\eeq
subject to the constraint
\beq
 A^{-2}_{\frac25} \chi_{-\frac25}^2 = 0. \label{singconst}
\eeq
Using the commutation relation $\{\Psi^{2},\Psi^{-2}\}\Phi^{0}$ given in
Appendix~\ref{appA}, with $m=-1,\,\,n=0$, one obtains:
\beq
\left( A^{2}_{\frac{7}{5}}A^{-2}_{-\frac{7}{5}}+A^{-2}_{\frac{2}{5}}
A^{2}_{-\frac{2}{5}} \right) \Phi^{0}=\frac{2\Delta_{2}}{c}\Delta_{\Phi^{0}}
\Phi^{0}.
\eeq
We observe that the constraint (\ref{singconst}) has the consequence of fixing
a matrix element at a lower level:
\beq
 A^{2}_{\frac{7}{5}}A^{-2}_{-\frac{7}{5}} \Phi^{0}=
 \frac{2\Delta_{2}}{c}\Delta_{\Phi^{0}}\Phi^{0}. \label{lowerlevel}
\eeq
Accepting Eqs.~(\ref{singconst}) and (\ref{lowerlevel})
we can eliminate the only state on level 2/5. We thus put
\beq
A^{2}_{-\frac{2}{5}}\Phi^{0}=0. \label{weput}
\eeq

We next demand degeneracy at level $3/5$ in the sector $q = \pm 1$. 
Once the level 2/5 is empty, there will be
only a single state on level 3/5, up to the $Z_2$ symmetry
$q \leftrightarrow -q$. We therefore set
\beq
 \chi_{-\frac35}^1 = A^1_{-\frac35} \Phi^0, \label{weset}
\eeq
subject to the constraint
\beq
 A^{-1}_{\frac35} \chi_{-\frac35}^1 = 0. \label{constr22}
\eeq
Using the commutation relation $\{\Psi^{-1},\Psi^{1}\}\Phi^{0}$ in
Appendix~\ref{appA}, with $n=1,\,\,m=0$, one obtains:
\beq
A^{1}_{\frac{3}{5}}A^{-1}_{-\frac{3}{5}} \Phi^{0}=
\frac{2\Delta_{1}}{c}\Delta_{\Phi^{0}}\Phi^{0}.
\eeq
The constraint (\ref{constr22}) requires that
\beq
\Delta_{\Phi^{0}} = 0. \label{dimidop}
\eeq
We have thus found the (trivial) scaling dimension of the identity
operator.

\subsection{Sector of the disorder operator $R_{a}$}

\subsubsection{Structure of the disorder modules}

The disorder operator (quintuplet) $R_{a}(z,\bar{z}), a=1,2,3,4,5$,
has non-abelian monodromy with respect to the chiral fields
$\Psi^{\pm 1}(z), \Psi^{\pm 2}(z)$.
This amounts to the decomposition of the local products $\Psi^{+1}(z)
R_{a}(0)$ and $\Psi^{+2}(z)R_{a}(0)$ into half-integer powers of $z$:
\bea
 \Psi^{+1}(z)R_{a}(0) &=& \sum_{n}\frac{1}{(z)^{\Delta_{1}+\frac{n}{2}}}
 A^{1}_{\frac{n}{2}}R_{a}(0), \label{mode1R} \\
 \Psi^{+2}(z)R_{a}(0) &=& \sum_{n}\frac{1}{(z)^{\Delta_{2}+\frac{n}{2}}}
 A^{2}_{\frac{n}{2}}R_{a}(0). \label{mode2R}
\eea
Because of the non-abelian monodromy of the disorder fields, the expansion
of the products $\Psi^{-1}(z)R_{a}(0)$ and $\Psi^{-2}(z)R_{a}(0)$ are 
related
to those in Eqs.~(\ref{mode1R})--(\ref{mode2R}), by an analytic continuation of
$z$ around 0 on both sides of these two equations.
One finds:
\bea
 \Psi^{-1}(z)R_{a}(0) &=& 
\sum_{n}\frac{(-1)^{n}}{(z)^{\Delta_{1}+\frac{n}{2}}}
 A^{1}_{\frac{n}{2}} \, {\sf U}R_{a}(0), \\
 \Psi^{-2}(z)R_{a}(0) &=& 
\sum_{n}\frac{(-1)^{n}}{(z)^{\Delta_{2}+\frac{n}{2}}}
 A^{2}_{\frac{n}{2}} \, ({\sf U})^2 R_{a}(0). \label{mode2R-}
\eea
Here ${\sf U}$ is a $5 \times 5$ matrix which rotates the index of the
disorder field backwards by one unit: ${\sf U} R_{a}(0) = R_{a-1}(0)$.
(Note that in Eq.~(\ref{mode2R-}) we have denoted the square of ${\sf U}$ by
$({\sf U})^2$ rather than by ${\sf U}^2$. This notation is intended to
avoid confusion with upper indices of operators which are abundant in our
presentation.)

The theory of disorder operators has been fully developed in
Ref.~[6] in the context of the first parafermionic conformal field
theory (with symmetry $Z_N$), and in Ref.~[3] in the context of the
second parafermionic theory (with symmetry $Z_3$).
As far as the general properties (products, analytic continuations)
of the disorder sector operators are concerned, the approach of
Refs.~[3,6] generalises directly to the present case.
In Appendix~\ref{appB} we present some details of the disorder sector which are
specific to the second $Z_5$ theory (the one treated in this paper).
In particular, the developments (\ref{mode1R})--(\ref{mode2R-}) are
justified and the commutation relations of the mode operators 
$A^{1}_{\frac{n}{2}}$ and $A^{2}_{\frac{n}{2}}$ in the disorder
sector are also given.

As compared to Figs.~\ref{fig4}--\ref{fig6}, the level structure of the 
modules of disorder operators is relatively simple. There are only integer and 
half-integer levels, and there exists zero modes for all the operators
$\{ \Psi^{q} \}$ acting on the quintuplet of disorder operators.
For the zero modes one should have, in general:
\bea
 A^{1}_{0}R_{a} &=& h_{1} \, ({\sf U})^{2}R_{a}, \label{Rzero1} \\
 A^{2}_{0}R_{a} &=& h_{2} \, ({\sf U})^{-1}R_{a}. \label{Rzero2}
\eea
The matrices $({\sf U})^2$ and $({\sf U})^{-1}$ turn the indices in 
accordance
with the multiplication rules of $Z_{5}$: we refer to Appendix~\ref{appB} for 
details.
The constants $h_{1}$ and $h_{2}$ are the eigenvalues of $R_{a}$ with 
respect
to $A^{1}_{0}$ and $A^{2}_{0}$. As in the case of the doublets, these
eigenvalues furnish an additional characterisation of the disorder
representations, in addition to the conformal dimension of $R_{a}$.
Actually,
it suffices to specify one of the eigenvalues, $h_1$ or $h_2$, since
the two are generally related, irrespective of the details of the
representation of a particular operator $R_{a}$. This can be seen by
applying the commutation relation (\ref{C25}) with $n=m=0$. We obtain
\beq
 2A^{1}_{0}A^{1}_{0}R_{a}(0)=\lambda^{1,1}_{2} 2^{\Delta_{2}-3}A^{2}_{0}
 R_{a}(0)+2^{-\Delta_{2}-2}
 \left( \kappa(0)+\frac{16\Delta_{1}}{c}\Delta_{R} \right)
 ({\sf U})^{-1}R_{a}(0),
\eeq
where $\kappa(n)$ is defined in Eq.~(\ref{C26}).
In particular, $\kappa(0)=-11/10$.
Taking into account Eqs.~(\ref{Rzero1})--(\ref{Rzero2}) one obtains
\beq
 2(h_{1})^{2}=\lambda^{1,1}_{2} 2^{\Delta_{2}-3}h_{2}
 +2^{-\Delta_{2}-2}
 \left( \kappa(0)+\frac{16\Delta_{1}}{c}\Delta_{R} \right). 
\label{relh1h2}
\eeq
Thus, for a given operator $R_{a}$, if $h_{1}$ is known then $h_{2}$ is
fixed by this equation.

\subsubsection{Degeneracy of the disorder modules}

As discussed above, the modules of the disorder operators consist of
integer and half-integer levels. The first descendents
of a given primary operator $R_{a}$ will therefore be found at level 1/2.
For a given value of the index $a$, there exist two states:
\bea
 \left( \chi^{(1)}_{a} \right)_{-\frac{1}{2}} &\equiv&
 A^{1}_{-\frac{1}{2}} ({\sf U})^{-2}R_{a}=
 ({\sf U})^{-2} A^{1}_{\frac{1}{2}}R_{a}, \label{Rstate1} \\
 \left( \chi^{(2)}_{a} \right)_{-\frac{1}{2}} &\equiv&
 A^{2}_{-\frac{1}{2}} {\sf U}R_{a}= {\sf U}A^{2}_{\frac{1}{2}}R_{a}.
 \label{Rstate2}
\eea
The matrices $({\sf U})^{-2}$ and ${\sf U}$ commute with the action of the
mode operators, $A^{1}_{-\frac{1}{2}}$ and $A^{2}_{-\frac{1}{2}}$.
They just compensate for the rotation of indices of the $R_{a}$ which are 
produced
by the action of $A^{1}_{-\frac{1}{2}}$ and $A^{2}_{-\frac{1}{2}}$.

In analogy with the computations for the doublet 1 operator (see above),
one could now try to produce one primary state at level 1/2, by making
a linear combination of the states (\ref{Rstate1})--(\ref{Rstate2}).
We shall, however, not follow this route. Instead, we shall require in
this case that both states, (\ref{Rstate1}) and (\ref{Rstate2}),
be primaries. Thus, the two of them will be eliminated eventually,
and there will be a complete degeneracy at level 1/2.

It should be remarked that in this process of looking for degeneracies 
at the lowest levels in the modules there is a certain choice, and the
way of performing the computations is not uniquely defined from the 
outset.
For some of the choices, the solutions found (and the formulae for the
dimensions in particular) will have to be classified as non-physical,
not matching with the theory that we are constructing. In general, the 
number of possible solutions is bigger then the number of physical operators
to be determined, and there is some redundancy. For instance,
in the case of the doublet 1 module, if we had required that both states 
at
level 2/5 be primaries, we would have found a non-physical value for the
conformal dimension of the doublet, a value which would have to be 
discarded.
Also, among the three solutions that we eventually found for the case of
doublet 1, see Eqs.~(\ref{soluD1a})--(\ref{soluD1c}), the first two will
turn out to be physical whilst the third one is not.

The distinction between physical and non-physical solutions shall be
done in the next section. All the solutions that we are getting at
present will serve in the next section to fix the form of the Kac
formula for dimensions and to fill the Kac table, positioning the
singlets, doublets and disorder operators correctly. In this analysis
some of the present solutions for $\Delta$ will turn out to be
inconsistent and will have be discarded.

Going back to the disorder operator $R_{a}$, the reason that we have 
decided to annihilate both states at level 1/2, Eqs.~(\ref{Rstate1})
and (\ref{Rstate2}), is that this yields physical solutions, consistent
with the eventual Kac formula. So in this particular case, the two
states which must be required to be degenerate in order
to fix both $h$ and $\Delta$ are situated on the {\em same} level.
This is not so in general. But for the fundamental disorder operator, with 
the
two required degeneracies situated at the first descendent levels, these
levels are both equal to 1/2.

In accordance with these remarks, we shall require that
\beq
 A^{1}_{\frac{1}{2}} \left( \chi^{(1)}_{a} \right)_{-\frac{1}{2}}=0,\quad
 A^{2}_{\frac{1}{2}} \left( \chi^{(1)}_{a} \right)_{-\frac{1}{2}}=0,
\eeq
\beq
 A^{1}_{\frac{1}{2}} \left( \chi^{(2)}_{a} \right)_{-\frac{1}{2}}=0,\quad
 A^{2}_{\frac{1}{2}} \left( \chi^{(2)}_{a} \right)_{-\frac{1}{2}}=0.
\eeq
This gives:
\beq
 A^{1}_{\frac{1}{2}}A^{1}_{-\frac{1}{2}}R_{a}=0,\quad
 A^{2}_{\frac{1}{2}}A^{1}_{-\frac{1}{2}}R_{a}=0, \label{assume1}
\eeq
\beq
 A^{1}_{\frac{1}{2}}A^{2}_{-\frac{1}{2}}R_{a}=0,\quad
 A^{2}_{\frac{1}{2}}A^{2}_{-\frac{1}{2}}R_{a}=0. \label{assume2}
\eeq
{}From the algebra (\ref{C25}), with $n=1, m=-1$, one then obtains:
\beq
 \left( 
A^{1}_{\frac{1}{2}}A^{1}_{-\frac{1}{2}}+\frac{2}{5}A^{1}_{0}A^{1}_{0}
 \right) R_a =
 \lambda^{1,1}_{2} 2^{\Delta_{2}-3}A_{0}^{2}R_{a}
 -2^{-\Delta_{2}-2} \left( \kappa(1)+\frac{16\Delta_{1}}{c}\Delta_{R} 
 \right)
 ({\sf U})^{-1}R_{a}. \label{disorderrel1}
\eeq
Imposing now that $A^{1}_{+\frac{1}{2}}A^{1}_{-\frac{1}{2}}R_{a}=0$,
and replacing  $A^{1}_{0},A^{2}_{0}$ by the corresponding eigenvalues,
Eqs.~(\ref{Rzero1})--(\ref{Rzero2}), we find the relation
\beq
 \frac{2}{5}(h_{1})^{2}=\lambda^{1,1}_{2} 2^{\Delta_{2}-3}h_{2}-
 2^{-\Delta_{2}-2} \left( \kappa(1)+\frac{16\Delta_{1}}{c}\Delta_{R} 
\right).
 \label{constr1}
\eeq
Here $\kappa(1)=-1/10$, from Eq.~(\ref{C26}).

Next, using the commutation relation (\ref{C28}), first with $n=1$, $m=-1$ and
then with $n=-1$, $m=1$, it is found that
\beq
 A^{1}_{\frac{1}{2}}A^{2}_{-\frac{1}{2}}R_{a}=
 A^{2}_{\frac{1}{2}}A^{1}_{-\frac{1}{2}}R_{a}=
 \left(-\frac{4}{5}h_{1}h_{2}+ 2^{\Delta_{2}-\Delta_{1}-2}\lambda^{2,1}_{-2}
 h_{2}- 2^{\Delta_{1}-\Delta_{2}-2}\lambda^{-2,1}_{-1}h_{1} \right)
 {\sf U}R_{a}. \label{disorderrel2}
\eeq
Using again Eqs.~(\ref{assume1})--(\ref{assume2}),
this gives a second relation:
\beq
 -\frac{4}{5}h_{1}h_{2}+ 2^{\Delta_{2}-\Delta_{1}-2}\lambda^{21}_{-2}
 h_{2}- 2^{\Delta_{1}-\Delta_{2}-2}\lambda^{-2,1}_{-1}h_{1}=0.
 \label{constr2}
\eeq
Finally, combining the commutation relations (\ref{C25}) and (\ref{C28}),
after some algebra it is found that
\bea
 A^{2}_{\frac{1}{2}}A^{2}_{-\frac{1}{2}}R_{a} &=&
 \frac{2^{-\Delta_{2}+3}}{\lambda^{1,1}_{-2}}
 \left[ \left(-\frac{4}{5}h_{2}+\frac{1}{4}(2)^{\Delta_{1}-\Delta_{2}-2}
 \lambda^{2,1}_{-1} \right) A^{1}_{\frac{1}{2}}A^{1}_{-\frac{1}{2}}
 ({\sf U})^{-1}R_{a} \right. \nn \\
 &+& \left. \left( 
\frac{12}{5}h_{1}-\frac{1}{3}(2)^{\Delta_{2}-\Delta_{1}-2}
 \lambda^{2,1}_{-2} \right) A^{1}_{\frac{1}{2}}A^{2}_{-\frac{1}{2}}
 ({\sf U})^{2} R_{a} \right]. \label{disorderrel3}
\eea
As a result, if the matrix elements $ A^{1}_{\frac{1}{2}}
A^{1}_{-\frac{1}{2}}R_{a}$ and 
$A^{1}_{\frac{1}{2}}A^{2}_{-\frac{1}{2}}R_{a}$
vanish, in this case all the matrix elements in
Eqs.~(\ref{assume1})--(\ref{assume2}) will vanish, as required.

Let us summarise the situation. The constraints expressing the
complete degeneracy at level 1/2 in the module of a disorder operator
$R_{a}$ are given by Eqs.~(\ref{constr1}) and (\ref{constr2}). In
addition, the zero mode eigenvalues $h_{1}$ and $h_{2}$ are related by
Eq.~(\ref{relh1h2}).  This set of three equations allows us to express
$h_{1}, h_{2}$ and $\Delta_{R}$ as functions of $c$, or as functions
of the related parameters, $p$ or $\alpha^{2}_{+}$, $\alpha^{2}_{-}$
defined in Eqs.~(\ref{soluD1par})--(\ref{CGparam}).

Solving Eqs.~(\ref{relh1h2}), (\ref{constr1}) and (\ref{constr2}) one 
finds
the following solutions for the conformal dimension $\Delta_{R}$:
\bea
 \Delta^{(1)}_{R} &=& \frac{1}{4}\frac{p+5}{p}, \qquad
 \Delta^{(2)}_{R}  =  \frac{1}{4}\frac{p-3}{p+2}, \\
 \Delta^{(3)}_{R} &=& \frac{1}{4}\frac{p^{2}+2p-5}{p(p+2)}.
\eea
Expressed in terms of $\alpha^{2}_{+}=(p+2)/p$ and 
$\alpha^{2}_{-}=p/(p+2)$,
these solutions take the form:
\bea
 \Delta^{(1)}_{R} &=& \frac{5}{8}\alpha^{2}_{+}-\frac{3}{8}, \qquad
 \Delta^{(2)}_{R}  =  \frac{5}{8}\alpha^{2}_{-}-\frac{3}{8}, 
\label{soluR1} \\
 \Delta^{(3)}_{R} &=& -\frac{5}{16}(\alpha^{2}_{+}+\alpha^{2}_{-})+
 \frac{7}{8}. \label{soluR3}
\eea

\section{Kac formula}
\label{sec3}

In Eq.~(\ref{SOcoset}) it was observed [5] that the central charge 
of
the second parafermionic theory corresponds to that of a coset model based 
on
the group $SO(N)$. We therefore consider it natural to look for a Kac 
formula
based on the weight lattice of the Lie algebra $B_{\frac{N-1}{2}}$ for 
$N$ odd, and $D_{\frac{N}{2}}$ for $N$ even. In the present case ($N=5$) 
the
relevant algebra will be $B_{2}$.

With this assumption the general form of the Kac formula will be taken as:
\beq
 \Delta_{(n_{1},n_{2})(n'_{1},n'_{2})}=
 \left( \vec{\beta}-\vec{\alpha}_{0} \right)^2
 -\vec{\alpha}_{0}^{2}+B,
 \label{Kac}
\eeq
where
\beq
 \vec{\beta}\equiv\vec{\beta}_{(n_{1},n_{2})(n'_{1},n'_{2})}=
 \left( \frac{1-n_{1}}{2}\alpha_{+}+\frac{1-n'_{1}}{2}\alpha_{-} \right)
 \vec{\omega}_{1} + \left( \frac{1-n_{2}}{2}\alpha_{+}+
 \frac{1-n'_{2}}{2}\alpha_{-} \right) \vec{\omega}_{2}. \label{betavec}
\eeq
Here $\vec{\omega}_{1},\vec{\omega}_{2}$ are the fundamental weights
for the representations of the algebra $B_{2}$, and
\bea
 \alpha_{+} &=& \sqrt{\frac{p+2}{p}},\qquad 
\alpha_{-}=-\sqrt{\frac{p}{p+2}}
 \label{alpha+-} \\
 \vec{\alpha}_{0} &=& \frac{\alpha_{+}+\alpha_{-}}{2}(\vec{\omega}_{1}+
 \vec{\omega}_{2}).
\eea
$B$ in Eq.~(\ref{Kac}) is the ``boundary term'', which will be zero for
singlets and will have to be defined for the doublet 1, doublet 2 and 
disorder
sectors (see below).

By assuming the operators to be associated with positions on a particular
lattice, we are implicitly assuming the existence of a Coulomb gas 
realisation
of the theory, with primary operators represented by vertex operators. 
This
explains the Coulomb gas form of Eq.~(\ref{Kac}) giving the dimensions
$\Delta_{(n_{1},n_{2})(n'_{1},n'_{2})}$.

The formula for the dimensions can be made more explicit
in different ways. For the present purpose of identifying the solutions 
for dimensions which has been found in the previous sections we shall
find it convenient to present it in the following form:
\bea 
 \Delta_{(n_{1},n_{2})(n'_{1},n'_{2})} &=&
 \frac{1}{4}(\vec{\omega}_{1},\vec{\omega}_{1})
 \left[ (n^{2}_{1}-1)\alpha^{2}_{+}+(n'^{2}_{1}-1)\alpha^{2}_{-} -
 2(n_{1}n'_{1}-1) \right] \nn \\
 &+& \frac{1}{4}(\vec{\omega}_{2},\vec{\omega}_{2})
 \left[ (n^{2}_{2}-1)\alpha^{2}_{+}+(n'^{2}_{2}-1)\alpha^{2}_{-} -
 2(n_{2}n'_{2}-1) \right] \label{Kac1} \\
 &+& \frac{1}{2}(\vec{\omega}_{1},\vec{\omega}_{2})
 \left[ (n_{1}n_{2}-1)\alpha^{2}_{+}+(n'_{1}n'_{2}-1)\alpha^{2}_{-} -
 (n_{1}n'_{2} +n'_{1}n_{2}-2) \right] + B. \nn
\eea

In the preceding sections we have directly computed the dimensions of
various operators, having modules which are degenerate on the first
descendent levels. The results are given in 
Eqs.~(\ref{soluD1a})--(\ref{soluD1c})
for case of doublet 1 operators, in Eq.~(\ref{soluD2a}) for the doublet 2, in
Eq.~(\ref{dimidop}) for the singlet, and finally in
Eqs.~(\ref{soluR1})--(\ref{soluR3})
for the disorder operator. Since these solutions of the degeneracy problem
are the simplest possible, the corresponding dimensions have to be 
looked for in the lower part (small positive $n$, $n'$) of
the table of dimensions, Eq.~(\ref{Kac1}).

First one should try to identify the operators by the coefficients of
$\alpha^{2}_{+}$ and $\alpha^{2}_{-}$. And if a consistent identification 
is
found, the next step will be to deduce the boundary terms from the
$\alpha^{2}_{\pm}$ independent terms in the formulae for the dimensions.

Proceeding in this way we have found that the dimensions (\ref{soluD1a}) 
for
the doublet 1 correspond, respectively, to
\beq
 \Delta_{(1,3)(1,1)}\quad\mbox{and}\quad \Delta_{(1,1)(1,3)}.
\eeq
The dimensions (\ref{soluD2a}) for the doublet 2 are those of
\beq
 \Delta_{(2,1)(1,1)}\quad\mbox{and}\quad \Delta_{(1,1)(2,1)},
\eeq
and the singlet dimension (\ref{dimidop}) is simply
\beq
 \Delta_{(1,1)(1,1)}.
\eeq
Finally, the dimensions (\ref{soluR1}) for the disorder operator
correspond to
\beq
 \Delta_{(1,2)(1,1)}\quad\mbox{and}\quad \Delta_{(1,1)(1,2)}.
\eeq

The values of the boundary terms are found to be the following:
\beq
 B_{S}=0,\quad B_{R}=\frac{1}{8},\quad B_{D^{1}}=\frac{1}{10},\quad
 B_{D^{2}}=\frac{3}{20}.
 \label{boundary}
\eeq
Here $S$, $R$, $D^{1}$ and $D^{2}$ denote, respectively, singlet, disorder
operator, doublet 1, and doublet 2.

The values of the scalar products $(\vec{\omega}_{a},\vec{\omega}_{b})$
which are verified by this identification procedure are the following:
\beq
 (\vec{\omega}_{1},\vec{\omega}_{1})=1,\qquad
 (\vec{\omega}_{2},\vec{\omega}_{2})=\frac{1}{2},\qquad
 (\vec{\omega}_{1},\vec{\omega}_{2})=\frac{1}{2}.
\eeq
They agree with the scalar products for the fundamental weights of the
algebra $B_{2}$. 

The remaining conformal dimensions found in the previons Section, namely
Eq.~(\ref{soluD1c}) for the doublet 1 and Eq.~(\ref{soluR3}) for the 
disorder
operator, do not fit the Kac formula (\ref{Kac1}). More precisely, the
dimension (\ref{soluD1c}) corresponds to an operator which is outside the
physical domaine of the Kac table.
This physical domain shall be defined in
the next section. And the dimension (\ref{soluR3}) cannot be obtained from 
the
formula (\ref{Kac1}) for integer values of the indices $n,n'$. We shall
therefore consider both of these dimensions as non-physical solutions of 
the
degeneracy problem. Speaking differently, the operators with these last
dimensions will not appear in the operator products of operators belonging 
to
the physical domain of the Kac table. For this reason they can be dropped.

\section{General theory}
\label{sec4}

In Section~\ref{sec3} we have fixed the positions in the Kac table
of the physical operators found by the direct degeneracy computations
of Section~\ref{sec2}.
Having identified these first few operators the problem is to fill out the
rest of the Kac table. In particular, we have to decide in general which
positions in the table should be occupied by
singlet, doublet 1, doublet 2 and disorder
operators, apart from the special cases already determined in the 
preceding
two sections. This problem amounts to assigning correctly the boundary 
terms (\ref{boundary}) to the Kac formula [see Eqs.~(\ref{Kac}) and 
(\ref{Kac1})].
More precisely, we should determine for which values of the indices
$(n_1,n_2)(n_1',n_2')$ the boundary term $B$ equals $B_S$, $B_R$,
$B_{D^1}$ or $B_{D^2}$ respectively.

\subsection{Labeling the sites of the Kac table}

As has already been said in the previous section, the general form
of the formula for the dimensions, Eq.~(\ref{Kac}), implies the existence
of a Coulomb gas representation for the present theory. We are not giving 
in this paper the corresponding explicit representations of the
operators. However, we shall use its geometrical significance, 
given in the form of
certain ``allowed moves'' on the lattice--Kac table. These moves
are assumed to be related to screening operators, which are the
vectors of simple roots of the algebra $B_{2}$. We remind that it
is on this algebra that the Kac formula [cf.~Eqs.~(\ref{Kac}) and 
(\ref{Kac1})]
is being built.

\begin{figure}
\begin{center}
 \leavevmode
 \epsfysize=350pt{\epsffile{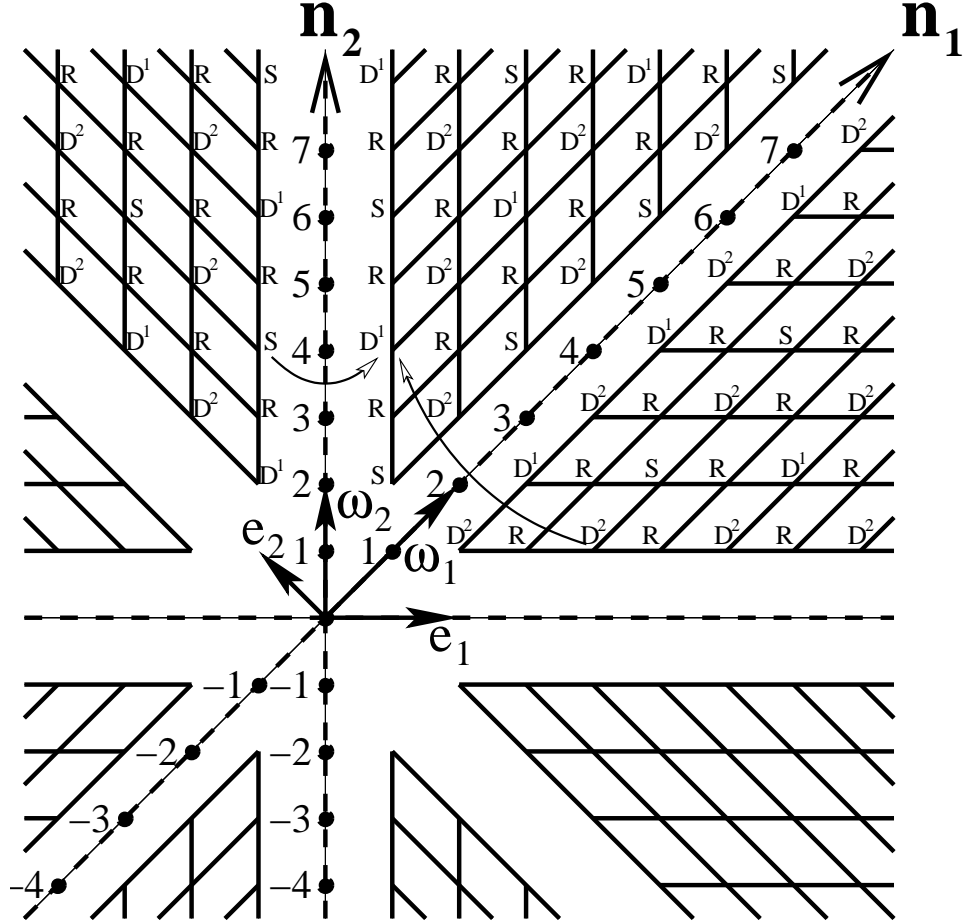}}
 \end{center}
 \protect\caption[3]{\label{fig7}Kac table of the $Z_5$ theory, based on 
the
 weight lattice of $B_2$. The lattice is spanned by (one half of) the
 fundamental weights $\vec{\omega}_{1}$, $\vec{\omega}_{2}$.
 We show here the ``basic layer'', i.e., the one
 with $(n_1',n_2')=(1,1)$. Each vertex of the lattice is labeled according
 to the nature of the associated operator ($S$, $D^1$, $D^2$, $R$).
 This labeling extends periodically throughout the lattice.
 The simple roots, $\vec{e}_{1}$ and $\vec{e}_{2}$,
 act as screening operators and are associated with lattice reflections
 (see text). The arrows give examples of pairs of operators which
 are linked by such reflections. The physical part of the spectrum is
 restricted to the wedge limited by the $n_1$ and $n_2$ axes.}
\end{figure}

The Kac table of the theory $Z_{5}$ is shown in Fig.~\ref{fig7}, together 
with
the vectors $\vec{e}_{1}$, $\vec{e}_{2}$ representing the screening 
operators
of the associated (assumed) Coulomb gas representation. The labeling of 
each
vertex of the lattice encodes the nature of the corresponding operator:
singlet ($S$), doublet 1 ($D^1$), doublet 2 ($D^2$) and disorder operator
($R$). The reasoning for the particular pattern shown in the figure will 
be
given below.

It should be observed that the Kac table of the theory $Z_{5}$ is in fact
four-dimensional, made of two-dimensional layers corresponding to the
$\alpha_{+}$ and $\alpha_{-}$ parts of Eq.~(\ref{Kac}). This separation on
the $\alpha_{+}$ and $\alpha_{-}$ parts is seen more explicitly from the
following form of the Kac formula:
\beq
 \Delta_{(n_{1},n_{2})(n'_{1},n'_{2})}=
 \left[ \left( \frac{n_{1}}{2}\vec{\omega}_{1} +
 \frac{n_{2}}{2}\vec{\omega}_{2} \right) \alpha_{+} +
 \left( \frac{n'_{1}}{2}\vec{\omega}_{1} +
 \frac{n'_{2}}{2}\vec{\omega}_{2} \right) \alpha_{-} \right]^{2}
 -\vec{\alpha}^{2}_{0}+B. \label{Kac2}
\eeq
In Fig.~\ref{fig7} only one layer is displayed, namely the one with either
$(n_1',n_2')=(1,1)$ or $(n_1,n_2)=(1,1)$. We shall refer
to the layers $(n_{1},n_{2})(1,1)$ or $(1,1)(n'_{1},n'_{2})$ as the
``basic layer'' of the Kac table, corresponding to either the
$\alpha_{+}$ or the $\alpha_{-}$ part of the spectrum. Such a layer is
basic in the sense that the other side has trivial indices.

Also shown on Fig.~\ref{fig7} are examples of reflections, marked
by arrows pointing from the domains on the left and on the right
into the physical domain, which is the upper-right corner of the
table (the wedge limited by the $n_1$ and $n_2$ axes). Just like in
Felder's resolution for minimal (Virasoro algebra) models [8],
these reflections, realised by integrated screenings,
indicate singular states (degeneracies) in the modules of physical
operators. In this sense the operators in the adjacent domains are
all ``ghosts'', which decouple from the physical operators in the operator
algebra, and in correlation functions. As usual, for the above reasoning
based on reflections to be justified, we must assume that the screenings
commute with the parafermionic algebra.

We can check these assumptions by looking at the low-lying
operators in the basic layer of the Kac table, cf.~Fig.~\ref{fig7}.
For the first operators of each type---$R$ (disorder), $D^{2}$ (doublet 
2),
$D^{1}$ (doublet 1) and $S$ (singlet)---the levels of degeneracy have been
defined in Section~\ref{sec2} by direct calculation. 
Alternatively, the levels of degeneracy could be found, if our assumptions
are correct, by taking the difference of the dimensions
$\Delta_{(n_{1}, n_{2})(n'_{1},n'_{2})}$, as given by Eq.~(\ref{Kac1}), 
for
two operators at the ends of the arrow representing a particular 
reflection in
Fig.~\ref{fig7}. 

For the operator $R$, positioned as $\Phi_{(1,2)(1,1)}$ or
$\Phi_{(1,1)(1,2)}$, both reflections gives the result $1/2$ for the
difference of the dimensions:
\bea
 \Delta_{(1,1)(-1,4)}-\Delta_{(1,1)(1,2)} &=& \frac{1}{2}, \label{diffdim1} \\
 \Delta_{(1,1)(3,-2)}-\Delta_{(1,1)(1,2)} &=& \frac{1}{2}.
\eea
This agrees with the finding of Section~\ref{sec2} that these disorder
operators are doubly degenerate at level $1/2$.

Next, for the operator $D^{1}$, at the location $\Phi_{(1,1)(1,3)}$, one 
finds
\bea
 \Delta_{(1,1)(-1,5)}-\Delta_{(1,1)(1,3)} &=& \frac{2}{5}, \\
 \Delta_{(1,1)(4,-3)}-\Delta_{(1,1)(1,3)} &=& \frac{4}{5}, \label{diffdim4}
\eea
according to Fig.~\ref{fig7}. This is in agreement with the fact that this
operator is degenerate at levels $2/5$ and $4/5$, as we have found in
Section~\ref{sec2}.

In a similar way, the two reflections for the operator $D^{2}$, located as
$\Phi_{(1,1)(2,1)}$, confirm the degeneracy at levels $1/5$ and $1$, as 
found
in Section~\ref{sec2}. And finally, in the case of the identity operator
$\Phi_{(1,1)(1,1)}=I$, which is of the $S$ type, the two reflections in
Fig.~\ref{fig7} give the differences of dimensions $2/5$ and $3/5$, in
accordance with the levels of degeneracy in the identity operator module 
as
dressed in Section~\ref{sec2}.

It is important to note that the labeling of
the ghost operator, as being $R$, $S$, $D^{1}$ or $D^{2}$ for every 
reflection, has to be in accordance with the nature
of the corresponding state in the module of the physical operator into
which the ghost operator is mapped. For instance, in the module of a
$D^1$ operator, cf.~Fig.~\ref{fig5}, a degenerate state at level $2/5$ is
a singlet, because that level belongs to the charge sector $q=0$. Likewise,
a doublet of degenerate states at level $4/5$ (charge sector $q=\pm2$)
corresponds to a doublet 2 operator.
Evidently, the differences of dimensions, as found in
Eqs.~(\ref{diffdim1})--(\ref{diffdim4}), are to be calculated with the
appropriate boundary terms [identified in Eq.~(\ref{boundary})] in the Kac
formula for $\Delta_{(n_{1},n_{2})(n'_{1}n'_{2})}$, cf.~Eq.~(\ref{Kac1}).
 
The disorder operators have to map among themselves, assuming
that the screening operators which realise the mappings (reflections) do 
not translate order to disorder.

So far we have identified the nature of the following operators, situated
at the low-lying corner of the physical domain shown in the table
of Fig.~\ref{fig7}:
\bea
 \Phi_{(1,1)(1,1)} &=& S=I, \\
 \Phi_{(1,1)(2,1)} &=& D^{2}, \\
 \Phi_{(1,1)(1,2)} &=& R, \\
 \Phi_{(1,1)(1,3)} &=& D^{1}.
\eea
Similar identifications apply on the side of $\alpha_{+}$, e.g.,
$\Phi_{(2,1)(1,1)}=D^{2}$ etc. We now show how to fill in the rest
of the basic layer, on the $\alpha_{-}$ side for instance, by making
use of simple fusion rules and reflections.

We first consider the operator product
\beq
 R \cdot D^{1}=\Phi_{(1,1)(1,2)}\cdot\Phi_{(1,1)(1,3)}.
 \label{opprod1}
\eeq
On the left-hand side, the result of the multiplication should be a 
disorder
operator. 

On the right-hand side, according to the Coulomb gas rules, the above
multiplication produces an operator, in the principal channel, with
\beq
 \vec{\beta}_{(1,1)(1,4)}=\vec{\beta}_{(1,1)(1,2)}+\vec{\beta}_{(1,1)(1,3)}.
 \label{vecadd1}
\eeq
Multiplication of
operators corresponds to addition of the corresponding vectors
$\vec{\beta}$ in Fig.~\ref{fig7}. The non-principal channels follow 
the principal one by shifts realised by the vectors
$-\vec{e}_{1}$ and $-\vec{e}_{2}$. But for the present purposes it
will be sufficient to follow the principal channel only.

At this point we should remark on a sign convention concerning the
orientation of the vectors in the Coulomb gas representation.
According to Eq.~(\ref{betavec}), the vectors $\vec{\beta}$ 
have a negative projection along the
fundamental weight vectors $\vec{\omega}_1$ and $\vec{\omega}_2$.
The usual convention in the Colomb gas representation analysis is to
change their sign, i.e., to orient them positively, as seen on 
Fig.~\ref{fig7}.
To compensate for this convention, the moves associated with the
screenings should then be realised by shifts in the direction {\em 
opposite}
to that of the simple roots, i.e., in the direction of the vectors
$-\vec{e}_{1}$ and $-\vec{e}_{2}$.
Without this convention for the inversion of directions,
in the graphical representation of Fig.~\ref{fig7}, the Kac table
should have been spanned by the vectors $-\vec{\omega}_{1}/2$ and
$-\vec{\omega}_{2}/2$, according to Eq.~(\ref{betavec}).
With this convention in mind, the graphical addition
[with respect to the origin $(1,1)(1,1)$] of the
lattice vectors in Fig.~\ref{fig7}, corresponding to the operators
$\Phi_{(1,1)(1,2)}$ and $\Phi_{(1,1)(1,3)}$ in Eq.~(\ref{opprod1}), is in
accordance with the additions of the vectors $\vec{\beta}$ in
Eq.~(\ref{vecadd1}).

As a result, Eq.~(\ref{opprod1}) leads us to conclude that the site
$(1,1)(1,4)$ of the Kac table is occupied by an operator of type $R$
(disorder).

Next, we consider multiplying the operator $D^{1}=\Phi_{(1,1)(1,3)}$
with itself:
\beq
 D^{1} \cdot D^{1}=\Phi_{(1,1)(1,3)}\cdot\Phi_{(1,1)(1,3)}. 
\label{opprod2}
\eeq
The right-hand side produces, in the principal channel, the operator
$\Phi_{(1,1)(1,5)}$. According to the left-hand side, this operator
has to be either of the $D^{2}$ or the $S$ type.
The result $D^{2}$ is obtained if both $D^{1}$ operators of the left-hand
side of Eq.~(\ref{opprod2}) belong to the $q=+1$ part of the doublet
(or if both have $q=-1$). However, if the $Z_5$ charge of the two
$D^{1}$ operators are opposite (one being $q=+1$ and the other $q=-1$),
the left-hand side produces a singlet ($S$) operator.

This ambiguity is due to the fact that the site in the Kac table
labeled $D^{1}$ is in fact the position of both members of the
doublet 1, viz.~operators with $q=\pm 1$. A similar remark applies to the
case when a $D^2$ operator participates in an operator product.
A site in the Kac table labeled $R$ accommodates the whole
quintuplet $\{R_{a}, a=1,...,5\}$ of disorder operators, all sharing a
particular value of the conformal dimension $\Delta$.

The ambiguity for assigning the correct label to the operator
$\Phi_{(1,1)(1,5)}$ is resolved by adding an argument based on 
reflections.
In fact, one can check that the horizontal reflection in
Fig.~\ref{fig7}---from the ghost site on the left, across the $n_{2}$ 
axis,
and to the site $(1,1)(1,5)$ on the right---has a ``bare gap'' of
\beq
 \Delta^{(0)}_{(1,1)(-1,7)}-\Delta^{(0)}_{(1,1)(1,5)}=\frac{1}{2}.
 \label{baregap}
\eeq
(Here, we have defined the ``bare'' dimensions as
$\Delta_{(n_1,n_2)(n_1',n_2')}-B$, i.e., by Eq.~(\ref{Kac1}) with the
boundary term being neglected.)

Suppose first that $\Phi_{(1,1)(1,5)}$ is a $D^{2}$, so that we shall
have to correct the second term in Eq.~(\ref{baregap}) by the boundary 
term
$B_{D^2} = 3/20$. One now examines in turn the three possible assignments
($D^{1},D^{2}$ or $S$) of a label for the ghost site at $(1,1)(-1,7)$,
each time correcting the first term in Eq.~(\ref{baregap}) by the 
corresponding
boundary term. In case of a consistent assignment, the right-hand side 
must
equal a level on which a ghost operator of the given nature can be a
submodule of the physical $D^2$ operator. According to Fig.~\ref{fig6},
these levels can be: $1/5 +$integer if the ghost is a $D^1$,
$3/5 +$integer if it is a $S$, and $1+$integer if it is a $D^2$.
It is easily verified that none of the three possible assignments
leads to a consistent result. Therefore, $\Phi_{(1,1)(1,5)}$ cannot be
a $D^{2}$ operator.

On the other hand, if $\Phi_{(1,1)(1,5)}$ is a singlet, and the 
calculation
of the gap in Eq.~(\ref{baregap}) is corrected by assuming a (ghost) 
operator
of type $D^{1}$ at the site $(1,1)(-1,7)$, then the corrected gap in
Eq.~(\ref{baregap}) will give $3/5$. It is seen from Fig.~\ref{fig4} that
a singlet operator can indeed accommodate a $D^1$ submodule at level 
$3/5$.
Moreover, $D^1$ is the only consistent labeling of the ghost operator.

The result of the above series of arguments is that the operator
$\Phi_{(1,1)(1,5)}$ is a singlet (and also that the ghost operator
$\Phi_{(1,1)(-1,7)}$ is a doublet 1).

At the other border of the physical domain, which is parallel to
the $n_{1}$ axis, by using a similar series of arguments one
concludes that the operator $\Phi_{(1,1)(3,1)}$, situated next to
$D^{2}=\Phi_{(1,1)(2,1)}$, can only be a singlet.
More precisely, in this case the bare gap for the reflection
across the $n_{1}$ axis reads:
\beq
 \Delta^{(0)}_{(1,1)(4,-1)}-\Delta^{(0)}_{(1,1)(3,1)}=\frac{1}{4}
\eeq
By correcting this formula by boundary terms, it can be checked that the
only labels that can be consistently assigned to the
sites $(1,1)(4,-1)$ and $(1,1)(3,1)$ are those given in Fig.~\ref{fig7}.

Having defined the positions of the first nontrivial singlets, i.e.,
the operators $\Phi_{(1,1)(3,1)}$ and $\Phi_{(1,1)(1,5)}$, we can
now extend the singlets over the whole lattice (the basic layer in
Fig.~\ref{fig7}). To this end, it suffices to multiply the singlets among
themselves. In the products they produce only singlets, and we
shall cover the lattice by adding the corresponding lattice
vectors. Put differently, the lattice vectors corresponding
to the operators $\Phi_{(1,1)(3,1)}$ and $\Phi_{(1,1)(1,5)}$ are
the fundamental ones for the sublattice of singlet operators.

Next, we can extend the labeling $D^{2}$ of the site $(1,1)(2,1)$
out over the entire lattice. This is done by multiplying 
$\Phi_{(1,1)(2,1)}$
by all possible singlets. As the singlet consists of only one
$q=0$ state, the argument based on addition of $Z_5$ charges involves
no ambiguity, and the result can only be another $D^2$ operator.
As a consequence, every fourth row parallel to the $n_{1}$ axis
(viz., the rows $1,5,9,\ldots$) become completely filled out
by an alternation of $S$ and $D^2$ operators; see Fig.~\ref{fig7}.

It is equally easy to extend the positions of disorder operators over
the whole lattice.

First, multiplying the operator $D^{1}=\Phi_{(1,1)(1,3)}$ by the basic
disorder operator $R=\Phi_{(1,1)(1,2)}$ we see that
$\Phi_{(1,1)(1,4)}$ must be another disorder operator.
Furthermore, the operators $\Phi_{(1,1),(2,2)}$ and $\Phi_{(1,1),(2,4)}$
are also disorder operators. This is seen by multiplying the first two
disorder operators, $\Phi_{(1,1)(1,2)}$ and $\Phi_{(1,1)(1,4)}$,
by the basic doublet, $D^2=\Phi_{(1,1)(2,1)}$.

Second, the above four disorder operators can be extended throughout the
lattice by multiplying them by all possible singlets. In all cases, the
result must be a disorder operator. We conclude that every second row
of the physical domain is occupied exclusively by disorder operators,
as shown on Fig.~\ref{fig7}.

At this point, only the rows $3,7,11,\ldots$ in the basic layer
remain to be determined. In fact, once the nature of the operators
$\Phi_{(1,1)(1,3)}$ and $\Phi_{(1,1),(2,3)}$ is determined, these
two operators can be extended throughout the entire lattice by
repeated multiplication by the two fundamental singlets, at positions
$(1,1)(3,1)$ and $(1,1)(1,5)$. The result will be that the rows in
question will realise an alternating sequence of these two operators.

The operator $\Phi_{(1,1)(1,3)}$ is, of course, already known: it is
the fundamental doublet 1 ($D^1$). It remains to define
$\Phi_{(1,1),(2,3)}$. Actually, by assuming the whole lattice---and not
just the physical domain---to be filled out in a homogeneous (periodic)
way, we could equivalently ask for the labeling of $\Phi_{(1,1)(2,-1)}$,
four rows below. Now, by consistency of reflections with the levels in the
module of the identity operator $S=I=\Phi_{(1,1)(1,1)}$, the ghost 
operator
$\Phi_{(1,1)(2,-1)}$ has to be $D^{2}$. It is easy to check that
this accounts for the degeneracy at level 2/5.
The identification of $\Phi_{(1,1)(2,3)}$ as a doublet 2 operator is
also confirmed by a direct degeneracy calculations given in
Appendix~\ref{appD}.

We thus reach the conclusion that the rows $3,7,11,\ldots$ in
the basic layer are occupied by an alternation of $D^1$ and $D^2$ 
operators,
as shown on Fig.~\ref{fig7}.

\underline{Conclusion:} By a series of arguments described above,
where we have combined reflections and simple fusion rules, we
have filled the whole lattice corresponding to the basic layer.
The result is shown in Fig.~\ref{fig7}.

The sites in the other layers, in which the indices of
$\Delta_{(n_{1},n_{2})(n'_{1}n'_{2})}$ are nontrivial on both the 
$\alpha_{+}$
and the $\alpha_{-}$ sides, must also be assigned $S$, $D^{1}$, $D^{2}$ 
and
$R$ labels. We now show how this can be accomplished by shifting the 
labels
assigned to the basic layer.

\begin{figure}
\begin{center}
 \leavevmode
 \epsfysize=350pt{\epsffile{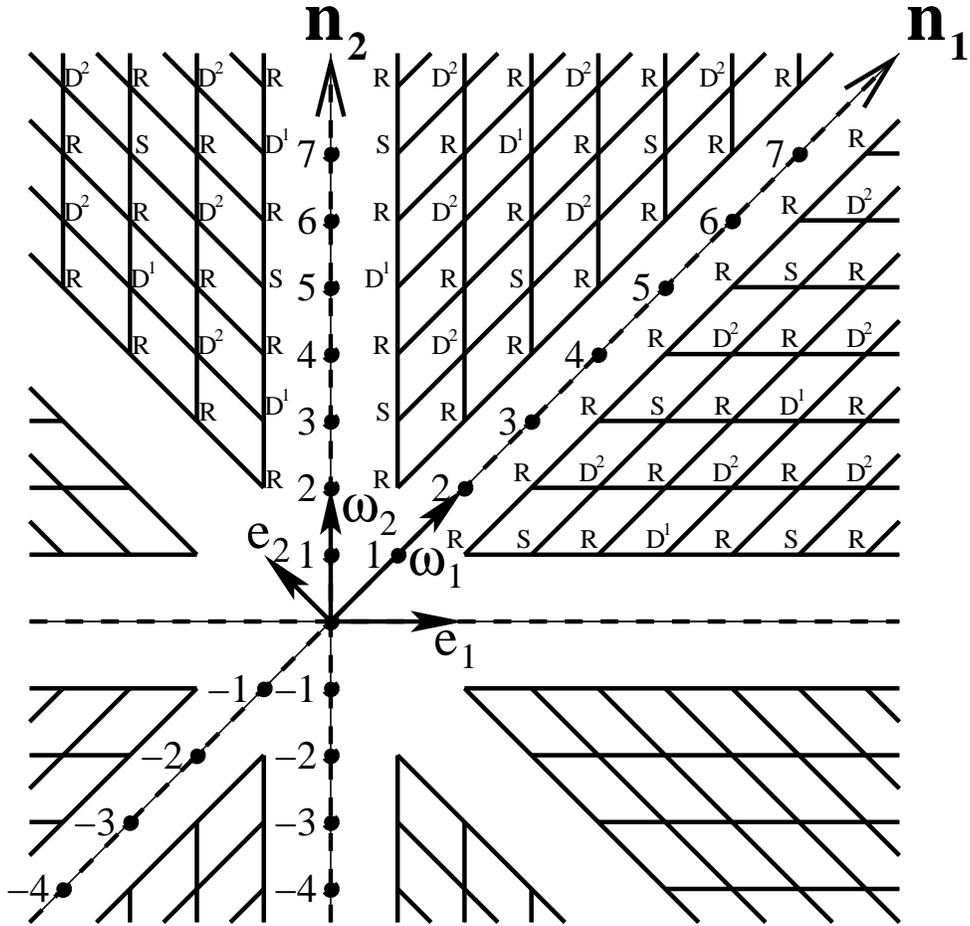}}
 \end{center}
 \protect\caption[3]{\label{fig8}Assingment of labels to the operators in 
the
 layer $\Phi_{(1,2)(n'_{1},n'_{2})}$. This is obtained by shifting the 
labels
 of Fig.~\ref{fig7}.}
\end{figure}

Let us assume, for definiteness, that the lattice in Fig.~\ref{fig7}
corresponds to the operators on the $\alpha_{-}$ side, i.e., the operators
$\Phi_{(1,1)(n'_{1},n'_{2})}$ whose indices on the $\alpha_{+}$ side are
trivial, $n_{1}=n_{2}=1$. Now consider, for instance, the operators
$\Phi_{(1,2)(n'_{1},n'_{2})}$ having the ``excitation'' $(1,2)$ on the
$\alpha_{+}$ side. In this case the assignment of $S$, $D^{1}$, $D^{2}$ 
and
$R$ labels to the corresponding two-dimensional lattice with coordinates
$(n'_{1},n'_{2})$ is simply obtained by a shift of the distribution of 
$S$,
$D^{1}$, $D^{2}$ and $R$ in the basic layer (see Fig.~\ref{fig7}) along 
the
vector $(1,2)$. The result is shown in Fig.~\ref{fig8}.

Put otherwise, the label of a generic operator 
$\Phi_{(n_1,n_2)(n_1',n_2')}$
only depends on the differences of indices, $(n_1-n_1',n_2-n_2')$.
Therefore, the label of $\Phi_{(n_1,n_2)(n_1',n_2')}$ coincides with that
of $\Phi_{(1,1),(n_1'+1-n_1,n_2'+1-n_2)}$. As this latter operator belongs
to the basic layer, depicted in Fig.~\ref{fig7}, the corresponding label
is known.

We remark that the above shifting argument could just as well have been
used to trivialise the indices on the $\alpha_-$ side. Thus, the label of
$\Phi_{(n_1,n_2)(n_1',n_2')}$ not only coincides with that of
$\Phi_{(1,1),(n_1'+1-n_1,n_2'+1-n_2)}$ but also with that of
$\Phi_{(n_1+1-n_1',n_2+1-n_2')(1,1)}$. Both these latter operators
belong to the basic layer, since the distinction between the $\alpha_+$
and $\alpha_-$ sides is just a convention. We can therefore conclude that
the labels of $\Phi_{(1,1),(1+\tilde{n}_1,1+\tilde{n}_2)}$ and
$\Phi_{(1,1),(1-\tilde{n}_1,1-\tilde{n}_2)}$ coincide, for any integers
$\tilde{n}_1$ and $\tilde{n}_2$.

The presentation described above---and in particular the notions of 
``basic
layer'' and ``other layers''---is of course just one particular way which 
we
have chosen to visualise the four-dimensional lattice of operators
$\Phi_{(n_{1},n_{2})(n'_{1},n'_{2})}$.

\subsection{Finite Kac tables for unitary theories}

All the discussion so far on the Kac table applies in general,
when the parameters $\alpha_{+}$ and $\alpha_{-}=-1/\alpha_{+}$ take
general values. As usual, when $\alpha^{2}_{+}$ takes rational
values, the Kac table becomes finite. For unitary theories%
\footnote{Unitarity is ensured by the existence of the corresponding
coset construction; see Eqs.~(\ref{SOcoset})--(\ref{paramp}).}
$\alpha^{2}_{+}=(p+2)/p$, cf.~Eqs.~(\ref{CGparam}) and (\ref{alpha+-}),
with $p$ being an integer. The physical operators' part of the Kac table
will then be delimited by:
\bea
 2\leq n'_{1}+n'_{2}\leq p+1, \nn\\
 2\leq n_{1}+n_{2}\leq p-1. \label{Kaclim}
\eea
This is similar to the Kac table of the conformal theory $WB_{2}$, 
defined,
among other $W$ theories, by Fateev and Lukyanov in Ref.~[9]. The
contents of this conformal theory ($WB_{2}$) and the one considered in the
present paper ($Z_{5}$, second solution) are completely different. Still, 
both
of them use the weight lattice of the algebra $B_{2}$ for their respective
representations. And the global properties of their representation 
lattices
(the Kac tables) are the same.

The delimitations of the Kac table of the theory $Z_{5}$, as given in
Eq.~(\ref{Kaclim}), can also be checked directly. To this end, we consider
reflections in the direction
\beq
 2\vec{e}_{2}+\vec{e}_{1}. \label{screen3}
\eeq
Using Eq.~(\ref{Kac1}) for $\Delta_{(n_{1},n_{2})(n'_{1},n'_{2})}$, 
together
with the appropriate boundary terms established above, one finds that the
operators above the line
\beq
 n'_{1}+n'_{2}=p+2 \label{refline}
\eeq
in Fig.~\ref{fig7} are all ghosts, which should decouple. Their dimensions
differ from those of their partners with respect to the reflections in the
line (\ref{refline}) by values which are consistent with the levels in the
corresponding modules. These reflections are realised by screenings that 
are
units of the vector (\ref{screen3}) [which is itself made out of three
screenings]. This is sufficient to decouple the operators above the
line (\ref{refline}). 

In giving this argument, we have referred to Fig.~\ref{fig7},
which is the basic layer of the Kac table. But the argument is true for
the other layers as well.

Finally, the remaining finite part of the Kac table, delimited by the
conditions in Eq.~(\ref{Kaclim}), possesses a further symmetry.
Using Eq.~(\ref{Kac1}) together with appropriate boundary terms, one can
check that the operation
\bea
 \left\{
 \begin{array}{ll}
 n'_{1}\rightarrow p+2-n'_{1}-n'_{2}, \qquad & n'_{2}\rightarrow n'_{2}, \\
 n_{1}\rightarrow p-n_{1}-n_{2},             &  n_{2}\rightarrow n_{2},
 \end{array}
 \right. \label{exsymm}
\eea
is a symmetry of $\Delta_{(n_{1},n_{2})(n'_{1},n'_{2})}$.

It is also a symmetry of the Kac table of $WB_{2}$ [9]. The basic
difference of the theory $Z_{5}$, from a very general point of
view, resides in its boundary terms for the different sectors.
The important point is that the symmetry (\ref{exsymm}), as well
as the decoupling of the operators outside of the domain (\ref{Kaclim}),
remains unbroken by the $S$, $D^{1}$, $D^{2}$, $R$ structure of the
lattice. Rather, it is consistent with it.

\begin{figure}
\begin{center}
 \leavevmode
 \epsfysize=170pt{\epsffile{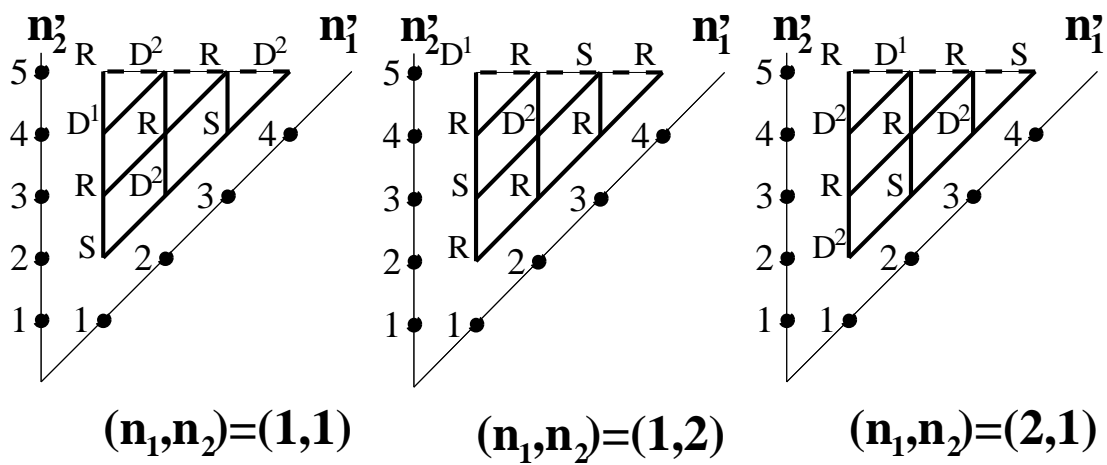}}
 \end{center}
 \protect\caption[3]{\label{fig9}Kac table of the the $p=4$ theory with
 $c=3/2$. There are three $\alpha_-$ layers, having a triangular shape,
 and each corresponding to fixed values of the $\alpha_+$ indices
 $(n_1,n_2)$. In the $(1,2)$ layer, the symmetry (\ref{exsymm})
 acts by reflecting the operators in a line that goes through the right
 angle of the triangle and the mid-point of the long side. In the
 $(1,1)$ and $(2,1)$ layers, the symmetry (\ref{exsymm}) acts by permuting
 these two layers and reflecting the triangle. The theory thus has a
 total of 16 primary operators. }
\end{figure}

Let us give a simple example of the series of theories that we have
constructed. According to the coset formula in
Eqs.~(\ref{SOcoset})--(\ref{paramp}), the first non-trivial $Z_{5}$ theory
should be that with $p=p_{\rm min}=4$, having $c=3/2$. Its Kac table,
which is made of three $\alpha_{-}$ layers, is shown in Fig.~\ref{fig9}.

The preceding theory, the one with $p=3$, is trivial. Its Kac table is
made of just one $\alpha_{-}$ layer, with all the operators being present
$(S,R,D^{1},D^{2})$ having the trivial scaling dimension, $\Delta=0$.
This could have been anticipated, because the central charge of this 
theory
is $c=0$, according to Eq.~(\ref{cparamp}).

\subsection{Characteristic equations}

In all the arguments so far, in building the Kac formula
and the Kac table, we have used arguments based on the Coulomb gas.
Its existance has been assumed, with given properties; we have
however not constructed the Coulomb gas explicitly.

Our assumptions (in particular for the fusion rules) could be supported,
at least partially, by
giving characteristic equations for conformal dimensions of
operators participating in a given fusion (operator algebra expansion).
These can be obtained from
differential equations for three-point correlation functions.

The differential equations, which are due to the degeneracy of the
modules, can be derived in a way analogous to that of
Ref.~[6], dealing with the $Z_{3}$ theory. Below we give
several examples of correlation functions and the resulting
characteristic equations for the dimensions of the operators. However,
we do not intend to dress an exhaustive list of such
equations. Rather, the examples discussed should be considered as an
additional demonstration on the consistency of our approach, and of
the theory.

In Appendix~\ref{appC} the correlation function
\beq
  \left \langle R^{(2)}(z_{2})\Phi^{q}(z_{3})R^{(1)}(z_{1}) \right \rangle
  \label{RPqR}
\eeq
is considered.
Here $R^{(1)}(z_{1})$ is a disorder operator whose module is
supposed to be completely degenerate at level 1/2, i.e.,
$R^{(1)}(z_{1})=\Phi_{(1,2)(1,1)} (z_{1})$ or
$\Phi_{(1,1)(1,2)}(z_{1})$. By $R^{(2)}(z_{2})$ we denote another,
generic disorder operator, and $\Phi^{q}(z_{3})$ with $q=0,\pm 1$
is a singlet or a doublet 1 operator. For the reasons given in
Appendix~\ref{appC}, the correlation functions with a doublet 2 operator
will be considered separately.

Using a method slightly different from that in Ref.~[6],
we have derived a differential equation for the function (\ref{RPqR});
see Appendix~\ref{appC}. Here we are concerned only with
the characteristic equation for dimensions which
results from it. One finds that the necessary condition for the
function (\ref{RPqR}) to be non-zero is given by the following equation
on the dimensions of the operators $R^{(1)}$, $R^{(2)}$ and $\Phi^{q}$:
\beq
 \Delta_{R^{(2)}}-\Delta_{\Phi^{q}}+\frac{1}{4} 
\left(1-\frac{2q^{2}}{5}\right)
 \Delta_{R^{(1)}}=
 (-1)^{q}\frac{5}{4}\frac{h^{(2)}_{1}}{h^{(1)}_{1}}\Delta_{R^{(1)}}.
 \label{char0}
\eeq
Here, $h^{(1)}_{1},\, h^{(2)}_{1}$ are the zero mode
eigenvalues of the operators $R^{(1)}$ and $R^{(2)}$ as defined in
Eq.~(\ref{Rzero1}).

In a similar way one derives analogous results
for the functions given below; see Appendix~\ref{appC}.
\begin{itemize}
 \item \underline{$\left \langle \Phi^{2}_{(2)}\Phi^{0}\Phi^{-2}_{(1)}
 \right \rangle$.}
 The module of the operator $\Phi^{-2}_{(1)}$ is supposed to be 
degenerate on
 levels 1/5 and 1, i.e., $\Phi^{-2}_{(1)}=\Phi_{(2,1)(1,1)}$ or
 $\Phi_{(1,1)(2,1)}$. Using these degeneracies one derives a
 differential equation, from which the following equation for the
 dimensions:
 \beq
  \Delta_{\Phi^{0}}-\Delta_{\Phi^{2}_{(2)}}-\frac{1}{4}\Delta_{\Phi^{2}_{(1)}}=
  -\frac{5}{4}\frac{h^{(2)}_{1,2}}{h^{(1)}_{1,2}}\Delta_{\Phi^{2}_{(1)}}
  \label{char1}
 \eeq
 is deduced. Here, $h^{(2)}_{1,2}$ and $h^{(1)}_{1,2}$ are the eigenvalues
 of the operators $\Phi^{2}_{(2)}$ and $\Phi^{-2}_{(1)}$ with
 respect to the zero modes $A^{1}_{0}, A^{-1}_{0}$, as defined in
 Eq.~(\ref{eig2}).
 \item \underline{$\left \langle \Phi^{2}_{(2)}\Phi^{1}\Phi^{2}_{(1)}
 \right \rangle.$}
 The module of the operator $\Phi^{2}_{(1)}$ is assumed to be degenerate 
on
 levels 1/5 and 1, i.e., $\Phi^{2}_{(1)}=\Phi_{(2,1)(1,1)}$ or
 $\Phi_{(1,1)(2,1)}$. The operator $\Phi^{2}_{(2)}$ is more
 general, since in this case we assume it to be degenerate once on
 level 1/5, whilst its second level of degeneracy can be general.
 {}From reflection arguments it is checked that an operator
 with such properties can be any one in the series:
 \beq
  \Phi^{2}_{(2)}=\Phi_{(2k,1)(1,1)} \mbox{ or }
  \Phi_{(1,1)(2k,1)},
 \eeq
 with $k=1,2,3,\ldots$. In this case the correlation function
 $\left \langle \Phi^{2}_{(2)}\Phi^{1}\Phi^{2}_{(1)} \right \rangle$
 leads to the following characteristic equation:
 \beq
  \Delta_{\Phi^{1}}=\Delta_{\Phi^{2}_{(2)}}+\frac{1}{2}\Delta_{\Phi^{2}_{(1)}}.
  \label{char2}
 \eeq
 \item \underline{$\left \langle 
\Phi^{1}_{(3)}\Phi^{1}_{(2)}\Phi^{-2}_{(1)}
 \right \rangle.$}
 Similarly, the module of the operator $\Phi^{-2}_{(1)}$ is assumed
 to be degenerate at levels 1/5 and 1. The operators
 $\Phi^{1}_{(3)}$ and $\Phi^{1}_{(2)}$ are general. The characteristic
 equation for this function turns ont to be very simple. This
 function is non-zero if:
 \beq
  \Delta_{\Phi^{1}_{(2)}}=\Delta_{\Phi^{1}_{(3)}}.
  \label{char3}
 \eeq
\end{itemize}

By analysing the characteristic equations given above it is verified
that they are in agreement with the general arguments based
on the Coulomb gas, i.e., the arguments that we have used in defining
the $S$, $D^{1}$, $D^{2}$, $R$ structure of the Kac table. They also
contain some more concrete information on the fusion rules of the theory.
This is the case of Eqs.~(\ref{char2}) and (\ref{char3}).
For instance, solving Eq.~(\ref{char2}), one finds that the fusion
of the operators:
\beq
 \Phi^{2}_{(1)}=\Phi_{(1,1)(2,1)} \mbox{ and }
 \Phi^{2}_{(2)}=\Phi_{(1,1)(2k,1)}
\eeq
produces, in particular, the operator
\beq
 \Phi^{1}=\Phi_{(1,1)(2k-1,3)}.
\eeq
This result is in agreement with the positioning of doublet 1 operators in
the third row in Fig.~\ref{fig7}. It is also in accordance with the 
addition of
vectors, corresponding to operators in the Coulomb gas picture,
and with the action of screenings. A move by $-\vec{e}_{1}$,
produced by the screening $\vec{e}_{1}$, will have to be used,
since the channel considered is not the principal one for the
Coulomb gas fusions.

The other two equations, (\ref{char0}) and (\ref{char1}), are of a
somewhat more general nature. Apart from the dimensions of the operators,
these equations also involve the zero mode
eigenvalues of the operators: $h^{(1)}_{1}$ and $h^{(2)}_{1}$ in
Eq.~(\ref{char0}), as well as $h^{(1)}_{1,2}$ and $h^{(2)}_{1,2}$ in
Eq.~(\ref{char1}).

The zero mode eigenvalues ($h_{1},h_{2,1},h_{1,2}$)
characterise the primary operators, and the corresponding
representations, of the parafermionic algebra on the same footing
as the $\Delta$, which are just the eigenvalues of the Virasoro
zero mode $L_{0}$. In particular, the theory possesses
examples of a couple of primary operators of the same nature,
disorder operators for instance, having the same values of $\Delta$
but different values of $h_{1}$. Such operators have to be
considered as being different.

Such examples were also present in the $Z_{3}$ theory of
Fateev-Zamolodchikov [6], in which the degeneracies on the
level of $\Delta$ are lifted by different values of $h_{1}$, in particular
in the disorder sector.

The presence of $h_{1}$ in Eq.~(\ref{char0}), and of $h_{1,2}$ in
Eq.~(\ref{char1}), complicates somewhat the analysis of the
corresponding fusion rules.

For some particular disorder or doublet operators, the correspondent value
$h_{1}$ or $h_{1,2}$ is known. This is the case for $h^{(1)}_{1}$ of
the operator $R^{(1)}$, in the correlation function
$\left \langle R^{(2)}\Phi^{q}R^{(1)} \right \rangle$.
Also, $h^{(1)}_{1,2}$ of the operator $\Phi^{2}_{(1)}$, in the other
correlation functions, is known from the degeneracy problem at level 1/5,
cf.~Eq.~(\ref{fixeig2}).

In the case of the function
$\left \langle R^{(2)}\Phi^{q}R^{(1)} \right \rangle$ there are
two independent channels, $q=0$ and $q=1$, to be considered in
Eq.~(\ref{char0}). If we set  $R^{(2)}=\Phi_{(1,2)(1,1)}$ or
$R^{(2)}=\Phi_{(1,1)(1,2)}$, all the $h_{1}$ in Eq.~(\ref{char0}) are 
known and
the dimension $\Delta_{\Phi^{q}}$ can be easily calculated for
each of the two channels.

\begin{figure}
\begin{center}
 \leavevmode
 \epsfysize=150pt{\epsffile{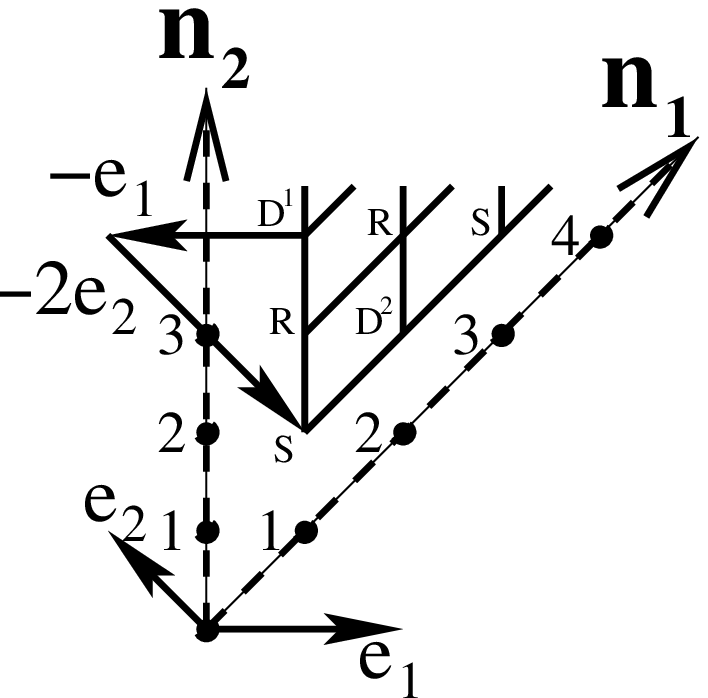}}
 \end{center}
 \protect\caption[3]{\label{figR1}The subdominant channel in the operator
 product $\Phi_{(1,2)(1,1)} \cdot \Phi_{(1,2)(1,1)}$ is obtained by action
 on the principal channel $\vec{\beta}_{(1,3)(1,1)}$ with the screenings
 $-\vec{e}_{1}-2\vec{e}_{2}$.}
\end{figure}

We then consider the functions:
\begin{itemize}
 \item \underline{$\left \langle 
\Phi_{(1,2)(1,1)}\Phi^{q}\Phi_{(1,2)(1,1)}
 \right \rangle$.} For the $q=0$ channel it follows from Eq.~(\ref{char0})
 that
 \beq
  \Delta_{\Phi^{0}}=0.
 \eeq
 This result was to be expected, as the two point function
 $\left \langle \Phi_{(1,2)(1,1)}\Phi_{(1,2)(1,1)} \right \rangle$ is 
 non-vanishing. More interesting is the $q=1$ channel for which
 Eq.~(\ref{char0}) gives
 \beq
  \Delta_{\Phi^{1}}=\Delta_{\Phi_{(1,3)(1,1)}}.
 \eeq
 This  result is in agreement with the fact that
 $\Phi_{(1,3)(1,1)}$ was deduced to be a doublet 1 ($q=1$),
 and with the Coulomb gas rules. Indeed, as previously discussed,
 we expect that the multiplication $\Phi_{(1,2)(1,1)} \cdot \Phi_{(1,2)(1,1)}$
 produces in the principal channel the operator with
 $\vec{\beta}_{(1,3)(1,1)}=\vec{\beta}_{(1,2)(1,1)}+\vec{\beta}_{(1,2)(1,1)}$;
 the $q=0$ channel follows the principal one by the shift
 $-\vec{e}_{1}-2\vec{e}_{2}$, as shown in Fig.~\ref{figR1}.\footnote{The 
 above
 result and the comments hold also for the the function
 $\left \langle \Phi_{(1,1)(1,2)}\Phi^{q}\Phi_{(1,1)(1,2)} \right 
\rangle$,
 provided that the indices $\alpha_{-}$ and  $\alpha_{+}$ are exchanged.}
 \item \underline{$\left \langle 
\Phi_{(1,1)(1,2)}\Phi^{q}\Phi_{(1,2)(1,1)}
 \right \rangle$.}
 The associated characteristic equation for the two channels is solved
 by:
 \bea
   \Delta_{\Phi^{0}} &=& \Delta_{\Phi_{(1,2)(1,2)}}, \nn \\
   \Delta_{\Phi^{1}} &=& \Delta_{\Phi_{(0,2)(1,2)}}.
 \eea
 Once again the correctness of the labeling of the lattice and
 of the Coulomb gas rules is confirmed.
\end{itemize}

It is interesting to notice that the operator $\Phi_{(0,2)(1,2)}$ is
not a physical one and, although decoupled from the theory, it is
admitted by Eq.~(\ref{char0}). This follows from the fact that the
characteristic equation does not give any information about the
structure constants of the operator algebra; in particular, these can
be equal to zero. In the case under consideration the
correlation function
$\left \langle \Phi_{(1,1)(1,2)}\Phi_{(0,2)(1,2)}\Phi_{(1,2)(1,1)}
\right \rangle$ is expected to vanish.

\subsection{General formula for the zero mode eigenvalues of disorder
operators}

In contrast to the particular cases considered above, the eigenvalues
$h$ are not known in general. The task thus becomes not only to fix
$\Delta$ but also $h^{(2)}_{1}$ of $R^{(2)}$ in Eq.~(\ref{char0})
and $h^{(2)}_{1,2}$ of $\Phi^{2}_{(2)}$ in Eq.~(\ref{char1}).

We focus in the following on Eq.~(\ref{char0}).
One of the two channels of this equation  could be used to define
$h^{(2)}_{1}$, by assuming a given value of $\Delta_{\Phi^{0}}$
for instance, chosen at a particular position in the Kac table.
The other channel, with $\Phi^{1}$, in which enters the same
$h^{(2)}_{1}$, could then serve to check for the presence of
$\Delta_{\Phi^{1}}$ at the appropriate position in the Kac table,
having the value calculated from the characteristic equation
(\ref{char0}). In this way some progress could be made in filling the Kac
table, independently of the Coulomb gas assumptions, partially at
least.

The problem could also be turned around. We could assume the
results of our analysis, based on the Coulomb gas arguments, in
which we have already completely defined the Kac formula for
the $\Delta$. We could then use instead the characteristic
equations to dress the general formula for the eigenvalues $h$.

We have succeeded in this direction for the disorder operators.
To this end, we have consider the product $R^{(1)}R^{(2)}$, where
$R^{(1)}=\Phi_{(1,2)(1,1)}$ and $R^{(2)}$ is again a general
disorder operator, $R^{(2)}=\Phi_{(n_{1},n_{2})(n_{1}',n_{2}')}$
with $n_{2}+n_{2}'$ odd.

\begin{figure}
\begin{center}
 \leavevmode
 \epsfysize=220pt{\epsffile{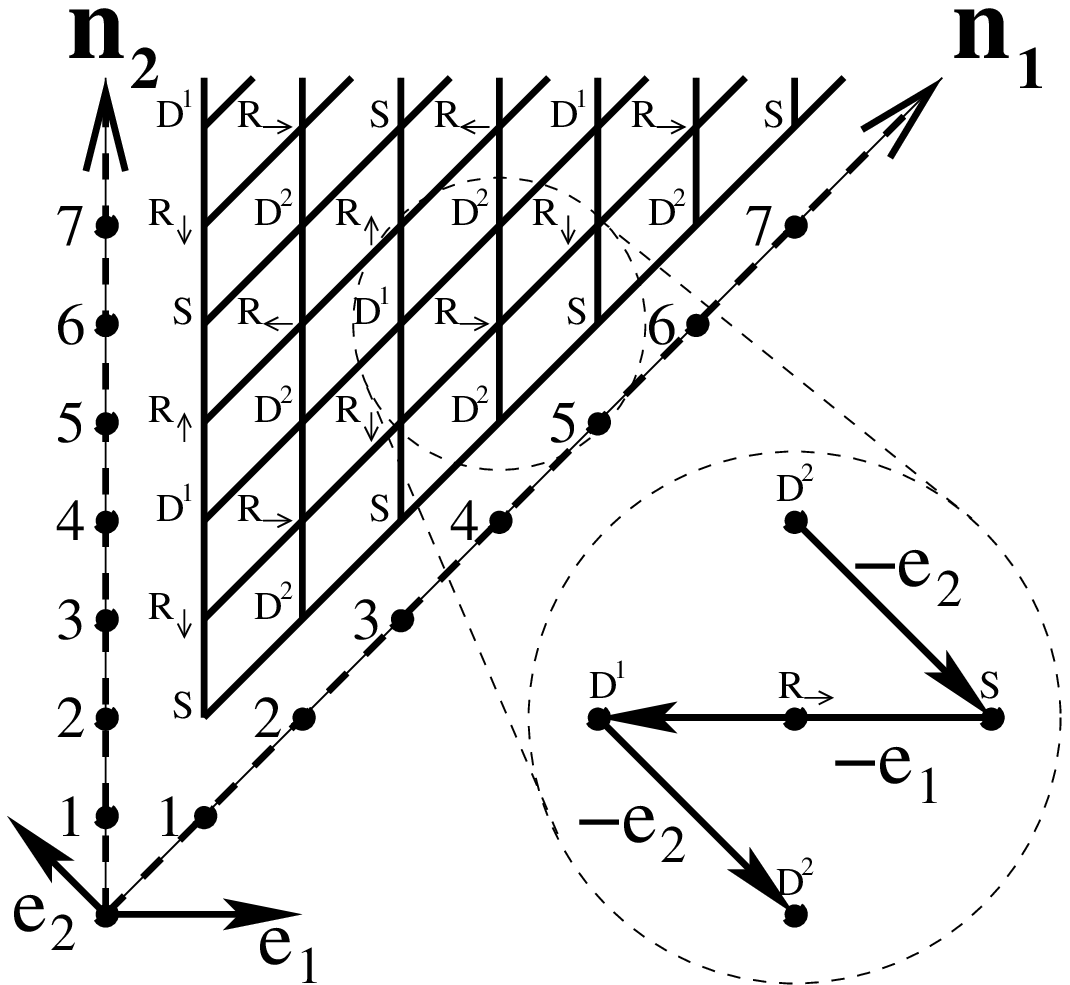}}
 \end{center}
 \protect\caption[3]{\label{figR2}Each disorder operator $R$ is surrounded
 by three doublets and one singlet operator, as shown in the inset.
 These latter four operators can be related to non-principal channels
 in the operator product $R^{(1)} \cdot R$, as discussed in the text.
 The disorder operators $R$ come in four varieties, according to their
 neighbouring doublet and singlet operators. The arrow labeling each $R$
 shows the relative position of the singlet operator. Alternatively, the
 arrows point away from the $D_1$ operators.}
\end{figure}

In Fig.~\ref{figR2} we show a layer of the Kac table with fixed
$\alpha_{+}$ indices. {}From the figure one sees that a disorder
operator $R^{(2)}$ is surrounded by four operators, at the positions:
$\vec{\beta}_{(n_{1}+1,n_{2})(n_{1}',n_{2}')}$,
$\vec{\beta}_{(n_{1}-1,n_{2})(n_{1}',n_{2}')}$,
$\vec{\beta}_{(n_{1}-1,n_{2}+1)(n_{1}',n_{2}')}$ and
$\vec{\beta}_{(n_{1}+1,n_{2}-1)(n_{1}',n_{2}')}$.
Among these four nearest-neighbour operators, two
are $D^{2}$ doublets, one is a $D^{1}$ doublet and one is a
singlet $S$. It is straightforward to verify that (see the inset of
Fig.~\ref{figR2}):
\bea
 \vec{\beta}_{(n_{1}+1,n_{2}-1)(n_{1}',n_{2}')} &=&
 \vec{\beta}_{(n_{1}+1,n_{2})(n_{1}',n_{2}')} - \vec{e}_{2}, \nn \\
 \vec{\beta}_{(n_{1}-1,n_{2}+1)(n_{1}',n_{2}')} &=&
 \vec{\beta}_{(n_{1}+1,n_{2})(n_{1}',n_{2}')} - \vec{e}_{2}-\vec{e}_{1},
 \nn \\
 \vec{\beta}_{(n_{1},n_{2}-1)(n_{1}',n_{2}')} &=&
 \vec{\beta}_{(n_{1}+1,n_{2})(n_{1}',n_{2}')}-2\vec{e}_{2}-\vec{e}_{1}.
\end{eqnarray}

According to the Coulomb gas rules, the operator
$\Phi_{(n_{1}+1,n_{2})(n_{1},n_{2}')}$ is produced in the
principal channel of the operator product expansion of $R^{(1)} \cdot 
R^{(2)}$
[indeed, $\vec{\beta}_{(n_{1}+1,n_{2})(n_{1}',n_{2}')}=
\vec{\beta}_{(1,2)(1,1)}+\vec{\beta}_{(n_{1},n_{2})(n_{1}',n_{2}')}$],
while the other three nearest-neighbour operators can be present in the
non-principal channels of the above expansion.

Therefore we can make the natural hypothesis that the doublets
$D^{1}$ and the singlet $S$, which are the nearest to $R^{2}$ in
the layer $(n_{1}',n_{2}')$, are present in the operator product expansion 
of
$R^{(1)} \cdot R^{(2)}$. As we showed previously, this hypothesis is
verified in the case when $R^{(2)}$ is taken to be completely degenerate
at level $1/2$.

We then assume that for $q=0$ (resp.~$q=1$),
$\Delta_{\Phi^{0}}$ (resp.~$\Delta_{\Phi^{1}}$) is equal
to the dimension of the nearest singlet $S$ (resp.~doublet $D^{1}$). The
eigenvalue $h^{(2)}_{1}$ can now be determined by solving the
correspondent equation.

Once $h^{(2)}_{1}$ has been fixed, for example using the $q=0$
channel, one can easily check that Eq.~(\ref{char0}) admits in the $q=1$
channel the nearest-neighbour doublet $D^{1}$. The above assumption is thus
perfectly consistent with the characteristic equation.

We can label quite naturally a disorder operator by
its relative position with respect to the nearest singlet.
As shown on Fig.~\ref{figR2}, four types of disorder operators can then
be identified, namely $R_{\uparrow}$, $R_{\downarrow}$,
$R_{\leftarrow}$ and $R_{\rightarrow}$.
The general formula for the eigenvalue $h^{X}_{1}$ of the disorder 
operator
$\Phi_{(n_{1},n_{2})(n_{1}',n_{2}')}$ of type $X$
($X=\ \uparrow,\downarrow,\leftarrow,\rightarrow$) turns out to be:
\bea
 h^{\downarrow}_{1}  &=& A\left[
 4\,(n_{1}+n_{2})\,p-4\,(n_{1}'+n_{2}')(p+2)+p+1\right], \nn \\
 h^{\uparrow}_{1}    &=& A\left[
 -4\,(n_{1}+n_{2})\,p+4\,(n_{1}'+n_{2}')(p+2)+p+1\right], \nn \\
 h^{\leftarrow}_{1}  &=& A\left[ 4\,n_{1}\,p-4\,n_{1}'(p+2)+p+1\right], \nn \\
 h^{\rightarrow}_{1} &=& A\left[ -4\,n_{1}\,p+4\,n_{1}'\,(p+2)+p+1\right],
 \label{4types}
\eea
where
\beq
 A= \frac{1}{80}\frac{\sqrt{5}\;2^{\frac{4}{5}}}{\sqrt{p-3}
 \,\sqrt{p+5}}.
\eeq
Notice that the symmetry (\ref{exsymm}) of the Kac table is respected by
Eq.~(\ref{4types}).

The above analysis could also be done by studying the fusion
$R^{(1)} \cdot R^{(2)}$ with $R^{(1)}=\Phi_{(1,1)(1,2)}$. It is easy to
verify that this procedure leads to the same results.

In the following we shall give a partial verification of
Eq.~(\ref{4types}) by determining explicitly the eigenvalue
$h_{1}$ for a subset of disorder operators.

\begin{figure}
\begin{center}
 \leavevmode
 \epsfysize=220pt{\epsffile{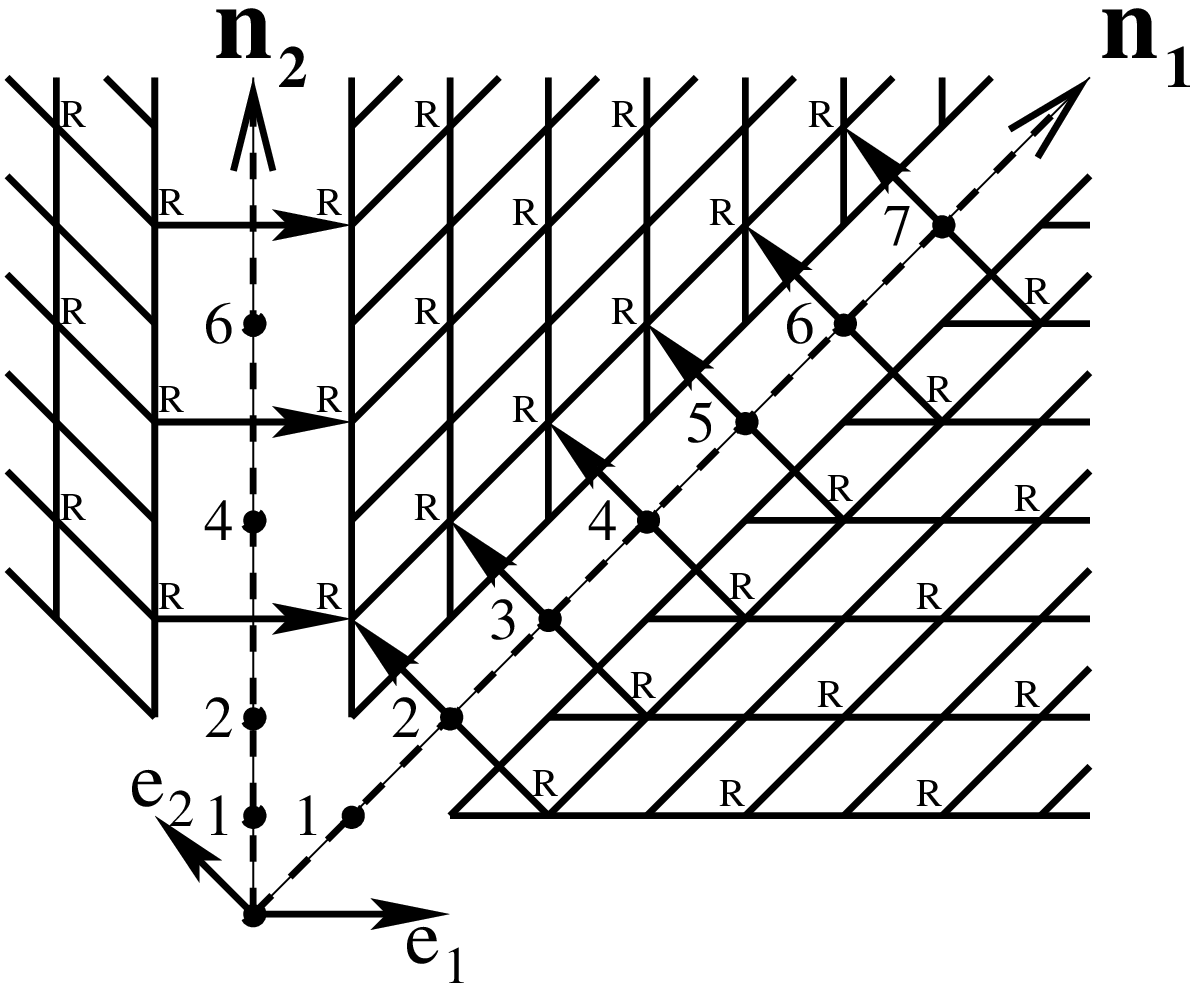}}
 \end{center}
 \protect\caption[3]{\label{figR3}The reflections shown on the figure
 are used to demonstrate that a subset of disorder operators belonging
 to the physical
 domain all have one degeneracy at level $1/2$. Apart from the fundamental
 disorder operator $\Phi_{(1,1)(1,2)}$, the other degeneracy is
 at a more descendent level than $1/2$.}
\end{figure}

As shown in Fig.~\ref{figR3}, where the particular layer
$(1,1)(n_{1}',n_{2}')$ is represented, there are two reflections, one by
the vector $-\vec{e}_{1}$ and the other by the vector $-2\vec{e}_{2}$,
which give the differences
\bea
 \Delta_{(1,n_{2})(-1,n_{2}'+2)}-\Delta_{(1,n_{2})(1,n_{2}')}&=&\frac{1}{2},\\
 \Delta_{(n_{1},1)(n_{1}'+1,-1)}-\Delta_{(n_{1}',1)(n_{1}',1)}&=&\frac{1}{2},
\eea
respectively for $R=\Phi_{(1,n_{2})(1,n_{2}')}$ (with $n_{1}'+n_{2}'$
even) and for $R=\Phi_{(n_{1},1)(n_{1}',2)}$.
These operators should therefore be partially degenerate at level $1/2$.

In Section~\ref{sec2} we required complete degeneracy at level $1/2$.
If now we demand only a partial degeneracy at this level, we do have one 
less
constraint. The eigenvalues  $h_{1}$ and $h_{2}$ will be
determined as functions of the dimension $\Delta_{R}$ and of the
central charge.
Then, if we use the value $\Delta_{R}$ provided by the Kac formula,
we have finally access to the expression of $h_{1}$ and $h_{2}$
for $R=\Phi_{(1,n_{2})(1,n_{2}')}$ (with $n_{1}'+n_{2}'$ odd) and
for $R=\Phi_{(n_{1},1)(n_{1}',2)}$.

At level $1/2$ of the module $R_{a}$ there are two states,
namely $\left( \chi^{(1)}_{a} \right)_{-\frac{1}{2}}$ and
$\left( \chi^{(2)}_{a} \right)_{-\frac{1}{2}}$,
as defined in Eqs.~(\ref{Rstate1}) and
(\ref{Rstate2}). We consider the linear combination
\beq
 \chi_{-\frac{1}{2}}= a(\chi^{(1)}_{a})_{-\frac{1}{2}}+
 b(\chi^{(2)}_{a})_{-\frac{1}{2}} \label{linicombo}
\eeq
which would be primary. As usual, we then require that
\beq
 A^{1}_{+\frac{1}{2}}\chi_{-\frac{1}{2}}=
 A^{2}_{+\frac{1}{2}}\chi_{-\frac{1}{2}}=0. \label{degconstr}
\eeq

Using Eq.~(\ref{linicombo}), one has from Eq.~(\ref{degconstr}) that
\bea
 aA^{1}_{\frac{1}{2}}A^{1}_{-\frac{1}{2}}R_{a}
 +bA^{1}_{\frac{1}{2}}A^{2}_{-\frac{1}{2}}R_{a}&=&0,\nn\\
 aA^{2}_{\frac{1}{2}}A^{1}_{-\frac{1}{2}}R_{a}
 +bA^{2}_{\frac{2}{5}}A^{2}_{-\frac{2}{5}}R_{a}&=& 0.
\eea
Defining now $\mu^{(i,j)}$ as
$A^{i}_{\frac{1}{2}}A^{j}_{-\frac{1}{2}}R_{a'}\equiv \mu^{(i,j)}
R_{a}$, the above system can be solved if
\beq
 \mu^{(1,1)}\mu^{(2,2)}-\mu^{(1,2)}\mu^{(2,1)}=0. \label{disorderdeg}
\eeq

The values of $\mu^{(i,j)}$ as a function of  $h_{1}$, $h_{2}$ and 
$\Delta_{R}$
can be extracted from Eqs.~(\ref{disorderrel1}), (\ref{disorderrel2}) and
(\ref{disorderrel3}). Applying then Eq.~(\ref{relh1h2}), it is 
straightforward
to see that Eq.~(\ref{disorderdeg}) results in a sixth-degree equation
in $h_{1}$.
Setting $\Delta_{R}=\Delta_{(1,n_{2})(1,n_{2}')}$ or
$\Delta_{R}=\Delta_{(n_{1}',1)(n_{1}',1)}$, this equation presents
six real solutions. Among these, one coincides with the value of
$h_{1}$ given by Eq.~(\ref{4types}).

In Section~\ref{sec2} we examined the degeneracies
at the first descendent levels of the modules, and we found a number of
solutions for the dimensions which we considered non-physical.
The case under consideration is analogous: for each of the above
operators, there is only one value of $h_{1}$ that fits with the
theory that we have built. The other solutions are rejected as
non-physical ones.

\section{Discussion}
\label{sec5}

In this paper we have constructed and analysed in some detail the
$Z_{5}$ parafermionic conformal theory, based on the second solution
for the corresponding parafermionic algebra [1].

Our principal result is the Kac table and the Kac formula of this
theory, i.e., the formula for conformal dimensions of primary
operators which realise degenerate representations of the
parafermionic algebra. This formula is given, in different forms, by
Eqs.~(\ref{Kac}), (\ref{Kac1}) and (\ref{Kac2}). Its boundary term $B$
is defined as follows:
\beq
 \begin{array}{llll}
 \mbox{Singlet sector:}   & B_S=0 \qquad &
 n_1'-n_1=0 \mbox{ mod } 2, \ \ \  & n_2'-n_2=0 \mbox{ mod } 4 \\
 \mbox{Doublet 1 sector:} & B_{D^1} = \frac{1}{10} \qquad &
 n_1'-n_1=0 \mbox{ mod } 2,        & n_2'-n_2=2 \mbox{ mod } 4 \\
 \mbox{Doublet 2 sector:} & B_{D^2} = \frac{3}{20} \qquad &
 n_1'-n_1=1 \mbox{ mod } 2,        & n_2'-n_2=0 \mbox{ mod } 2 \\
 \mbox{Disorder sector:} & B_{R} = \frac{1}{8} \qquad &
 n_1'-n_1=0 \mbox{ mod } 1,        & n_2'-n_2=1 \mbox{ mod } 2 \\
 \end{array} \label{complfilling}
\eeq
Graphically, the distribution of operators belonging to different
sectors is depicted in Fig.~\ref{fig7} in the case of the basic layer,
$(1,1)(n'_{1},n'_{2})$. This figure actually contains the full 
information,
since the other layers, with $(n_1,n_2) \neq (1,1)$, are obtained by
translating the labels in the basic layer, cf.~Eq.~\ref{complfilling}.

In analogy with the theory $Z_{3}$ [3], we find it natural to
assume that the infinite set of $Z_{5}$ theories that we have defined,
labeled by the parameter $p=4,5,6,\ldots$, describes higher multicritial
points in statistical models with $Z_{5}$ symmetry.

\begin{figure}
\begin{center}
 \leavevmode
 \epsfysize=90pt{\epsffile{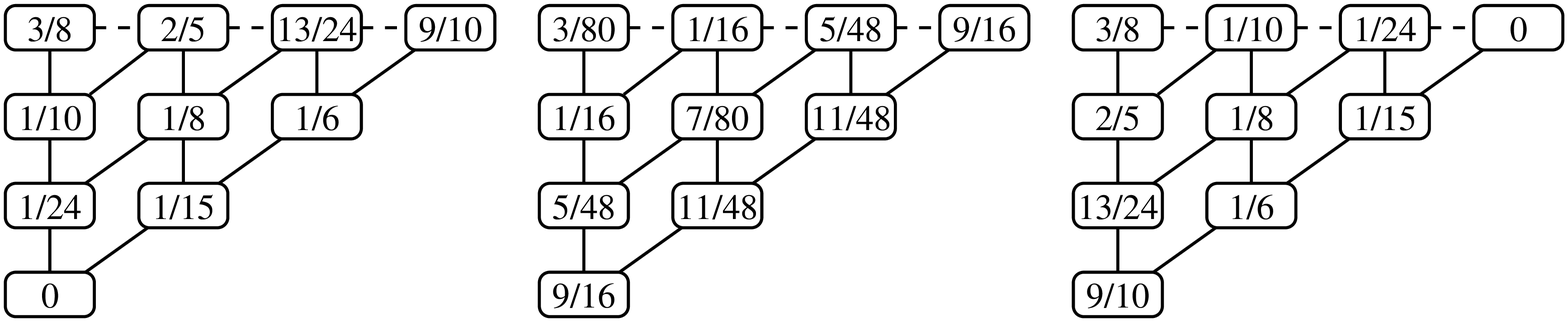}}
 \end{center}
 \protect\caption[3]{\label{fig10}Conformal dimensions $\Delta$ of fields
 in the $Z_5$ theory with $p=4$ and $c=3/2$, that are primary with respect
 to the parafermionic currents. The conventions for the 
$(n_1,n_2)(n_1',n_2')$
 labels are as in Fig.~\ref{fig7}.}
\end{figure}

The first and the simplest member in this set is the theory with $p=4$,
having $c=\frac{3}{2}$, which we have already mentioned as an example
in the previous section. The three layers of its Kac table are given
in Fig.~\ref{fig9}. The explicit values of the conformal dimensions of
its 16 different primary operators are shown in Fig.~\ref{fig10}.
By primary, we here mean primary with respect to the parafermionic
currents. Note that these operators are linked by the symmetry 
(\ref{exsymm}),
which has been explained in graphical terms in the caption of 
Fig.~\ref{fig9}.

\begin{figure}
\begin{center}
 \leavevmode
 \epsfysize=200pt{\epsffile{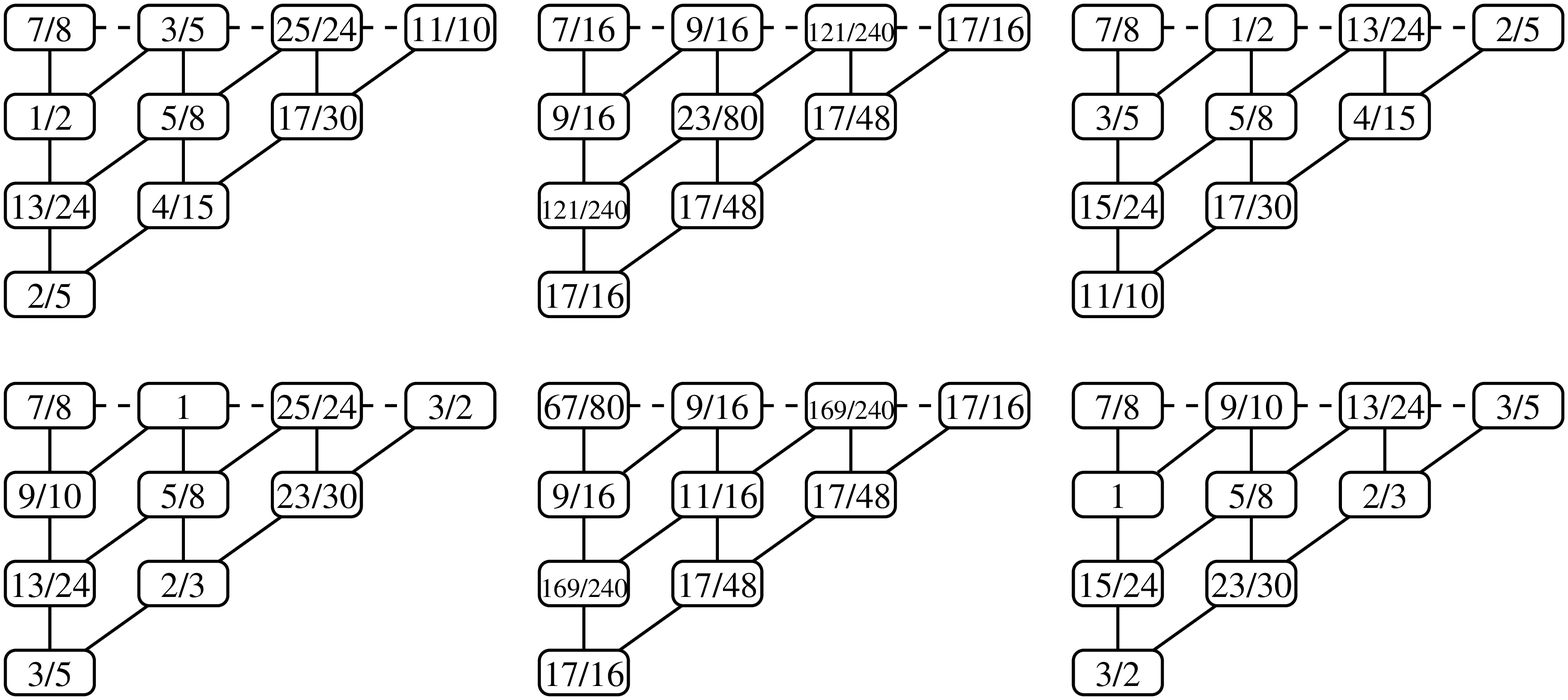}}
 \end{center}
 \protect\caption[3]{\label{fig11}Conformal dimensions of fields
 that are primary with respect to the Virasoro subalgebra (but descendents
 on the first two levels with respect to the parafermionic currents).}
\end{figure}

It should be added that the first and second descendents of the
operators in Fig.~9 are still Virasoro algebra primaries. Their
dimensions are obtained from those in Fig.~\ref{fig10} by adding:
\beq
 \begin{array}{lll}
 \frac{2}{5} \mbox{ and } \frac{3}{5} & \mbox{for} & S, \\
 \frac{2}{5} \mbox{ and } \frac{4}{5} & \mbox{for} & D^{1}, \\
 \frac{1}{5} \mbox{ and } \frac{3}{5} & \mbox{for} & D^{2}, \\
 \frac{1}{2}                          & \mbox{for} & R. \\
 \end{array}
\eeq
These numbers are is in accordance with the structure of the modules,
as shown in Figs.~\ref{fig4}--\ref{fig6}. Explicit values for the
resulting conformal dimensions are shown in Fig.~\ref{fig11}.

One could conjecture that it should be possible to suitably modify the
$Z_N$ symmetric lattice models defined in Refs.~[1,2],
which realise the first solution for $Z_{N}$ parafermionic theories,
so as to drive them to multicriticality while preserving $Z_N$ symmetry.
If so, the tricritical point corresponding to the theory $Z_{5}$, with 
$p=4$,
would be the simplest example to be discovered.

{}From a theoretical point of view, another principal result of our
paper is that the $B_{n}$ algebra lattices accommodate the
representations of conformal theories with $Z_{N}$ symmetry. So far,
those lattices were known to be relevant only for the $W$ theories
$WB_{n}$ [9]. In the present paper we have given the demonstration
of this result for the particular case of $B_{2}$, corresponding to
the $Z_{5}$ theory.

In our next paper [4] we shall give the generalisation to $B_{n}$,
with $n>2$ general, corresponding to the $Z_{N}$ theory with $N=2n+1$.
This series is the second solution of the $Z_{N}$ parafermionic theory
with $N$ odd.
The case of $Z_{N}$ with $N$ even has to be studied separately. It has
new features, not present in the theories with $N$ odd. The Kac table
for these theories should be built on $D_{n}$ lattices, with $N=2n$.

{\bf Acknowledgments:} We would like to thank P.~Degiovanni, V.~A.~Fateev,
and P.~Mathieu for very useful discussions. We are particularly grateful to
M.-A.~Lewis, F.~Merz, and M.~Picco for their collaboration during the initial
stages of this project.

\appendix

\section{Mode operator commutation relations in the sectors of singlet, 
doublet 1 and doublet 2 operators}
\label{appA}

In general, the mode operators are defined by integrals of the type
\bea
 A_{-\delta+m}\Phi(0) &=& \frac{1}{2\pi i}
 \oint {\rm d}z \,(z)^{\Delta_{\Psi}-\delta+m-1}\Psi(z)\Phi(0), \\
 \Psi(z)\Phi(0) &=& 
\sum_{m}\frac{1}{(z)^{\Delta_{\Psi}-\delta+m}}A_{-\delta+m}
 \Phi(0),
\eea
where $\delta$ represents the first gap in the corresponding module.
Explicit definitions for the mode operators in the various sectors
have been given in
Eqs.~(\ref{mode10})--(\ref{mode20}) for the singlet sector,
in Eqs.~(\ref{mode1-1})--(\ref{mode21}) for the doublet 1,
and in Eqs.~(\ref{mode1-2})--(\ref{mode22}) for the doublet 2.

The algebra of the mode operators, in the form of commutation relations,
can be calculated by combining these definitions with the algebra of
the chiral operators, Eqs.~(\ref{ope1})--(\ref{ope2}).

\subsection{An example: Computation of 
$\{\Psi^{1},\Psi^{1}\}\Phi^{-1}$}

In analogy with the methods used in Refs.~[1,3],
the commutation relation of the modes $A^{1}_{\mu}$ in the
sector of $\Phi^{-1}$ can be derived from the integral:
\bea
 I=\frac{1}{(2\pi i)^{2}}\oint_{C^{1}_{0}} {\rm d}z' \oint_{C_{0}}{\rm d}z 
\,
 (z')^{\Delta_{1}-\frac{2}{5}+n-1} (z)^{\Delta_{1}-\frac{2}{5}+m-1}\nn\\
 \times (z'-z)^{2\Delta_{1}-\Delta_{2}-1} \Psi^{1}(z')\Psi^{1}(z)
 \Phi^{-1}(0)
 \label{Iexample}
\eea
In this integral the fractions $-2/5$ in the powers of $z'$ and $z$
are in correspondence with the levels of the first descendents
in the module of $\Phi^{-1}$; see Fig.~\ref{fig5}. They will be the first
operators, the dominant ones, in the operator product expansion of
$\Psi^{1}(z')\Phi^{-1}(0)$ and $\Psi^{1}(z)\Phi^{-1}(0)$, respectively.

Let us also comment on the presence of the factor
$(z'-z)^{2\Delta_{1}-\Delta_{2}-1}$. In general, an operator product
$\Psi^{q'}(z') \Psi^{q}(z)$ must be accompanied by a factor of
of $(z'-z)^{\Delta_{q'}+\Delta_{q}-\Delta_{q'+q}}$ to compensate for the
abelian monodromy of the two chiral fields, i.e., the phase factor which
is produced when one of the fields, say $\Psi^{q'}(z')$, is analytically
continued around the other one, $\Psi^q(z)$. To pick up the relevant
terms in the operator product expansions (\ref{ope1})--(\ref{ope2}) we
shall need an additional integer shift $p$, making the total factor
$(z'-z)^{\Delta_{q'}+\Delta_{q}-\Delta_{q'+q}-p}$.

In order to touch all terms appearing explicitly in
Eqs.~(\ref{ope1})--(\ref{ope2}) one should take
$p=2$ when $q'+q \neq 0$ and $p=3$ when $q'+q = 0$. In the latter case
we pick up the stress energy operator $T(z)$ in the expansion, thus making
connection with the Virasoro algebra.

There is one exception to this rule. Namely, in the case of a product of
identical operators, such as $\Psi^{1}(z')\Psi^{1}(z)$ above, keeping
the derivative term $\partial \Psi^2(z)$ in Eq.~(\ref{ope1}) does not
bring new information into the commutation relation. For this reason
we shall take the shift $p=1$ when $q=q'$.

Treating the integral (\ref{Iexample}) in the same way as it has been done
in Ref.~[1,3], one obtains the commutation relation in the form
of an infinite sum:
\beq 
 \sum^{\infty}_{l=0} D^{l}_{-\frac{1}{5}}[A^{1}_{-\frac{3}{5}+n-l}
 A^{1}_{-\frac{2}{5}+m+l}+A_{-\frac{3}{5}+m-l}^{1}A^{1}_{-\frac{2}{5}
 +n+l}]\Phi^{-1}(0)=\lambda^{1,1}_{2}A^{2}_{n+m-1}\Phi^{-1}(0).
 \label{commexample}
\eeq
Here the coefficients $D^{l}_{\alpha}$ are defined by the expansion
\beq
 (1-x)^{\alpha}=\sum^{\infty}_{l=0}D^{l}_{\alpha}x^{l},
\eeq
and in general $\alpha=\Delta_{k'}+\Delta_k-\Delta_{k'+k}-p$.
Note that the sign between the terms on the left-hand side of the
commutation relation (\ref{commexample}) is determined by the parity of 
$p$.
Thus, it is plus for $p=1,3$ and minus for $p=2$.

\subsection{List of commutation relations}

Proceeding in a way similar to the above example, one can obtain
the algebra of the various mode operators, in the various sectors
of representation fields. For convenience, we give the complete
list of algebras in Table~\ref{tab1}. When referring to these relations
we shall use a shorthand notation of the type $\{\Psi^k,\Psi^l\}\Phi^m$,
which is reproduced to the left of each entry in the table.

Presumably not all of the commutation relations given in Table~\ref{tab1}
are needed for the degeneracy calculations, which gave the results 
presented
in Section~\ref{sec2}. But we have found that a large number of them are
needed. So we have chosen to present the whole list, which grew on itself
in various trials of preliminary calculations, without giving extra effort
to reduce it somewhat.

\begin{deluxetable}{lrcl}
 \rotate
\tablecolumns{4}
\tablewidth{0pc} 
\tablecaption{Commutation relations for the parafermionic mode operators
$\Psi^k$, in the basis of singlet and doublet representation fields
$\Phi^m$.} 
\tablehead{}
\startdata 


$\{\Psi^{1},\Psi^{1}\}\Phi^{-2}$ &
$\sum^{\infty}_{l=0} D^{l}_{-\frac{1}{5}}[A^{1}_{-\frac{2}{5}+n-l}
A^{1}_{-\frac{1}{5}+m+l}+A^{1}_{-\frac{2}{5}+m-l}A^{1}_{-\frac{1}{5}+
n+l}]\Phi^{-2}$ &
$=$ &
$\lambda^{1,1}_{2}A^{2}_{-\frac{3}{5}+n+m}\Phi^{-2}$ \\

$\{\Psi^{1},\Psi^{1}\}\Phi^{-1}$ &
$\sum^{\infty}_{l=0} D^{l}_{-\frac{1}{5}}[A^{1}_{-\frac{3}{5}+n-l}
A^{1}_{-\frac{2}{5}+m+l}+A_{-\frac{3}{5}+m-l}^{1}A^{1}_{-\frac{2}{5}
+n+l}]\Phi^{-1}$ &
$=$ &
$\lambda^{1,1}_{2}A^{2}_{n+m-1}\Phi^{-1}$ \\

$\{\Psi^{1},\Psi^{1}\}\Phi^{0}$ &
$\sum^{\infty}_{l=0} D^{l}_{-\frac{1}{5}}[A^{1}_{-\frac{4}{5}+n-l}
A^{1}_{-\frac{3}{5}+m+l}+A^{1}_{-\frac{4}{5}+m-l}A^{1}_{-\frac{3}{5}+n+l}]
\Phi^{0}$ &
$=$ &
$\lambda^{1,1}_{2}A^{2}_{-\frac{2}{5}+n+m-1}\Phi^{0}$ \\

$\{\Psi^{1},\Psi^{1}\}\Phi^{1}$ &
$\sum^{\infty}_{l=0} D^{l}_{-\frac{1}{5}}[A^{1}_{n-l-1}
A^{1}_{-\frac{4}{5}+m+l}+A^{1}_{m-l-1}A^{1}_{-\frac{4}{5}+
n+l}]\Phi^{1}$ &
$=$ &
$\lambda^{1,1}_{2}A^{2}_{-\frac{4}{5}+n+m-1}\Phi^{1}$ \\

$\{\Psi^{1},\Psi^{1}\}\Phi^{2}$ &
$\sum^{\infty}_{l=0} D^{l}_{-\frac{1}{5}}[A^{1}_{-\frac{1}{5}+n-l}
A^{1}_{m+l}+A^{1}_{-\frac{1}{5}+m-l}A^{1}_{n+l}]\Phi^{2}$ &
$=$ &
$\lambda^{1,1}_{2}A^{2}_{-\frac{1}{5}+n+m}\Phi^{2}$ \\


$\{\Psi^{1},\Psi^{-1}\}\Phi^{0}$ &
$\sum^{\infty}_{l=0} D^{l}_{\frac{1}{5}}[A^{1}_{-\frac{2}{5}+n-l}
A^{-1}_{-\frac{3}{5}+m+l}+A^{-1}_{-\frac{2}{5}+m-l}A^{1}_{-\frac{3}{5}
+n+l}]\Phi^{0} $ &
$=$ &
$\{\frac{1}{2}n(n-1)\delta_{n+m-1,0}+
\frac{2\Delta_{1}}{c}L_{n+m-1}\}\Phi^{0}$ \\

$\{\Psi^{1},\Psi^{-1}\}\Phi^{1}$ &
$\sum^{\infty}_{l=0} D^{l}_{\frac{1}{5}}[A^{1}_{-\frac{3}{5}+n-l}
A^{-1}_{-\frac{2}{5}+m+l}+A^{-1}_{-\frac{1}{5}+m-l}A^{1}_{-\frac{4}{5}
+n+l}]\Phi^{1}$ &
$=$ &
$\left\{ \frac{1}{2}(n-\frac{1}{5})(n-\frac{6}{5})
\delta_{n+m-1,0}+\frac{2\Delta_{1}}{c}L_{n+m-1} \right\} \Phi^{1}$ \\

$\{\Psi^{1},\Psi^{-1}\}\Phi^{2}$ &
$\sum^{\infty}_{l=0} D^{l}_{\frac{1}{5}}[A^{-1}_{n-l}
A^{1}_{m+l}+A^{1}_{-\frac{4}{5}+m-l+1}A^{-1}_{-\frac{1}{5}
+n+l}]\Phi^{2}$ &
$=$ &
$\left\{ \frac{1}{2}(n+\frac{2}{5})(n-\frac{3}{5})
\delta_{n+m,0}+\frac{2\Delta_{1}}{c}L_{n+m} \right\} \Phi^{2}$ \\


$\{\Psi^{1},\Psi^{2}\}\Phi^{-2}$ &
$\sum^{\infty}_{l=0} D^{l}_{-\frac{2}{5}}[A^{1}_{-\frac{3}{5}+n-l}
A^{2}_{-\frac{3}{5}+m+l}-A^{2}_{m-l-1}A^{1}_{-\frac{1}{5}+n+l}]
\Phi^{-2}$ &
$=$ &
$\lambda^{1,2}_{-2}\frac{2n-m}{3}A^{-2}_{-\frac{1}{5}+n+m-1}\Phi^{-2}$ \\

$\{\Psi^{1},\Psi^{2}\}\Phi^{-1}$ &
$\sum^{\infty}_{l=0} D^{l}_{-\frac{2}{5}}[A^{1}_{-\frac{4}{5}+n-l}
A^{2}_{m+l}-A^{2}_{-\frac{2}{5}+m-l}A^{1}_{-\frac{2}{5}+n+l}]\Phi^{-1}$ &
$=$ &
$\lambda_{-2}^{1,2}\frac{2n-m-1}{3}A^{-2}_{-\frac{4}{5}+n+m}\Phi^{-1}$ \\

$\{\Psi^{1},\Psi^{2}\}\Phi^{0}$ &
$\sum^{\infty}_{l=0} D^{l}_{-\frac{2}{5}}[A^{2}_{-\frac{4}{5}+n-l}
A^{1}_{-\frac{3}{5}+m+l}-A^{1}_{m-l-1}A^{2}_{-\frac{2}{5}+
n+l}]\Phi^{0}$ &
$=$ &
$\lambda_{-2}^{1,2}\frac{n-2m+1}{3}A^{-2}_{-\frac25+n+m-1} \Phi^0$ \\

$\{\Psi^{1},\Psi^{2}\}\Phi^{1}$ 
&$\sum^{\infty}_{l=0} D^{l}_{-\frac{2}{5}}[A^{2}_{-\frac{1}{5}+n-l-1}
A^{1}_{-\frac{4}{5}+m+l}-A^{1}_{-\frac{1}{5}+m-l-1}A^{2}_{-\frac{4}{5}+
n+l}]\Phi^{1}$ &
$=$ &
$\lambda_{-2}^{1,2}\frac{n-2m+1}{3}A^{-2}_{n+m-2} \Phi^1$ \\

$\{\Psi^{1},\Psi^{2}\}\Phi^{2}$ &
$\sum^{\infty}_{l=0} D^{l}_{-\frac{2}{5}}[A^{1}_{-\frac{2}{5}+n-l}
A^{2}_{-\frac{1}{5}+m+l}-A^{2}_{-\frac{3}{5}+m-l}A^{1}_{n+l}]
\Phi^{2}$ &
$=$ &
$\lambda^{1,2}_{-2}\frac{2n-m}{3}A^{-2}_{-\frac{3}{5}+n+m}\Phi^{2}$ \\


$\{\Psi^{1},\Psi^{-2}\}\Phi^{-2}$ &
$\sum^{\infty}_{l=0} D^{l}_{\frac{2}{5}}[A^{-2}_{-\frac{4}{5}+n-l+1}
A^{1}_{-\frac{1}{5}+m+l}-A^{1}_{-\frac{4}{5}+m-l+1}A^{-2}_{-\frac{1}{5}+n+l}]
\Phi^{-2}$ &
$=$ &
$\lambda^{1,-2}_{-1}\frac{n-3m}{4}A^{-1}_{n+m}\Phi^{-2}$ \\

$\{\Psi^{1},\Psi^{-2}\}\Phi^{-1}$ &
$\sum^{\infty}_{l=0} D^{l}_{\frac{2}{5}}[A^{-2}_{-\frac{2}{5}+n-l}
A^{1}_{-\frac{2}{5}+m+l}-A^{1}_{m-l}A^{-2}_{-\frac{4}{5}+n+l}]
\Phi^{-1}$ &
$=$ &
$\lambda^{1,-2}_{-1}\frac{n-3m}{4}A^{-1}_{-\frac{4}{5}+n+m}\Phi^{-1}$ \\

$\{\Psi^{1},\Psi^{-2}\}\Phi^{0}$ &
$\sum^{\infty}_{l=0} D^{l}_{\frac{2}{5}}[A^{-2}_{n-l}
A^{1}_{-\frac{3}{5}+m+l}-A^{1}_{-\frac{1}{5}+m-l}A^{-2}_{-\frac{2}{5}+
n+l}]\Phi^{0}$ &
$=$ &
$\lambda_{-1}^{1,-2}\frac{n-3m+1}{4}A^{-1}_{-\frac{3}{5}+n+m}\Phi^{0}$ \\

$\{\Psi^{1},\Psi^{-2}\}\Phi^{1}$ &
$\sum^{\infty}_{l=0} D^{l}_{\frac{2}{5}}[A^{1}_{-\frac{2}{5}+n-l}
A^{-2}_{m+l}-A^{-2}_{-\frac{3}{5}+m-l+1}A^{1}_{-\frac{4}{5}+
n+l}]\Phi^{1}$ &
$=$ &
$\lambda_{-1}^{1,-2}\frac{3n-m-2}{4}A^{-1}_{-\frac{2}{5}+n+m}\Phi^{1}$ \\

$\{\Psi^{1},\Psi^{-2}\}\Phi^{2}$ &
$\sum^{\infty}_{l=0} D^{l}_{\frac{2}{5}}[A^{1}_{-\frac{3}{5}+n-l+1}
A^{-2}_{-\frac{3}{5}+m+l}-A^{-2}_{-\frac{1}{5}+m-l}A^{1}_{n+l}]\Phi^{2}$ &
$=$ &
$\lambda^{1,-2}_{-1}\frac{3n-m+1}{4}A^{-1}_{-\frac{1}{5}+n+m}\Phi^{2}$ \\


$\{\Psi^{2},\Psi^{2}\}\Phi^{-2}$ &
$\sum^{\infty}_{l=0} D^{l}_{\frac{11}{5}}[A^{2}_{-\frac{2}{5}+n-l+2}
A^{2}_{-\frac35+m+l}+A^{2}_{-\frac25+m-l+2}A^{2}_{-\frac35+n+l}]\Phi^{-2}$ 
&
$=$ &
$\lambda_{-1}^{2,2}A^{-1}_{n+m+1}\Phi^{-2}$ \\

$\{\Psi^{2},\Psi^{2}\}\Phi^{-1}$ &
$\sum^{\infty}_{l=0} D^{l}_{\frac{11}{5}}[A^{2}_{-\frac{4}{5}+n-l+3}
A^{2}_{m+l}+A^{2}_{-\frac{4}{5}+m-l+3}A^{2}_{n+l}]\Phi^{-1}$ &
$=$ &
$\lambda_{-1}^{2,2}A^{-1}_{-\frac{4}{5}+n+m+3}\Phi^{-1}$ \\

$\{\Psi^{2},\Psi^{2}\}\Phi^{0}$ &
$\sum^{\infty}_{l=0} D^{l}_{\frac{11}{5}}[A^{2}_{-\frac{1}{5}+n-l+2}
A^{2}_{-\frac{2}{5}+m+l}+A^{2}_{-\frac{1}{5}+m-l+2}A^{2}_{-\frac{2}{5}+n+l}]
\Phi^{0}$ &
$=$ &
$\lambda^{2,2}_{-1}A^{-1}_{-\frac{3}{5}+n+m+2}\Phi^{0}$ \\

$\{\Psi^{2},\Psi^{2}\}\Phi^{1}$ &
$\sum^{\infty}_{l=0} D^{l}_{\frac{11}{5}}[A^{2}_{-\frac{3}{5}+n-l+2}
A^{2}_{-\frac{4}{5}+m+l}+A^{2}_{-\frac{3}{5}+m-l+2}A^{2}_{-\frac{4}{5}+n+l}]
\Phi^{1}$ &
$=$ &
$\lambda^{2,2}_{-1}A^{-1}_{-\frac{2}{5}+n+m+1}\Phi^{1}$ \\

$\{\Psi^{2},\Psi^{2}\}\Phi^{2}$ &
$\sum^{\infty}_{l=0} D^{l}_{\frac{11}{5}}[A^{2}_{n-l+2}
A^{2}_{-\frac15+m+l}+A^{2}_{m-l+2}A^{2}_{-\frac15+n+l}]\Phi^{2}$ &
$=$ &
$\lambda_{-1}^{2,2}A^{-1}_{-\frac15+n+m+2}\Phi^{2}$ \\


$\{\Psi^{2},\Psi^{-2}\}\Phi^{0}$ &
$\sum^{\infty}_{l=0} D^{l}_{\frac{9}{5}}[A^{2}_{-\frac35+n-l+2}
A^{-2}_{-\frac25+m+l}+A^{-2}_{-\frac{3}{5}+m-l+2}A^{2}_{-\frac{2}{5}
+n+l}]\Phi^{0} $ &
$=$ &
$\left\{ \frac{1}{2}n(n+1)
\delta_{n+m+1,0}+\frac{2\Delta_{2}}{c}L_{n+m+1} \right\} \Phi^{0}$ \\

$\{\Psi^{2},\Psi^{-2}\}\Phi^{1}$ &
$\sum^{\infty}_{l=0} D^{l}_{\frac{9}{5}}[A^{2}_{n-l+1}
A^{-2}_{m+l}+A^{-2}_{-\frac{1}{5}+m-l+2}A^{2}_{-\frac{4}{5}
+n+l}]\Phi^{1} $ &
$=$ &
$\left\{ \frac{1}{2}(n+\frac{3}{5})(n-\frac{2}{5})
\delta_{n+m+1,0}+\frac{2\Delta_{2}}{c}L_{n+m+1} \right\} \Phi^{1}$ \\

$\{\Psi^{2},\Psi^{-2}\}\Phi^{2}$ &
$\sum^{\infty}_{l=0} D^{l}_{\frac{9}{5}}[A^{2}_{-\frac25+n-l+2}
A^{-2}_{-\frac35+m+l}+A^{-2}_{-\frac{4}{5}+m-l+2}A^{2}_{-\frac{1}{5}
+n+l}]\Phi^{2} $ &
$=$ &
$\left\{ \frac{1}{2}(n+\frac65)(n+\frac15)
\delta_{n+m+1,0}+\frac{2\Delta_{2}}{c}L_{n+m+1} \right\} \Phi^{2}$ \\


$\{T,\Psi\}\Phi$ &
$[L_{n} A_{-\delta+m} - A_{-\delta+m} L_{n}]\Phi$ &
$=$ &
$[(\Delta_{\Psi}-1)n+\delta-m]A_{-\delta+n+m}\Phi$ \\

$\{T,T\}\Phi$ &
$[L_n L_m - L_m L_n] \Phi$ &
$=$ &
$\left\{ (n-m)L_{n+m} + \frac{c}{12} n(n^2-1) \delta_{n+m,0} \right\} 
\Phi$

\label{tab1}

\enddata 
\end{deluxetable}

\subsection{Details of the degeneracy calculation for doublet 1}

For the calculation of degeneracy in the module of a doublet 1 operator,
as presented in Section~\ref{sec:D1}, various matrix elements are needed.
We here show how these can be obtained from the commutation relations 
given
in Table~\ref{tab1}.

For the degeneracy at level $2/5$ one first needs the matrix elements
$A^{1}_{\frac{2}{5}}A^{1}_{-\frac{2}{5}}\Phi^{-1}$ and
 $A^{-1}_{\frac{2}{5}}A^{1}_{-\frac{2}{5}}\Phi^{-1}$.

\begin{itemize}

 \item \underline{$A^{1}_{\frac{2}{5}}A^{1}_{-\frac{2}{5}}\Phi^{-1}$.}
 {}From the commutation relation
 $\{\Psi^1,\Psi^1\}\Phi^{-1}$ with $n=1$, $m=0$ one finds:
 \bea
  A^{1}_{\frac{2}{5}}A^{1}_{-\frac{2}{5}}
  \Phi^{-1} &=&\lambda^{1,1}_{2}A^{2}_{0}\Psi^{-1}, \\
  A^{1}_{\frac{2}{5}}A^{1}_{-\frac{2}{5}}
  \Phi^{-1} &=& \lambda^{1,1}_{2}h_{2,1}\Phi^{1}.
 \eea
 We recall that $h_{2,1}$ is the zero mode eigenvalue.

 \item\underline{$A^{-1}_{\frac{2}{5}}A^{1}_{-\frac{2}{5}}\Phi^{-1}$.}
 The commutation relation
 $\{\Psi^{1},\Psi^{-1}\}\Phi^{1}$ with $n=1$, $m=0$ yields:
 \beq
 A^{-1}_{\frac{2}{5}}A^{1}_{-\frac{2}{5}}
 \Phi^{-1}= \left( -\frac{2}{25}+
 \frac{2\Delta_{1}}{c}\Delta_{\Phi^{1}} \right) \Phi^{-1}.
 \eeq
\end{itemize}

Next, for the degeneracy on level 4/5, one needs four different
matrix elements, namely
$A^{-1}_{\frac{4}{5}}A^{1}_{-\frac{4}{5}}\Phi^{1}$,
$A^{-1}_{\frac{4}{5}}A^{-2}_{-\frac{4}{5}}\Phi^{-1}$,
$A^{2}_{\frac{4}{5}}A^{1}_{-\frac{4}{5}}\Phi^{1}$, and
$A^{2}_{\frac{4}{5}}A^{-2}_{-\frac{4}{5}}\Phi^{-1}$.

\begin{itemize}

 \item\underline{$A^{-1}_{\frac{4}{5}}A^{1}_{-\frac{4}{5}}\Phi^{1}$.}
 The commutation relation
 $\{\Psi^{1},\Psi^{-1}\}\Phi^{1}$ with $n=0$, $m=1$ reads
 \beq
  A^{1}_{\frac{4}{5}}A^{-1}_{-\frac{4}{5}}\Phi^{-1}=
  \left( \frac{3}{25}+\frac{2\Delta_{1}}{c}\Delta_{\Phi^{1}}
  \right) \Phi^{-1}
 \eeq

 \item\underline{$A^{-1}_{\frac{4}{5}}A^{-2}_{-\frac{4}{5}}\Phi^{-1}$.}
 This is found from the commutation relation
 $\{\Psi^1,\Psi^2\}\Phi^1$ with $n=0$, $m=2$:
 \bea
  A^{1}_{\frac{4}{5}}A^{2}_{-\frac{4}{5}}\Phi^{1} &=&
  \lambda^{2,1}_{-2}A^{-2}_{0}\Phi^{1}=\lambda^{2,1}_{-2}h_{2,1}\Phi^{-1},
  \nn\\
  A^{-1}_{\frac{4}{5}}A^{-2}_{-\frac{4}{5}}\Phi^{-1}
  &=&\lambda^{2,1}_{-2}h_{2,1}\Phi^{1}.
 \eea

 \item\underline{$A^{2}_{\frac{4}{5}}A^{1}_{-\frac{4}{5}}\Phi^{1}$.}
 We use the commutation relation
 $\{\Psi^1,\Psi^2\}\Phi^1$ with $n=2$, $m=0$:
 \bea
  A^{2}_{\frac{4}{5}}A^{1}_{-\frac{4}{5}}\Phi^{1} &=&
  \lambda^{2,1}_{-2}A^{-2}_{0}\Phi^{1}=\lambda^{2,1}_{-2}h_{2,1}\Phi^{-1},
  \nn\\
  A^{2}_{\frac{4}{5}}A^{1}_{-\frac{4}{5}}\Phi^{1}
  &=& \lambda^{2,1}_{-2}h_{2,1}\Phi^{-1}
 \eea

 \item\underline{$A^{2}_{\frac{4}{5}}A^{-2}_{-\frac{4}{5}}\Phi^{-1}$.}
 The commutation relation
 $\{\Psi^{2},\Psi^{-2}\}\Phi^{1}$ with $n=0$, $m=-1$ produces:
 \beq
  A^{2}_{\frac{4}{5}}A^{-2}_{-\frac{4}{5}}\Phi^{-1}+
  A^{-2}_{1}A^{2}_{-1}\Phi^{-1}-\frac{9}{5}
  A^{-2}_{0}A^{2}_{0}\Phi^{-1}=\left( \frac{-3}{25}+\frac{2\Delta_{2}}{c}
  \Delta_{\Phi^{1}} \right) \Phi^{-1}
  \label{B31}
 \eeq
 This introduces yet another matrix element. To complete the calculation
 one thus has to compute $A^{-2}_{1}A_{-1}^{2}\Phi^{-1}$. 

 \item\underline{$A^{-2}_{1}A^{2}_{-1}\Phi^{-1}$}
 {}From the commutator
 $\{\Psi^1,\Psi^1\}\Phi^{-1}$ with $n=0$, $m=0$ we get:
 \beq
  2A^{1}_{-\frac{3}{5}}A^{1}_{-\frac{2}{5}}\Phi^{-1} =
  \lambda^{1,1}_{2}A^{2}_{-1}\Phi^{-1}.
  \label{B32}
 \eeq
 Similarly, with $n=1$, $m=1$:
 \beq
  2A^{-1}_{\frac{2}{5}}A^{-1}_{\frac{3}{5}}(\Phi^{1})_{-1}=
\lambda^{1,1}_{2}A^{-2}_{1}(\Phi^{1})_{-1}
  \label{B33}
 \eeq
 where $(\Phi^{1})_{-1} \equiv A^{2}_{-1}\Phi^{-1}$. Combining
 Eqs.~(\ref{B32})--(\ref{B33}) one obtains:
 \beq
  A^{-2}_{1}A^{2}_{-1}\Phi^{-1}=\frac{4}{(\lambda^{2}_{11})^{2}}
  A^{-1}_{\frac{2}{5}}A^{-1}_{\frac{3}{5}}
  A^{1}_{-\frac{3}{5}}A^{1}_{-\frac{2}{5}}\Phi^{-1}.
  \label{B34}
 \eeq
On the right-hand side, the product 
$A^{-1}_{\frac{3}{5}}A^{1}_{-\frac{3}{5}}$
acts on the state 
$A^{1}_{-\frac{2}{5}}\Phi^{-1}\equiv(\Phi^{0})_{-\frac{2}{5}}$.
This can be simplified through the use of the commutator
$\{\Psi^1,\Psi^{-1}\}\Phi^0$ with $n=0$, $m=1$:
\beq
A^{-1}_{\frac{3}{5}}A^{1}_{-\frac{3}{5}}(\Phi^{0})_{-\frac{2}{5}}-
\frac{1}{5}A^{-1}_{-\frac{2}{5}}A^{1}_{\frac{2}{5}}(\Phi^{0})_{-\frac{2}{5}}
+A^{1}_{-\frac{2}{5}}A^{-1}_{\frac{2}{5}}
(\Phi^{0})_{-\frac{2}{5}}=\frac{2\Delta_{1}}{c}
(\Delta_{\Phi^{1}}+\frac{2}{5})(\Phi^{0})_{-\frac{2}{5}}.
\eeq
When substituting this relation into Eq.~(\ref{B34}) it is found that:
\bea
 A^{-2}_{1}A^{2}_{-1}\Phi^{1} &=& \frac{4}{(\lambda^{1,1}_{2})^{2}}
 \left[ 
\frac{1}{5}A^{-1}_{\frac{2}{5}}A^{-1}_{-\frac{2}{5}}A^{1}_{\frac{2}{5}}
 A^{1}_{-\frac{2}{5}}\Phi^{-1}-A^{-1}_{\frac{2}{5}}A^{1}_{-\frac{2}{5}}
 A^{-1}_{\frac{2}{5}}A^{1}_{-\frac{2}{5}}\Phi^{-1}+ \right. \nn \\
 & & \left. \frac{2\Delta_{1}}{c} \left( \Delta_{\Phi^{1}}+
 \frac{2}{5} \right) A^{-1}_{\frac{2}{5}}A^{1}_{-\frac{2}{5}}
 \Phi^{-1} \right] \nn \\
 &=& \frac{4}{(\lambda^{1,1}_{2})^{2}} \left[ \frac{1}{5}
 (\lambda^{1,1}_{2})^{2}(h_{1,2})^{2}- \left( -\frac{2}{25}+
 \frac{2\Delta_{1}}{c}\Delta_{\Phi^{1}} \right)^{2} + \right. \nn \\
 & & \left. \frac{2\Delta_{1}}{c} \left( \Delta_{\Phi^{1}}+
 \frac{2}{5} \right) \left( -\frac{2}{25}+\frac{2\Delta_{1}}{c}
 \Delta_{\Phi^{1}} \right) \right] \Phi^{-1} \nn \\
 &=& \frac{4}{(\lambda^{1,1}_{2})^{2}} \left[ -\frac{4}{5}y^{2}+
 \frac{2\Delta_{1}}{c}(\Delta_{\Phi^{1}}+\frac{2}{5})y \right] \Phi^{-1},
 \label{B36}
\eea
where we have defined
\beq
 y=\lambda^{1,1}_{2}h_{2,1}= -\frac{2}{25}+\frac{2\Delta_{1}}{c}
 \Delta_{\Phi^{1}}.
\eeq
We have here taken into account the degeneracy at level 2/5 which
produced Eq.~(\ref{fixeig1}). We have taken the solution of this equation
with the sign $+$.

Substituting now Eq.~(\ref{B36}) for $A^{-2}_{1}A^{2}_{-1}\Phi^{1}$
into Eq.~(\ref{B31}), and replacing also $A^{-2}_{0}A^{2}_{0}$ by its
eigenvalue $(h_{(2,1)})^{2}$, one obtains finally for $A^{2}_{\frac{4}{5}}
A^{-2}_{-\frac{4}{5}}\Phi^{-1}$ the expression (\ref{matrx4}).

\end{itemize}

Having completed the computation of the necessary matrix elements,
we conclude by giving some additional comments on the procedure of
degenerating the doublet 1 module.

At level 2/5, we have initially degenerated the state 
$\chi^0_{-\frac{2}{5}}$;
see Eq.~(\ref{singD1a}). The coefficients $a$, $b$ have to satisfy the 
equation
\beq
 \cases{a\mu^{(1,1;-1)}+b\mu^{(1,-1;1)}=0,\cr
 a\mu^{(-1,1;-1)}+b\mu^{(-1,-1;1)}=0.\cr}
 \label{B38}
\eeq
The degeneracy condition is:
\beq
 (\mu^{(1,1;-1)})^{2}=(\mu^{(1,-1;1)})^{2},
\eeq
which is Eq.~(\ref{detD1}). If we choose the solution with the sign $+$,
then by Eq.~(\ref{B38}) $a=-b$, and the state
\beq
\chi^0_{-\frac{2}{5}}=A^{1}_{-\frac{2}{5}}\Phi^{-1}-A^{-1}_{-\frac{2}{5}}
\Phi^{1}
\eeq
will be primary. We put it equal to zero, resulting into the equality:
\beq
 A^{1}_{-\frac{2}{5}}\Phi^{-1}=A^{-1}_{-\frac{2}{5}}\Phi^{1}.
\eeq
At this level there will presently remain one state, instead of two, which 
is:
\beq
 \tilde{\chi}^0_{-\frac{2}{5}}=
 A^{1}_{-\frac{2}{5}}\Phi^{-1}+A^{-1}_{-\frac{2}{5}}
\Phi^{1}=2A^{1}_{-\frac{2}{5}}\Phi^{-1}=2A^{-1}_{-\frac{2}{5}}\Phi^{1}.
\eeq

Consider next level 4/5 in the module of the doublet 1, 
cf.~Fig.~\ref{fig5}.
It would appear that, on the side $q=+2$ for instance, we could produce
three states: one with the arrow going down from the summit $\Phi^{-1}$,
another with the arrow going down from the other summit $\Phi^{1}$,
and a last state which is descendent from
the unique remaining state at level 2/5. More precisely, these three
states can be written:
\beq
 A^{1}_{-\frac{4}{5}}\Phi^{1}, \qquad
 A^{-2}_{-\frac{4}{5}}\Phi^{-1}, \qquad \mbox{and} \qquad
 A^{2}_{-\frac{2}{5}}A^{1}_{-\frac{2}{5}}\Phi^{-1}=
 A^{2}_{-\frac{2}{5}}A^{-1}_{-\frac{2}{5}}\Phi^{1}.
 \label{B43}
\eeq
But in fact these states are linearly dependent. Namely, taking $n=m=0$ in
the commutation relation $\{\Psi^1,\Psi^2\}\Phi^{-1}$ one gets:
\bea
 A^{1}_{-\frac{4}{5}}A^{2}_{0}\Phi^{-1}-
 A^{2}_{-\frac{2}{5}}A^{1}_{-\frac{2}{5}}\Phi^{-1} &=&
 -\frac13 \lambda^{1,2}_{-2} A^{-2}_{-\frac{4}{5}}\Phi^{-1}, \nn \\
 h_{2,1}A^{1}_{-\frac{4}{5}}\Phi^{-1}-
 A^{2}_{-\frac{2}{5}}A^{1}_{-\frac{2}{5}}\Phi^{-1} &=&
 -\frac13 \lambda^{1,2}_{-2} A^{-2}_{-\frac{4}{5}}\Phi^{-1}.
\eea
This last relation allows to present the third state in (\ref{B43}) as
a linear combination of the first two states.

In an analogous way one could check that it suffices to demand
that the state $\chi^2_{-\frac{4}{5}}$ in Eq.~(\ref{singD1b}) be annihilated by
$A^{-1}_{+\frac{4}{5}}$ and by $A^{2}_{+\frac{4}{5}}$, resulting
in the matrix elements in Eqs.~(\ref{matrx1})--(\ref{matrx4}).
The annihilation by $A^{-2}_{+\frac{2}{5}}$
\beq
A^{-2}_{+\frac{2}{5}}\chi^2_{-\frac{4}{5}}=0,
\eeq
which is equivalent to the condition
\beq
A^{-1}_{\frac{2}{5}}A^{-2}_{\frac{2}{5}}\chi^2_{-\frac{4}{5}}=0,
\eeq
will then follow automatically by the commutation relations of the
algebra. 

\section{Sector of disorder operators}
\label{appB}

\subsection{Fusion rules and analytic continuations}

To discuss the sector of disorder operators, it is useful to recall
first the content and the multiplication rules of the group $D_{5}$,
which is the symmetry of the theory.

This group contains the rotation elements of $Z_{5}$
\beq
 1, \Omega^{1}, \Omega^{2}, \Omega^{-2}, \Omega^{-1},
 \label{C1}
\eeq
where $\Omega^{1}$ denotes the basic rotation through $+2\pi/5$.
In addition, $D_{5}$ possesses five $Z_{2}$ type reflection elements
\beq
 \left\{ Z^{(a)}_{2}, a=1,2,3,4,5 \right\},
 \label{C2}
\eeq
whose respective reflection axes are shown in Fig.~\ref{fig1}.

In conformal field theory the $Z_{5}$ elements (\ref{C1}) are represented
by the chiral fields:
\beq
 1, \Psi^{1}(z), \Psi^{2}(z), \Psi^{-2}(z), \Psi^{-1}(z),
\eeq
and the $Z_{2}$ elements (\ref{C2}) by the disorder operators:
\beq
 \left\{ R_{a}(z,\bar{z}), a=1,2,3,4,5 \right\}.
\eeq

In order to match the multiplication rules of $D_{5}$, we have to assume
the following fusion rules on the level of the operator algebra, 
\beq
 \Psi^{q}\Psi^{q'} \sim \Psi^{q+q'}. \label{C5}
\eeq
with $q=0,\pm1,\pm2$. The addition of the $Z_5$ ``charges'' $q$ and $q'$ 
is
defined modulo 5. Moreover, for the fusion of a parafermion and a
disorder field:
\bea
 \Psi^{1}R_{a}\sim R_{a-2},&\quad& \Psi^{-1}R_{a}\sim R_{a+2}\nn\\
 \Psi^{2}R_{a}\sim R_{a+1},&\quad& \Psi^{-2}R_{a}\sim R_{a-1}\nn\\
 R_{a}\Psi^{1}\sim R_{a+2},&\quad& R_{a}\Psi^{-1}\sim R_{a-2}\nn\\
 R_{a}\Psi^{2}\sim R_{a-1},&\quad& R_{a}\Psi^{-2}\sim R_{a+1} \label{C6}
\eea
and for the fusion of two disorder fields
\bea
 R_{a+1}R_{a} \sim \Psi^{2},&\quad& R_{a}R_{a+1}\sim\Psi^{-2}\nn\\
 R_{a+2}R_{a} \sim \Psi^{-1},&\quad& R_{a}R_{a+2}\sim\Psi^{1}\nn\\
 R_{a+3}R_{a} \sim \Psi^{1}, &\quad& \mbox{etc.}, \label{C7}
\eea
where once again the indices $a$ are considered modulo 5.

The group $D_{5}$ is non-commutative. In the field theory this
amounts to non-abelian analytic continuation rules of $\Psi^{q}(z)$
around $R_{a}(z',\bar{z}')$.
The definition of these rules is not unique. To obtain consistent
rules one has to fix a certain convention, which is then to be
kept throughout the theory.

For instance, according to Eq.~(\ref{C6}),
$\Psi^{1}(z)R_{1}(0)\sim R_{4}(0)$.
When $\Psi^{1}(z)$ in continued around $R_{1}(0)$, i.e.,
inversing the order of the two operators,
either $\Psi^{1}$ or $R_{1}$ has to be changed to keep
the same result for the product. The choice is either
\beq
 R_{2}(0)\Psi^{1}(z) \label{C8}
\eeq
or
\beq
 R_{1}(0)\Psi^{-1}(z), \label{C9}
\eeq
both of which produce $R_{4}(0)$, according to the fusion rules 
(\ref{C6}).

As will become clear shortly, other possible choices where the indices
of both $R$ and $\Psi$ are changed (but with the product maintained as
$R_{4}$) are not justified by the corresponding lattice realisation
of the operators $R$ and $\Psi$.

\begin{figure}
\begin{center}
 \leavevmode
 \epsfysize=90pt{\epsffile{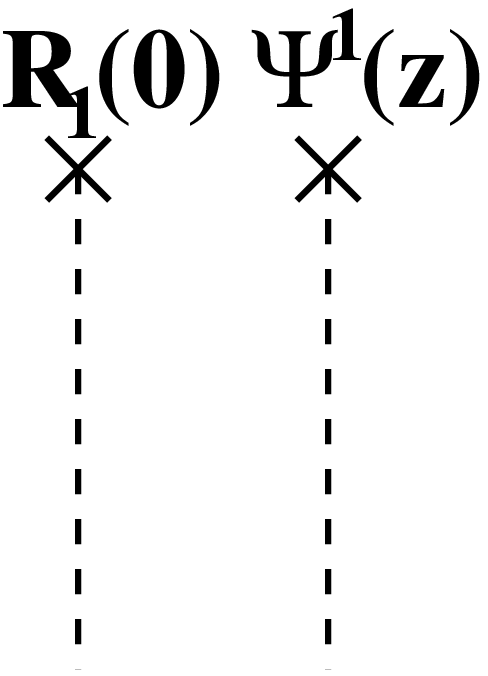}}
 \end{center}
 \protect\caption[3]{\label{figC1}Reference position of the tails (branch
 cuts) of a pair of mutually non-local operators, $R_1(0)$ and 
$\Psi^1(z)$.}
\end{figure}

The choice will be made in the following way. As shown on 
Fig.~\ref{figC1},
the operators $R$ and $\Psi$ are assumed to have tails (branch cuts) 
attached
to them. These tails go to the point at infinity in a reference direction,
which has to be chosen in the same way for all operators.
The tails are due to mutual non-locality of the corresponding operators.
In the continuum theory, the tails correspond to the cut lines of
branching points, if we assume for instance that we are dealing with a
particular correlation function in which $\Psi^{1}(z)R_{1}(0)$ are 
present,
together with some other operators.

More concrete information is contained in the lattice realisation of 
parafermions and disorder operators. As is well-known from the studies of
duality transformations of lattice spin systems with discrete symmetries, 
the
tails of non-local objects like $\Psi$ and $R$ are characterised by the 
corresponding group elements. In the example of Fig.~\ref{figC1}, these
group elements will the $\Omega^{1}$ rotation element for the tail of
$\Psi^{1}(z)$ and the $Z^{(1)}_{2}$ reflection element for the
tail of $R_{1}(0)$.

This type of construction is originally due to Kadanoff and Ceva 
[7],
who worked in the context of the two-dimensional Ising model. The concepts
were further generalised to other statistical models in numerous papers in
the 1970s.

\begin{figure}
\begin{center}
 \leavevmode
 \epsfysize=110pt{\epsffile{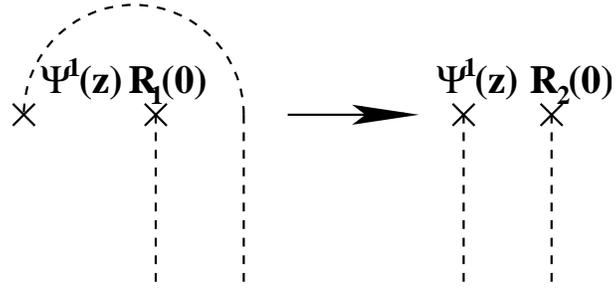}}
 \end{center}
 \protect\caption[3]{\label{figC2}Continuing the operator $\Psi^1(z)$
 above $R_1(0)$. In the process, the latter operator becomes $R_2(0)$.}
\end{figure}

Now, consider moving $\Psi^{1}(z)$ in Fig.~\ref{figC1} so as to place it 
on the
other side of $R_{1}(0)$. This can be realised in two steps. First, we 
move
$\Psi^{1}(z)$ around $R_{1}(0)$ in such a way that its tail passes above
that of $R_{1}(0)$. This configuration is shown in the left part of
Fig.~\ref{figC2}. Next, we pull the tail of $\Psi^{1}(z)$
in order to put it into the reference position (with all tails going 
straight
down). By doing so, we shall have to pass the tail of $\Psi^{1}$ across
the operator $R_{1}(0)$. As a result, $R_{1}(0)$ changes its index
and becomes $R_{2}(0)$, according to the action of the $\Omega^{1}$
group element---characterising the tail of $\Psi^{1}(z)$---on the operator
$R_{1}(0)$. The final configuration is shown on the right part of
Fig.~\ref{figC2}.

In algebraic equations the positioning of the operators is the opposite
of the one shown on the figures, i.e., left to right. The process of the
continuation described above, and shown in Figs.~\ref{figC1}--\ref{figC2},
thus corresponds to:
\beq
 \Psi^{1} \Uparrow R_{1}: \qquad
 \Psi^{1}(z)R_{1}(0) \analcont R_{2}(0)\Psi^{1}(z). \label{C10}
\eeq
(The symbols $\Psi^{1} \Uparrow R_{1}$ are to be read ``continuation of
$\Psi^{1}$ {\em above} $R_{1}$''.)
It could also be remarked that the ordering in algebraic equations of
mutually non-local operators correspond to the ordering of the tails
of these operators; compare Eq.~(\ref{C10}) and 
Figs.~\ref{figC1}--\ref{figC2}.

\begin{figure}
\begin{center}
 \leavevmode
 \epsfysize=110pt{\epsffile{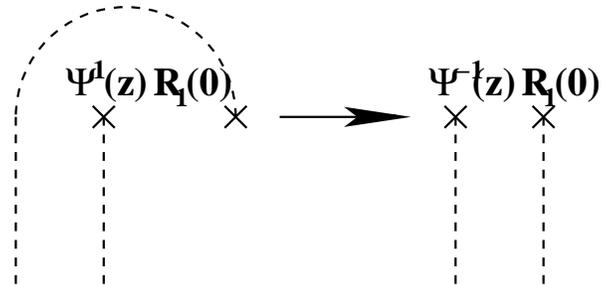}}
 \end{center}
 \protect\caption[3]{\label{figC3}Continuing the operator $\Psi^1(z)$
 below $R_1(0)$. In the process, the former operator becomes 
$\Psi^{-1}(z)$.}
\end{figure}

If we continue $\Psi^{1}(z)$ in Fig.~\ref{figC1} so that it passes {\em 
below}
$R_{1}(0)$ the result will be different. The process is shown in
Fig.~\ref{figC3}. First, as $\Psi^{1}(z)$ moves below $R_{1}(0)$ it pushes
away the tail of $R_{1}(0)$. This situation is shown on the left part of
Fig.~\ref{figC3}. Next, the tail of $R_{1}(0)$ is pulled across
$\Psi^{1}(z)$ changing it to $\Psi^{-1}(z)$, according to the action of 
$Z^{(1)}_{2}$ reflections on $\Psi^{1}(z)$. The tails can now be taken to
the reference position, as shown in the right part of Fig.~\ref{figC3}.
Algebraically this corresponds to:
\beq
 \Psi^{1} \Downarrow R_{1}: \qquad
 \Psi^{1}(z)R_{1}(0) \analcont R_{1}(0)\Psi^{-1}(z). \label{C11}
\eeq
(The symbols $\Psi^{1} \Downarrow R_{1}$ are to be read ``continuation of
$\Psi^{1}$ {\em below} $R_{1}$''.)
We see that the two possible products in Eqs.~(\ref{C8}) and (\ref{C9})
correspond to the two different ways of continuing $\Psi^{1}(z)$ around
$R_{1}(0)$, namely above [Eq.~(\ref{C10})] and below [Eq.~(\ref{C11})].

We mentioned above that different conventions for defining the analytic
continuations are possible. More precisely, these conventions have to
do with the reference positions chosen for the tails. Our convention,
which fixes the rest, is to pull the tails straight down, as a
reference position.

Another possible convention would be to stretch the tails upwards.
With this convention, the results of the continuations in
Eqs.~(\ref{C10})--(\ref{C11}) would be different, but equivalent if the
alternative convention is pursued throughout the calculations.

With the ``all tails down'' convention being chosen, we can now complete 
the
list of fusions given in Eqs.~(\ref{C5})--(\ref{C7}) with the 
corresponding
analytic continuation rules (monodromies) as
$\Psi^{q}(z)$ turns around the disorder operators.
These rules are the following:
\bea
 \Psi^{q} \Uparrow R_{a}: &\qquad&
 \Psi^{q}R_{a} \analcont R_{a+q}\Psi^{q}, \label{C12} \\
 \Psi^{q} \Downarrow R_{a+q}: &\qquad&
 R_{a+q}\Psi^{q} \analcont \Psi^{-q}R_{a+q}, \label{C13} \\
 \Psi^{q} \Downarrow R_{a}: &\qquad&
 \Psi^{q}R_{a} \analcont R_{a}\Psi^{-q}, \\
 \Psi^{-q} \Uparrow R_{a}: &\qquad&
 R_{a}\Psi^{-q} \analcont \Psi^{-q}R_{a+q}. \label{C15}
\eea

It should be mentioned that in all these continuations, 
Eqs.~(\ref{C10})--(\ref{C15}), one has to add phase factors---which have
been suppressed in the above equation---that correspond to the abelian 
part
of non-locality of the operators. Consider for instance the product
$\Psi^{1}(z)R_{1}(0)$. According to the fusion rules (\ref{C6}) we should 
have:
\beq
 \Psi^{1}(z)R_{1}(0)\sim\frac{1}{(z)^{\Delta_{1}}}R_{4}(0)+\ldots. 
\label{C16}
\eeq
When $\Psi^{1}(z)$ is turned around $R_{1}(0)$ in the
positive trigonometric direction, cf.~Eq.~(\ref{C10}) and 
Fig.~\ref{figC2},
the result of the continuation as given by Eq.~(\ref{C10}) must be 
multiplied
by an additional phase factor of $\exp(-i\pi\Delta_{1})$. This phase 
factor
compensates for the phase factor which is produced on the right-hand side
of Eq.~(\ref{C16}) by the prefactor $1/(z)^{\Delta_{1}}$ upon the analytic
continuation of $z$ around $0$.

This means that the complete form of the continuation in Eq.~(\ref{C10})
reads:
\beq
 \Psi^{1} \Uparrow R_{1}: \qquad
 \Psi^{1}(z)R_{1}(0) \analcont {\rm e}^{-i\pi\Delta_{1}} 
R_{2}(0)\Psi^{1}(z).
\eeq
With this convention for the continuation of operators, there will be no
extra phase factors in the fusion rules in Eqs.~(\ref{C5})--(\ref{C7}).
On the other hand, it is understood that in the continuation rules
given by Eqs.~(\ref{C12})--(\ref{C13}), the corresponding phase factors
are to be added in the calculations, even if they are not shown. We have
chosen not to show the phase factors explicitly in
Eqs.~(\ref{C12})--(\ref{C13}) in order to focus on the 
non-abelian nature of the 
continuations.

For the abelian non-locality, as for two $\Psi$ operators in the product
\beq
 \Psi^{q}(z)\Psi^{q'}(0),
\eeq
the indices do not change when $\Psi^{q}(z)$ is continued around
$\Psi^{q'}(0)$. However, it must be remembered that the result of
the continuation still picks up the corresponding abelian phase factor.

\subsection{Mode operators and commutation relations}

We now turn to the definition of mode operators of the $\Psi$
in the disorder sector, and to the calculation of their commutation
relations.

According to the continuation rules (\ref{C12})--(\ref{C13}),
the operator $\Psi^{q}(z)$
has to be turned twice around $R_{a}(0)$ in order to obtain again the 
initial
operators. To match this twofold non-locality of $\Psi^{q}(z)$
with respect to $R_{a}(0)$, the product $\Psi^{q}(z)R_{a}(0)$ has
to be expanded in integer and half-integer powers of $z$. In other words,
it should be a series in powers of $\sqrt{z}$. The general form of this
series is given in Eqs.~(\ref{mode1R})--(\ref{mode2R-}), which defines the
mode operators $\{A^{1}_{\frac{n}{2}}\}$ and $\{A^{2}_{\frac{n}{2}}\}$.
Alternatively, and in accordance with these expansions, the mode operators
could be defined by the integrals:
\bea
 A^{1}_{\frac{n}{2}}R_{a}(0) &=& \frac{1}{4\pi i}\oint_{C_{0}}{\rm d}z
 (z)^{\Delta_{1}
 +\frac{n}{2}-1}\Psi^{1}(z)R_{a}(0), \\
 A^{2}_{\frac{n}{2}}R_{a}(0) &=& \frac{1}{4\pi i}\oint_{C_{0}}{\rm d}z
 (z)^{\Delta_{2}
 +\frac{n}{2}-1}\Psi^{2}(z)R_{a}(0), \\
 (-1)^{n}A^{1}_{\frac{n}{2}}UR_{a}(0) &=& \frac{1}{4\pi i}\oint_{C_{0}}
 {\rm d}z(z)^{\Delta_{1}+\frac{n}{2}-1}\Psi^{-1}(z)R_{a}(0), \\
 (-1)^{n}A^{2}_{\frac{n}{2}}(U)^{2}R_{a}(0) &=&
 \frac{1}{4\pi i}\oint_{C_{0}}{\rm d}z(z)^{\Delta_{2}
 +\frac{n}{2}-1}\Psi^{-2}(z)R_{a}(0).
\eea

\begin{figure}
\begin{center}
 \leavevmode
 \epsfysize=110pt{\epsffile{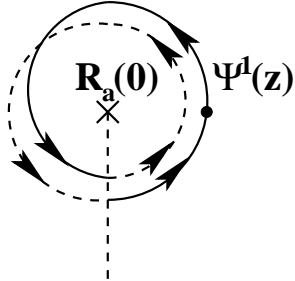}}
 \end{center}
 \protect\caption[3]{\label{figC4}Integration contours used in defining 
the
 parafermionic mode operators in the disorder basis.}
\end{figure}

The integrals over $z$ are to be performed over the double contours around
the operator $R_{a}(0)$, as shown on Fig.~\ref{figC4}, exactly as 
described
in Refs.~[3,6]. Commutation relations for the parafermionic 
mode
operators in the disorder basis can be obtained by a route similar to the
one followed in Refs.~[3,6], by considering the integrals:
\bea
 I_{11} &=& \frac{1}{(4\pi i)^{2}}\oint_{C'_{0}}{\rm d}z'\oint_{C_{0}}
 {\rm d}z(z')^{\Delta_{1}+\frac{n}{2}-1}(z)^{\Delta_{1}+\frac{m}{2}-1} 
\nn\\
 &\times& (\sqrt{z'}-\sqrt{z})^{2\Delta_{1}-\Delta_{2}-1}
 (\sqrt{z'}+\sqrt{z})^{2\Delta_{1}-3}\Psi^{1}(z')\Psi^{1}(z)R_{a}(0), \\
 I_{21} &=& \frac{1}{(4\pi i)^{2}}\oint_{C'_{0}}{\rm d}z'\oint_{C_{0}}
 {\rm d}z(z')^{\Delta_{2}+\frac{n}{2}-1}(z)^{\Delta_{1}+\frac{m}{2}-1} \nn 
\\
 &\times& (\sqrt{z'}-\sqrt{z})^{\Delta_{1}-2}(\sqrt{z'}+
 \sqrt{z})^{\Delta_{2}-2}\Psi^{2}(z')\Psi^{1}(z)R_{a}(0).
\eea
In these integrals, the double contour $C_{0}$ is initially inside the 
double
contour $C'_{0}$. By deforming this configuration of integration contours,
i.e., taking $C_{0}$ outside $C'_{0}$ and picking up additional terms
[3,6], one can derive the commutation relations given below.

\begin{itemize}
 \item \underline{$\{ \Psi^{1},\Psi^{1} \} R_{a}(0)$.}
 {}From the integral $I_{11}$ one obtains
 \bea 
  \sum^{\infty}_{l=0}D^{l}_{\alpha\beta} \left( A^{1}_{\frac{n-l}{2}}
  A^{1}_{\frac{m+l}{2}} \right. &+& \left.
  A^{1}_{\frac{m-l}{2}}A^{1}_{\frac{n+l}{2}} \right) R_{a}(0) =
  \lambda^{1,1}_{2} 2^{\Delta_{2}-3}A^{2}_{\frac{n+m}{2}}R_{a}(0)
  \label{C25} \\
  &+& (-1)^{n} 2^{-\Delta_{2}-2} \left[ \kappa(n)\delta_{n+m,0} +
  \frac{16\Delta_{1}}{c} L_{\frac{n+m}{2}} \right] ({\sf U})^{-1}R_{a}(0), 
\nn
 \eea
 where $\alpha=2\Delta_{1}-\Delta_{2}-1=-\frac{1}{5}$ and
 $\beta=2\Delta_{1}-3=\frac{1}{5}$. Furthermore, we have
 \beq
  \kappa(n)=(2\Delta_{1}+n-1)(2\Delta_{1}+n-2)-(2\Delta_{1}
  +n-1)(\Delta_{2}+1)+\frac{(\Delta_{2}+1)(\Delta_{2}+2)}{4} \label{C26}
 \eeq
 and the coefficients $D^{l}_{\alpha\beta}$ are defined as:
 \beq
  (1-x)^{\alpha}(1+x)^{\beta}=\sum^{\infty}_{l=0}D^{l}_{\alpha\beta}x^{l}.
  \label{C27}
 \eeq
 Finally, we recall that the matrix ${\sf U}$ is a $5 \times 5$ matrix 
which
 rotates the index of the disorder field backwards by one unit:
 ${\sf U} R_{a}(0) = R_{a-1}(0)$.
 \item \underline{$\{\Psi^{2},\Psi^{1}\}R_{a}(0)$.}
 Similarly, starting from the integral $I_{21}$ one can derive the 
 following commutation relation:
 \bea 
  \sum^{\infty}_{l=0}D^{l}_{\alpha'\beta'} \left( A^{2}_{\frac{n-l}{2}}
  A^{1}_{\frac{m+l}{2}} \right. &-& \left. A^{1}_{\frac{m-l}{2}}
  A^{2}_{\frac{n+l}{2}} \right) R_{a}(0) = 
  (-1)^{n+m} 2^{\Delta_{2}-\Delta_{1}-2}\lambda^{2,1}_{-2}
  \frac{n-2m}{3}A^{2}_{\frac{n+m}{2}}({\sf U})^{2}R_{a}(0) \nn \\
  &+& 
(-1)^{m} 2^{\Delta_{1}-\Delta_{2}-2}\lambda^{-2,1}_{-1}\frac{n-3m}{4}
  A^{1}_{\frac{n+m}{2}}({\sf U})^{-1}R_{a}(0), \label{C28}
 \eea
 where $\alpha'=\Delta_{1}-2=-\frac{2}{5}$ and
 $\beta'=\Delta_{2}-2=\frac{2}{5}$.
 The coefficients $D^{l}_{\alpha'\beta'}$ have the same meaning as in
 Eq.~(\ref{C27}); $\lambda^{2,1}_{-2}$ and $\lambda^{-2,1}_{-1}$
 are the structure constants of the parafermionic algebra (\ref{ope1}).
\end{itemize} 

The commutation relation $\{\Psi^{2},\Psi^{2}\}R_{a}(0)$ could be
derived in a similar way, but it turns out not to be useful in the
calculations of degeneracies in the modules. The relations
$\{\Psi^{1},\Psi^{1}\}R_{a}(0)$ and $\{\Psi^{2},\Psi^{1}\}R_{a}(0)$
given above are sufficient.

\section{Differential and characteristic equations for the
correlation functions
$\left \langle R^{(2)}\Phi^{q}R^{(1)} \right \rangle$,
$\left \langle \Phi^{2}_{(2)}\Phi^{0}\Phi^{-2}_{(1)} \right \rangle$,
$\left \langle \Phi^{2}_{(2)}\Phi^{1}\Phi^{2}_{(1)} \right \rangle$ and
$\left \langle \Phi^{1}_{(3)}\Phi^{1}_{(2)}\Phi^{-2}_{(1)} \right 
\rangle$.}
\label{appC}

We first show how to compute the differential and characteristic equations
for the
correlation function $\left \langle R^{(2)}\Phi^{q}R^{(1)} \right \rangle$
in the sectors $q=0,\pm1$.

\subsection{$\left \langle 
R^{(2)}_{4}(z_{2})\Phi^{1}(z_{3})R^{(1)}_{1}(z_{1})
\right \rangle$}

It is assumed that the module of $R^{(1)}_{1}(z_{1})$ is completely 
degenerate
at level $\frac{1}{2}$. This implies that:
\beq
 A^{1}_{-\frac{1}{2}}R^{(1)}=A^{2}_{-\frac{1}{2}}R^{(1)}=0.
\eeq
Using this fact in the commutation relations (\ref{C25}) and (\ref{C28}),
with $n=m=-1$, one finds the equation
\beq
 A^{1}_{-1}R^{(1)}_{3}=gL_{-1}R^{(1)}_{1}. \label{aD2}
\eeq
The coefficient of proportionality $g$ in Eq.~(\ref{aD2}) is explicitly
determined by this algebraic calculation. But it will also be defined
independently by the differential equation for the function
$\left \langle R^{(2)}_{4}(z_{2})\Phi^{1}(z_{3})R^{(1)}_{1}(z_{1})
\right \rangle$, which follows from Eq.~(\ref{aD2}), i.e., by assuming 
only
the proportionality between the operators on the left- and right-hand 
sides.
It can be checked that the two ways of defining the coefficient
$g$ are consistent. For this reason, and in order to avoid giving
additional details, we shall leave Eq.~(\ref{aD2}) as it is, with $g$
undetermined. As just stated, the differential equation will eventually
determine $g$, and in a form which is much simpler than that obtained
from the direct calculation of $g$ by the algebra.

{}From Eq.~(\ref{aD2}) it follows that the function
$\left \langle R^{(2)}\Phi^{1}R^{(1)} \right \rangle$ satisfies the 
equation:
\beq
 \left \langle R^{(2)}_{4}(z_{2})
 \Phi^{1}(z_{3})(A^{1}_{-1}R^{(1)}_{3}(z_{1})) \right \rangle =
 g\frac{\partial}{\partial z_{1}} \left \langle R^{(2)}_{4}(z_{2})
 \Phi^{1}(z_{3})R^{(1)}_{1}(z_{1}) \right \rangle. \label{aD3}
\eeq
It remains to reexpress the function on the left-hand side in terms of the
initial function $\left \langle R^{(2)}\Phi^{1}R^{(1)} \right \rangle$.

\begin{figure}
\begin{center}
 \leavevmode
 \epsfysize=110pt{\epsffile{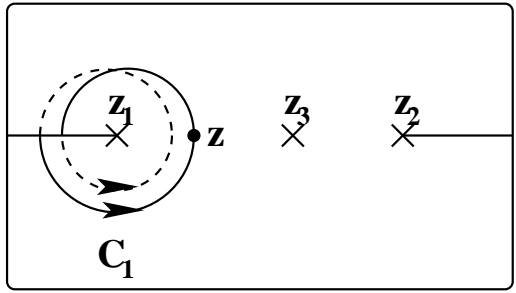}}
 \end{center}
 \protect\caption[3]{\label{figD1}Initial integration contour used in the
 definition of the integral (\ref{aD4}).}
\end{figure}

For this purpose we shall consider the integral
\bea
 I_{1}&=&\frac{1}{4\pi i}\oint_{C_{1}}{\rm d}z(z-z_{1})^{\Delta_{1}-2}
 \left \langle 
R^{(2)}_{4}(z_{2})\Phi^{1}(z_{3})\Psi^{1}(z)R^{(1)}_{3}(z_{1})
 \right \rangle \nn\\
 &\times& \left( 1-\sqrt{\eta} \right)^{\Delta_{1}-\frac{4}{5}+1}
 \left( 1+\sqrt{\eta} \right)^{\Delta_{1}-
 \frac{2}{5}+1}(z_{3}-z)^{-1}(z_{2}-z)^{\Delta_{1}+1}, \label{aD4}
\eea
which has been defined in terms of the anharmonic ratio
\beq
 \eta=\sqrt{\frac{(z-z_{1})z_{23}}{z_{31}(z_{2}-z)}} \label{aD5}
\eeq
with $z_{ij} \equiv z_{i}-z_{j}$.
The contur $C_{1}$ is shown in Fig.~\ref{figD1}. In order to return to the
original Riemann sheet, $C_1$ must wind around the point $z_{1}$ twice.
This is so because of the twofold monodromy of $\Psi^{1}(z)$ with respect 
to
$R^{(1)}_{3}(z_{1})$ in the correlation function in Eq.~(\ref{aD4}).

The extra factors in the integral (\ref{aD4}) are chosen to make the whole
expression in the integral a uniform function of $z$ on the two-sheet
Riemann surface. The passage between the two sheets is made by crossing
the branch cuts shown in Fig.~\ref{figD1}. Apart from this general 
condition,
there is a number of extra conditions which fix the powers of the various
factors in Eq.~(\ref{aD4}) completely. These extra conditions are:
\begin{enumerate}
 \item to get a leading contribution coming from the integral over the
 contour $C_{1}$ in Eq.~(\ref{aD4}), cf.~Fig.~\ref{figD1}, that is 
proportional
 to the function $\left \langle R^{(2)}_{4}(z_{2})
 \Phi^{1}(z_{3})\left(A^{1}_{-1}R^{(1)}_{3}(z_{1})\right) \right \rangle$;
 \item to get a contribution from the integral around the point $z_{2}$
 [under a transformation of the contours which will be given below]
 that is proportional to the function
 $\left \langle R^{(2)}\Phi^{1}R^{(1)} \right \rangle$;
 \item to avoid contributions from the contours around the point $z_{3}$, 
which would
 involve correlation functions with descendent operators in the module of
 $\Phi^{1}(z_{3})$;
 \item to ensure that the integrals over the contours at infinity vanish.
\end{enumerate}

\begin{figure}
\begin{center}
 \leavevmode
 \epsfysize=190pt{\epsffile{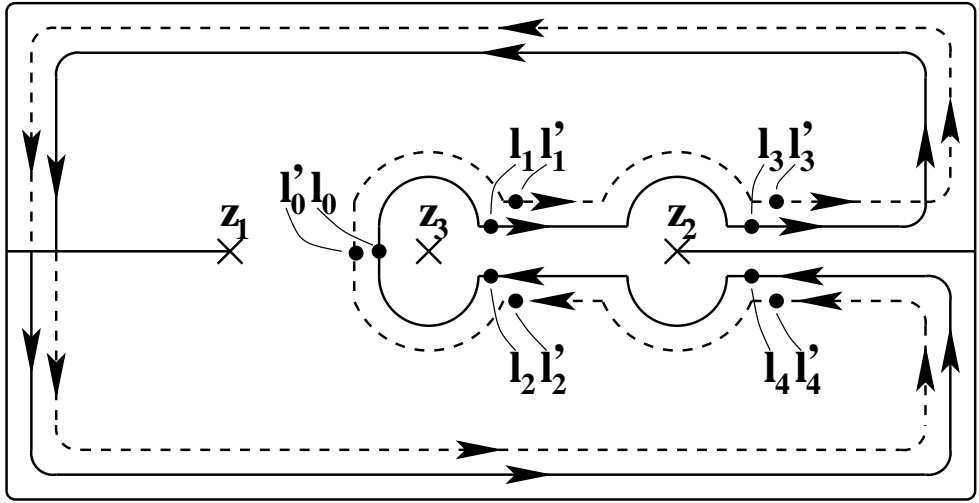}}
 \end{center}
 \protect\caption[3]{\label{figD2}Deformation of the integration contour
 of Fig.~\ref{figD1}.}
\end{figure}

Once the above conditions are met, we can take advantage of transforming
the contour $C_{1}$ in Fig.~\ref{figD1} into the
configuration shown in Fig.~\ref{figD2}.

The initial position of $z$ is denoted $l_{0}$ in Fig.~\ref{figD2}. In 
this
position there are no extra phases or sign factors, and the operators 
assume
their initial configuration, as in the correlation function in 
Eq.~(\ref{aD4}).
To get the correct expression of the integrand in Eq.~(\ref{aD4}) for 
other
parts of the contour [these parts carry various labels $l_k$ in
Fig.~\ref{figD2}], the extra factors and the operators in the correlation
function in Eq.~(\ref{aD4}) have to be modified with respect to the 
initial
configuration $l_{0}$. This modification takes the form of an analytic
continuation for the extra factors, and is done according to the rules
describled in Appendix~\ref{appB} for the operators. It should be observed
that the whole operation is somewhat delicate.

\begin{figure}
\begin{center}
 \leavevmode
 \epsfysize=120pt{\epsffile{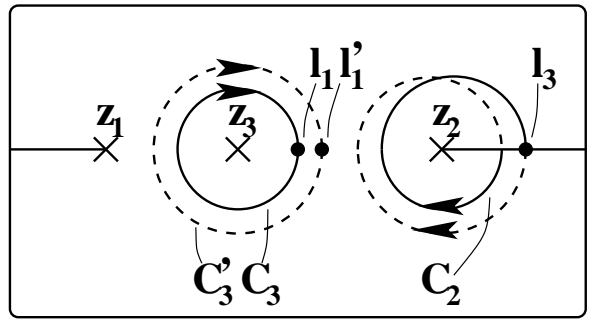}}
 \end{center}
 \protect\caption[3]{\label{figD3}Final integration contours for the
 evaluation of Eq.~(\ref{aD4}).}
\end{figure}

When this analysis is done one finds that there is a partial cancellation 
of
the integration paths in Fig.~\ref{figD2}, and that the integrals over the
paths at infinity vanish (as they should, according to the above 
conditions).
One is then finally left with the contours of integration shown in
Fig.~\ref{figD3}. The positions $l_{1}$, $l'_{1}$ and $l_{3}$ of the 
variable
$z$ are chosen to be the initial ones for the corresponding integrands.

In this way one finds that the integral $I_{1}$ in Eq.~(\ref{aD4}),
cf.~Fig.~\ref{figD1}, can be decomposed as
\beq
 I_{1}=I_{3}+I'_{3}+I_{2}, \label{aD6}
\eeq
where $I_{3}$, $I'_{3}$ and $I_{2}$ are respectively the integrals over 
the
contours $C_{3}$, $C'_{3}$ and $C_{2}$ in Fig.~\ref{figD3}.

One may check that the integrals $I_{3}$ and $I'_{3}$ vanish. In fact,
\bea
 I_{3} &\propto& \oint_{C_{3}}{\rm d}z \, \left \langle
 R^{(2)}_{4}(z_{2}) \left( \Psi^{1}(z)\Phi^{1}(z_{3}) \right)
 R^{(1)}_{3}(z_{1}) \right \rangle (z-z_{3})^{\Delta_{1}-\frac{4}{5}} \nn 
\\
 &\propto& \left \langle R^{(2)}_{4}(z_{2}) \left(
 A^{1}_{+\frac{1}{5}}\Phi^{1}(z_{3}) \right) R^{(1)}_{3}(z_{1})
 \right \rangle = 0.
\eea
In a similar fashion we find that
\bea
 I'_{3} &\propto& \oint_{C'_{3}}{\rm d}z \, \left \langle
 R^{(2)}_{4}(z_{2}) \left( \Psi^{-1}(z)\Phi^{1}(z_{3}) \right)
 R^{(1)}_{4}(z_{1}) \right \rangle (z-z_{3})^{\Delta_{1}-\frac{2}{5}} \nn 
\\
 &\propto& \left \langle R^{(2)}_{4}(z_{2}) \left(
 A^{-1}_{+\frac{3}{5}}\Phi^{1}(z_{3}) \right) R^{(1)}_{4}(z_{1})
 \right \rangle = 0.
\eea
For the integral $I_{2}$ one finds:
\bea
 I_{2} &=& \frac{1}{4\pi i}\oint_{C_{2}^{\ast}}{\rm d}z \
 (z-z_{2})^{\Delta_{1}-1} \frac{(z_{21})^{\Delta_{1}}z_{23}}{(z_{31})^{2}}
 \left \langle \left( \Psi^{1}(z)R^{(2)}_{3}(z_{2}) \right) 
\Phi^{1}(z_{3}) 
 R^{(1)}_{3}(z_{1}) \right \rangle \nn \\
 &=& \frac{(z_{21})^{\Delta_{1}}z_{23}}{(z_{31})^{2}}
 \left \langle \left( A^{1}_{0}R^{(2)}_{3}(z_{2}) \right)
 \Phi^{1}(z_{3}))R^{(1)}_{3}(z_{1}) \right \rangle \nn \\
 &=& h^{(2)}_{1}\frac{(z_{21})^{\Delta_{1}}z_{23}}{(z_{31})^{2}}
 \left \langle R^{(2)}_{1}(z_{2})\Phi^{1}(z_{3}))R^{(1)}_{3}(z_{1})
 \right \rangle.
\eea
Here, $C^{\ast}_{2}$ is the contour $C_{2}$ with the orientation reversed,
and $h^{(2)}_{1}$ is the zero mode eigenvalue which has been defined in
Section~\ref{sec2} of the main text.

Finally, after some expansions of the extra factors in Eq.~(\ref{aD4}),
one finds the following result for the initial integral $I_{1}$:
\bea
 I_{1} &=& \frac{(z_{21})^{\Delta_{1}+1}}{(z_{31})}
 \left\{ \left \langle R^{(2)}_{4}(z_{2})\Phi^{1}(z_{3}) \left(
 A^{1}_{-1}R^{(1)}_{3}(z_{1}) \right) \right \rangle \right. \nn\\
 &+& \left. \left( -\frac{23}{25}\frac{1}{z_{31}}-
 \frac{17}{25}\frac{1}{z_{21}} \right)
 h^{(1)}_{1} \, \left \langle R^{(2)}_{4}(z_{2})\Phi^{1}(z_{3})
 R^{(1)}_{1}(z_{1}) \right \rangle \right\}.
\eea
According to Eq.~(\ref{aD6}),
\beq
 I_{1}=I_{2}
\eeq
because $I_{3}$ and $I'_{3}$ vanish. Substituting the results given above 
for
the integrals $I_{1}$ and $I_{2}$ one obtains the equation:
\bea
 \left \langle R^{(2)}_{4}(z_{2})\Phi^{1}(z_{3}) \left(
 A^{1}_{-1}R^{(1)}_{3}(z_{1}) \right) \right \rangle &=&
 \left[ \frac{1}{z_{31}} \left( \frac{23}{25}h^{(1)}_{1}+h^{(2)}_{1} 
\right) +
 \frac{1}{z_{21}} \left( \frac{17}{25}h^{(1)}_{1}-h^{(2)}_{1} \right)
 \right ] \nn \\
 &\times& \left \langle 
R^{(2)}_{4}(z_{2})\Phi^{1}(z_{3})R^{(1)}_{1}(z_{1})
 \right \rangle. \label{aD12}
\eea
In getting this result we have use the identity
\beq
 \left \langle R^{(2)}_{4}\Phi^{1}R^{(1)}_{1} \right \rangle =
 \left \langle R^{(2)}_{1}\Phi^{1}R^{(1)}_{3} \right \rangle,
\eeq
which is valid because the two functions differ by a $Z_{5}$ rotation of the
indices, by two units, of the operators $R^{(2)}$ and $R^{(1)}$. This
rotation is a symmetry of the theory.

Substituting the expression (\ref{aD12}) into the left-hand side of 
Eq.~(\ref{aD3}) one obtains the differential equation for the function
$\left \langle R^{(2)}\Phi^{1}R^{(1)} \right \rangle$:
\bea
 g\frac{\partial}{\partial z_{1}} \left \langle
 R^{(2)}_{4}(z_{2})\Phi^{1}(z_{3})R^{(1)}_{1}(z_{1}) \right \rangle &=&
 \left[ \frac{1}{z_{31}} \left( \frac{23}{25}h^{(1)}_{1}+h^{(2)}_{1} 
\right)
 +\frac{1}{z_{21}}
 \left( \frac{17}{25}h^{(1)}_{1}-h^{(2)}_{1} \right) \right] \nn\\
 &\times& \left \langle 
R^{(2)}_{4}(z_{2})\Phi^{1}(z_{3})R^{(1)}_{1}(z_{1})
 \right \rangle. \label{aD14}
\eea

The characteristic equation for dimensions of the operators can be 
obtained
by substituting the functional form of the correlation function
$\left \langle R^{(2)}_{4}\Phi^{1}R^{(1)}_{1} \right \rangle$, which is
uniquely determined by conformal invariance:
\bea
 & & \left \langle R^{(2)}_{4}(z_{2})\Phi^{1}(z_{3})R^{(1)}_{1}(z_{1})
 \right \rangle \nn \\
 & & = \frac{\rm const.}{(z_{23})^{\Delta_{R^{(2)}}+\Delta_{\Phi^{1}}-
 \Delta_{R^{(1)}}}(z_{12})^{\Delta_{R^{(2)}}+\Delta_{R^{(1)}}-
 \Delta_{\Phi^{1}}}(z_{13})^{\Delta_{\Phi^{1}}+\Delta_{R^{(1)}}-
 \Delta_{R^{(2)}}}}.
\eea
This gives:
\bea
 & & g \left( \frac{\Delta_{R^{(2)}}+\Delta_{R^{(1)}}-
 \Delta_{\Phi^{1}}}{z_{21}} +
 \frac{\Delta_{\Phi^{1}}+\Delta_{R^{(1)}}-\Delta_{R^{(2)}}}{z_{31}} 
\right)
 \nn\\
 & & = \frac{1}{z_{31}} \left( \frac{23}{25}h^{(1)}_{1}+h^{(2)}_{1} 
\right) +
 \frac{1}{z_{21}} \left( \frac{17}{25}h^{(1)}_{1}-h^{(2)}_{1} \right),
\eea
and finally:
\bea
 g(\Delta_{R^{(2)}}+\Delta_{R^{(1)}}-\Delta_{\Phi^{1}}) &=&
 \frac{17}{25}h^{(1)}_{1}-h^{(2)}_{1} \label{aD17} \\
 g(\Delta_{\Phi^{1}}+\Delta_{R^{(1)}}-\Delta_{R^{(2)}}) &=&
 \frac{23}{25}h^{(1)}_{1}+h^{(2)}_{1}. \label{aD18}
\eea
{}From these equations it follows that
\beq
 h^{(1)}_{1}=\frac{5}{4}g\Delta_{R^{(1)}},
\eeq
or
\beq
 g=\frac{4}{5}\frac{h^{(1)}_{1}}{\Delta_{R^{(1)}}}. \label{aD20}
\eeq
In this way we have determined the value of the constant of 
proportionality
in Eq.~(\ref{aD2}), as has been discussed at the beginning of this
demonstration.

The second equation that is obtained from the system
(\ref{aD17})--(\ref{aD18}) reads
\beq
 g \left( \Delta_{R^{(2)}}-\Delta_{\Phi^{1}}+\frac{3}{20}\Delta_{R^{(1)}}
 \right) = -h^{(2)}_{1}.
\eeq
Substituting the value of $g$ determined in Eq.~(\ref{aD20}), one can 
present
this equation in the form:
\beq
 \Delta_{R^{(2)}}-\Delta_{\Phi^{1}}+\frac{3}{20}\Delta_{R^{(1)}}=
 -\frac{5}{4}\frac{h^{(2)}_{1}}{h^{(1)}_{1}}\Delta_{R^{(1)}}. \label{aD22}
\eeq

\subsection{$\left \langle 
R^{(2)}_{1}(z_{2})\Phi^{0}(z_{3})R^{(1)}_{1}(z_{1})
\right \rangle$}

In a similar way as has been described above, but starting from the 
function
$\left \langle R^{(2)}_{1}\Phi^{0}R^{(1)}_{1} \right \rangle$ and the 
integral:
\bea
 I_{1} &=& \frac{1}{4\pi i}\oint_{C_{1}}{\rm d}z \, 
(z-z_{1})^{\Delta_{1}-2}
 \left \langle 
R^{(2)}_{1}(z_{2})\Phi^{0}(z_{3})\Psi^{1}(z)R^{(1)}_{3}(z_{1})
 \right \rangle \nn \\
 &\times& (1-\sqrt{\eta})^{\Delta_{1}-\frac{3}{5}+1}
 (1+\sqrt{\eta})^{\Delta_{1}-
 \frac{3}{5}+1}(z_{3}-z)^{-1}(z_{2}-z)^{\Delta_{1}+1}, \label{aD23}
\eea
while still using Eq.~(\ref{aD2}), one obtains the following
differential equation:
\bea
 g\frac{\partial}{\partial z_{1}} \left \langle 
 R^{(2)}_{1}(z_{2})\Phi^{0}(z_{3})R^{(1)}_{1}(z_{1}) \right \rangle &=&
 \left[ \frac{1}{z_{31}} \left( h^{(1)}_{1}-h^{(2)}_{1} \right) +
 \frac{1}{z_{21}} \left( \frac{3}{5}h^{(1)}_{1}+h^{(2)}_{1} \right) 
\right]
 \nn\\
 &\times& \left \langle 
R^{(2)}_{1}(z_{2})\Phi^{0}(z_{3})R^{(1)}_{1}(z_{1})
 \right \rangle. \label{aD24}
\eea
Using the same procedure as for the function
$\left \langle R^{(2)}_{4}\Phi^{1}R^{(1)}_{1} \right \rangle$,
cf.~Eq.~(\ref{aD14}), one gets from Eq.~(\ref{aD24}) the following
characteristic equation for the dimensions:
\beq
 \Delta_{R^{(2)}}-\Delta_{\Phi^{0}}+\frac{1}{4}\Delta_{R^{(1)}}=
 \frac{5}{4}\frac{h^{(2)}_{1}}{h^{(1)}_{1}}\Delta_{R^{(1)}} \label{aD25}
\eeq
Eqs.~(\ref{aD22}) and (\ref{aD25}) can be cast in the joint form
\beq
 \Delta_{R^{(2)}}-\Delta_{\Phi^{q}}+\frac{1}{4}
 \left( 1-\frac{2q^{2}}{5} \right) \Delta_{R^{(1)}}=
 (-1)^{q}\frac{5}{4}\frac{h^{(2)}_{1}}{h^{(1)}_{1}}\Delta_{R^{(1)}}.
\eeq
This last equation summarises the characteristic equations for the 
function
\beq
 \left \langle R^{(2)}(z_{2})\Phi^{q}(z_{3})R^{(1)}(z_{1}) \right \rangle
\eeq
with $q=0,\pm 1.$

For this same function with $q=2$, we have not yet succeded in defining
the appropriate integral which would verify the conditions analogous to
those given above after Eqs.~(\ref{aD4})--(\ref{aD5}).
Still, we have made some progress in the study of other correlation 
functions
containing the doublet 2 operator $\Phi^{\pm 2}$.

\subsection{$\left \langle
\Phi^{2}_{(2)}(z_{3})\Phi^{0}(z_{2})\Phi^{-2}_{(1)}(z_{1}) \right 
\rangle$}

Here the module of the operator $\Phi^{-2}_{(1)}(z_{1})$ is assumed to be
degenerate at levels $\frac{1}{5}$ and 1. Then, according to the analysis 
in 
Section~\ref{sec2}, the following relation holds:
\beq
 A^{1}_{-1}\Phi^{2}_{(1)}=\frac{\mu_{12}}{\mu_{11}}L_{-1}\Phi^{-2}_{(1)},
 \label{aD28}
\eeq
where $\mu_{11}$ and $\mu_{12}$ are the matrix elements defined in
Eqs.~(\ref{matele11})--(\ref{matele12}):
$\mu_{11}=2\Delta_{\Phi^{2}_{(1)}}$ and
$\mu_{12}=\frac{5}{8}h^{(1)}_{1,2}$.

In this case, using Eq.~(\ref{aD28}) and the integral
\bea
 I_{1} &=& \frac{1}{2\pi i}\oint_{C_{1}}{\rm d}z \,
 \left \langle \Phi^{2}_{(2)}(z_{3})\Phi^{0}(z_{2})\Psi^{1}(z)
 \Phi^{2}_{(1)}(z_{1}) \right \rangle \nn \\
 &\times& (z-z_{1})^{\Delta_{1}-2}(z_{2}-z)^{\Delta_{1}-
 \frac{3}{5}}(z_{3}-z)^{\Delta_{1}-1}, \label{aD29}
\eea
one derives the following differential equation:
\bea
 \frac{\mu_{12}}{\mu _{11}} \frac{\partial}{\partial {z_{1}}} \left 
\langle
 \Phi^{2}_{(2)}(z_{3})\Phi^{0}(z_{2}) \Phi^{-2}_{(1)}(z_{1}) \right 
\rangle
 &=& \left[ \frac{1}{z_{21}} \left( h^{(1)}_{1,2}-h^{(2)}_{1,2} \right)+
 \frac{1}{z_{31}} \left( \frac{3}{5}h^{(1)}_{1,2}+h^{(2)}_{1,2} \right)
 \right ] \nn\\
 &\times& \left \langle \Phi^{2}_{(2)}(z_{3})\Phi^{0}(z_{2})
 \Phi^{-2}_{(1)}(z_{1}) \right \rangle. \label{aD30}
\eea
It could be remarked that the transformation of contours in the integral
(\ref{aD29}), needed to derive the right-hand side in Eq.~(\ref{aD30}),
is much simpler than the transformation employed for the integrals
(\ref{aD4}) and (\ref{aD23}) containing correlation functions
with disorder operators.

{}From Eq.~(\ref{aD30}) one gets the following characteristic equation:
\beq
\Delta_{\Phi^{0}}-\Delta_{\Phi^{2}_{(2)}}-\frac{1}{4}\Delta_{\Phi^{2}_{(1)}}=
-\frac{5}{4}\frac{h^{(2)}_{1,2}}{h^{(1)}_{1,2}}\Delta_{\Phi_{(1)}^{2}}
\eeq

According to Eq.~(\ref{fixeig2}), the eigenvalue $h^{(1)}_{1,2}$
is defined by the equation
\beq
 \left( h^{(1)}_{1,2} \right)^{2} =
 -\frac{3}{25}+\frac{2\Delta_{1}}{c}\Delta_{\Phi^{2}_{(1)}}.
\eeq
But the other eigenvalue, $h^{(2)}_{1,2}$, for the general doublet 2 
operator
$\Phi^{2}_{(2)}(z_{2})$ remains unknown and has to be defined together 
with
the dimensions $\Delta_{\Phi^{2}_{(2)}}$ and $\Delta_{\Phi^{0}}$. 

\subsection{$\left \langle
\Phi^{2}_{(2)}(z_{3})\Phi^{1}(z_{2})\Phi^{2}_{(1)}(z_{1}) \right \rangle$}

Here the module of $\Phi^{2}_{(1)}(z_{1})$ is assumed to be degenerate on
levels $\frac{1}{5}$ and 1. We also impose a restriction, 
in this particular case, on the operator
$\Phi^{2}_{(2)}(z_{3})$: its module is assumed to be degenerate at level
$\frac{1}{5}$ while the second level of degeneracy remains general.
This is the case for the operators
$\Phi_{(2k,1)(1,1)}$ and $\Phi_{(1,1)(2k,1)}$, with $k=1,2,3,\ldots$,
as can be shown using the method of reflections discussed in
Section~\ref{sec4} of the main text.

Using Eq.~(\ref{aD28}) and the integral
\bea
 I_{1} &=& \frac{1}{2\pi i}\oint_{C_{1}}{\rm d}z \,
 \left \langle \Phi^{2}_{(2)}(z_{3})\Phi^{1}(z_{2})\Psi^{-1}(z)
 \Phi^{-2}_{(1)}(z_{1}) \right \rangle \nn \\
 &\times& (z-z_{1})^{\Delta_{1}-2}(z_{2}-z)^{\Delta_{1}-
 \frac{2}{5}}(z_{3}-z)^{\Delta_{1}-\frac{1}{5}-1},
\eea
one derives in this case the differential equation
\bea
 & & \frac{\mu_{12}}{\mu_{11}} \frac{\partial}{\partial_{z_{1}}}
 \left \langle \Phi^{2}_{(2)}(z_{3})\Phi^{1}(z_{2})\Phi^{2}_{(1)}(z_{1})
 \right \rangle \nn \\
 & & = \left( \frac{6}{5}\frac{1}{z_{21}}+\frac{2}{5}\frac{1}{z_{31}} 
\right)
 h^{(1)}_{1,2} \left \langle \Phi^{2}_{(2)}(z_{3})\Phi^{1}(z_{2})
 \Phi^{2}_{(1)}(z_{1}) \right \rangle
\eea
and the following characteristic equation
\beq
 \Delta_{\Phi^{1}}=\Delta_{\Phi^{2}_{(2)}}+\frac{1}{2}\Delta_{\Phi^{2}_{(1)}}.
\eeq

\subsection{$\left \langle \Phi^{1}_{(3)}(z_{3})\Phi^{1}_{(2)}(z_{2})
\Phi^{-2}_{(1)}(z_{1}) \right \rangle$}

As for the previous two correlation functions, the module of the operator
$\Phi^{-2}_{(1)}(z_{1})$ is assumed to be degenerate at levels 
$\frac{1}{5}$
and 1, so that the relations (\ref{aD28}) apply. In this case the 
appropriate
integral is
\bea
 I_{1} &=& \frac{1}{2\pi i}\oint_{C_{1}}{\rm d}z \, 
(z-z_{1})^{\Delta_{1}-2}
 \left \langle \Phi^{1}_{(3)}(z_{3})\Phi^{1}_{(2)}(z_{2})\Psi^{1}(z)
 \Phi^{2}_{(1)}(z_{1}) \right \rangle \nn\\
 &\times& (z_{3}-z)^{\Delta_{1}-\frac{4}{5}}(z_{2}-z)^{\Delta_{1}-
 \frac{4}{5}}.
\eea
The differential equation in this case is of the form
\bea
 & & \frac{\mu_{12}}{\mu_{11}} \frac{\partial}{\partial_{z_{1}}}
 \left \langle \Phi^{1}_{(3)}(z_{3})\Phi^{1}{(2)}(z_{2})
 \Phi^{-2}_{(1)}(z_{1}) \right \rangle \nn\\
 & & = \frac{4}{5} \left( \frac{1}{z_{31}}+\frac{1}{z_{21}} \right)
 h^{(1)}_{1,2} \left \langle \Phi^{1}_{(3)}(z_{3})\Phi^{1}_{(2)}(z_{2})
 \Phi^{-2}_{(1)}(z_{1}) \right \rangle,
\eea
and the characteristic equation is
\beq
 \Delta_{\Phi^{1}_{(2)}}=\Delta_{\Phi^{1}_{(3)}}.
\eeq

\section{Imposing degeneracy at levels $3/5$ and $1$ for the doublet 2 module}
\label{appD}

A second solution of the degeneracy calculation for the doublet 2 can
be obtained by choosing to degenerate at levels $3/5$ and $1$.
In particular, we omit degenerating at level $1/5$ and leave the two
mutually conjugate states on this level.

It then appears that one can descend to level $3/5$ in two independent 
ways,
\beq
 A^1_{-\frac25}A^1_{-\frac15}\Phi^{-2} \mbox{ and }
 A^{-1}_{-\frac25}A^{-1}_{-\frac15}\Phi^2.
\eeq
Assuming this to be so we could search for a singular state at level $3/5$
in the form
\beq
 \chi^0_{-\frac35} = a A^1_{-\frac25}A^1_{-\frac15}\Phi^{-2} +
                   b A^{-1}_{-\frac25}A^{-1}_{-\frac15}\Phi^2. \label{D2}
\eeq

But actually, to be sure, we should check that acting with the zero-mode
at level $1/5$ does not lead to further independent states. To this end,
we examine the combination $A^2_0 A^1_{-\frac15} \Phi^{-2}$ (and its
conjugate). Using $\{\Psi^1,\Psi^2\}\Phi^{-2}$ with $n=0$, $m=1$ we find
\beq
 A^2_0 A^1_{-\frac15} \Phi^{-2} = \frac13 \lambda_{-2}^{1,2}
 A^{-2}_{-\frac15} \Phi^{-2}. \label{D3}
\eeq
Furthermore, from $\{\Psi^1,\Psi^1\}\Phi^2$ with $n=1$, $m=0$ we have
\beq
 A^{-2}_{-\frac15} \Phi^{-2} = \frac{2h}{\lambda_2^{1,1}}
 A^{-1}_{-\frac15} \Phi^{2}. \label{D4}
\eeq
Combining Eqs.~(\ref{D3}) and (\ref{D4}) we obtain the desired identity:
\beq
 A_0^2 A_{-\frac15}^1 \Phi^{-2} = \frac{2 h \lambda_{-2}^{1,2}}
 {3 \lambda_2^{1,1}} A^{-1}_{-\frac15} \Phi^2, \label{D5}
\eeq
which implies that the most general state at level $3/5$ indeed has the
form (\ref{D2}).

To obtain degeneracy at level $3/5$ we now
define the following matrix elements:
\bea
 \mu_{11} (A^1_{-\frac15} \Phi^{-2}) &=&
 A^{-1}_{\frac25} A^1_{-\frac25} A^1_{-\frac15} \Phi^{-2}, \\
 \mu_{12} (A^1_{-\frac15} \Phi^{-2}) &=&
 A^{-1}_{\frac25} A^{-1}_{-\frac25} A^{-1}_{-\frac15} \Phi^2, \\
 \mu_{21} (A^{-1}_{-\frac15} \Phi^{2}) &=&
 A^{1}_{\frac25} A^{1}_{-\frac25} A^{1}_{-\frac15} \Phi^{-2}, \\
 \mu_{22} (A^{-1}_{-\frac15} \Phi^{2}) &=&
 A^{1}_{\frac25} A^{-1}_{-\frac25} A^{-1}_{-\frac15} \Phi^{2}.
\eea
By charge conjugation symmetry, $\mu_{11}=\mu_{22}$ and 
$\mu_{12}=\mu_{21}$.
The degeneracy condition therefore becomes
\beq
 (\mu_{11})^2 = (\mu_{21})^2. \label{D10}
\eeq
The explicit computation of $\mu_{11}$ and $\mu_{12}$ reads as follows:
\begin{itemize}
\item \underline{$\mu_{11}$.} {}From $\{\Psi^1,\Psi^{-1}\}\Phi^{1}$ with
$n=0$, $m=1$ we obtain:
\beq
 \mu_{11} (A^1_{-\frac15} \Phi^{-2}) =
 \left[ -\frac2{25} + \frac{16}{5c} \left( \Delta+\frac15 \right)
 \right] A^1_{-\frac15} \Phi^{-2} -
 A^1_{-\frac15} \left( A^{-1}_{\frac15} A^1_{-\frac15} \right) \Phi^{-2},
\eeq
where we have denoted $\Delta \equiv \Delta_{\Phi^2}$ for brevity.
This calls for the evaluation of a further matrix element defined by
\beq
 \mu_0 \Phi^2 = A^{-1}_{\frac15} A^1_{-\frac15} \Phi^{-2},
\eeq
which is easily found from $\{\Psi^1,\Psi^{-1}\}\Phi^{2}$ with
$n=m=0$. The result is
\beq
 \mu_0 = -\frac{3}{25} + \frac{16 \Delta}{5c} - h^2, \label{D13}
\eeq
where $h \equiv h_{1,2}$ denotes the eigenvalue of the zero mode.
Combining this, we obtain
\beq
 \mu_{11} = \frac{1}{25} \left(1 + \frac{16}{c} \right) + h^2. \label{D14}
\eeq

\item \underline{$\mu_{12}$.} We shall need the identity
\beq
 A^1_{-\frac25} A^1_{-\frac15} \Phi^{-2} = \frac12 \lambda_2^{1,1}
 A^2_{-\frac35} \Phi^{-2}, \label{D15}
\eeq
which follows from $\{\Psi^1,\Psi^1\}\Phi^{-2}$ with $n=1$, $m=0$.
In particular, using Eq.~(\ref{D15}),
\beq
 \mu_{21} \left( A^{-1}_{-\frac15} \Phi^{2} \right) =
 \frac12 \lambda_2^{1,1} A^{-1}_{\frac25} A^{-2}_{-\frac35} \Phi^2.
\eeq
This can be rewritten by use of $\{\Psi^1,\Psi^2\}\Phi^{-2}$ with
$n=1$, $m=0$:
\beq
 \mu_{21} \left( A^{-1}_{-\frac15} \Phi^{2} \right) =
 \frac13 \lambda_2^{1,1} \lambda_{-2}^{1,2} A^2_{-\frac15} \Phi^2,
\eeq
and from Eq.~(\ref{D4}) we finally obtain
\beq
 \mu_{21} = \frac23 h \lambda_{-2}^{1,2}. \label{D18}
\eeq
\end{itemize}
 The condition (\ref{D10}), with the ingredients
(\ref{D14}) and (\ref{D18}), presents two pairs of solutions, 
$h_{(1)}^{(\pm)}$ and
 $h_{(2)}^{(\pm)}$, which are fixed as a function
 of $c$ only (there is no dependence on $\Delta$).
 Using the parametrisation (\ref{cparamp}) one has :
\bea
h_{(1)}^{(\pm)}&=&\pm \frac{\sqrt{5}}{5}\frac{(p-1)}{\sqrt{(p+5)(p-3)}}, 
\label{h1}\\
h_{(2)}^{(\pm)}&=&\pm \frac{\sqrt{5}}{5}\frac{(p+3)}{\sqrt{(p+5)(p-3)}}. 
\label{h2}
\eea

Up to and including level $3/5$ there is now a total of five states, i.e.,
one state in each charge sector. In particular, one can descend via these
states to the $q=2$
sector at level $1$ in five ways. However, we shall now show that only 
four of these ways are linearly independent.

To this end, consider $\{\Psi^1,\Psi^2\}(\Phi^{-1})_{-\frac15}$
with $n=m=0$, where we act on the state
$(\Phi^{-1})_{-\frac15} \equiv A^1_{-\frac15}\Phi^{-2}$:
\beq
 \left[A^1_{-\frac45} A^2_0 - A^2_{-\frac25} A^1_{-\frac25} \right]
 A^1_{-\frac15} \Phi^{-2} = -\frac13 \lambda_{-2}^{1,2}
 A^{-2}_{-\frac45} A^1_{-\frac15} \Phi^{-2}.
\eeq
Keeping in mind Eq.~(\ref{D5}), we conclude that the state
\beq
 A^2_{-\frac25} \left( A^1_{-\frac25} A^1_{-\frac15} \Phi^{-2} \right) 
\equiv
 A^2_{-\frac25} \left( A^{-1}_{-\frac25} A^{-1}_{-\frac15} \Phi^2 \right)
\eeq
is linearly dependent on $A^1_{-\frac45} A^{-1}_{-\frac15} \Phi^2$ and
$A^{-2}_{-\frac45} A^1_{-\frac15} \Phi^{-2}$. In other words, descending
to level $1$ through the state at level $3/5$ can be ignored in the
degeneracy computation.

We have also verified that the four (conjugate) ways of
descending to the $q=-2$ sector at level $1$, followed
by action with the operator $A^{-1}_0$, do not give further independent states
in the $q=2$ sector at level $1$.

We therefore search for a singular state at level $1$ of the form
\beq
 \chi^2_{-1} = \alpha A^{-1}_{-1} \Phi^{-2} +
             \beta A^{-2}_{-\frac45} A^1_{-\frac15} \Phi^{-2} +
             \gamma A^1_{-\frac45} A^{-1}_{-\frac15} \Phi^2 +
             \delta L_{-1} \Phi^2.
\eeq
Defining the following 16 matrix elements
\bea
 \mu_{11} \Phi^{-2} &=& A^1_1 A^{-1}_{-1} \Phi^{-2}, \\
 \mu_{12} \Phi^{-2} &=& A^1_1 A^{-2}_{-\frac45} A^1_{-\frac15} \Phi^{-2}, \\
 \mu_{13} \Phi^{-2} &=& A^1_1 A^1_{-\frac45} A^{-1}_{-\frac15} \Phi^2, \\
 \mu_{14} \Phi^{-2} &=& A^1_1 L_{-1} \Phi^2, \\
 \mu_{21} \left( A^1_{-\frac15} \Phi^{-2} \right) &=&
  A^2_{\frac45} A^{-1}_{-1} \Phi^{-2}, \\
 \mu_{22} \left( A^1_{-\frac15} \Phi^{-2} \right) &=&
  A^2_{\frac45} A^{-2}_{-\frac45} A^1_{-\frac15} \Phi^{-2}, \\
 \mu_{23} \left( A^1_{-\frac15} \Phi^{-2} \right) &=&
  A^2_{\frac45} A^1_{-\frac45} A^{-1}_{-\frac15} \Phi^2, \\
 \mu_{24} \left( A^1_{-\frac15} \Phi^{-2} \right) &=&
  A^2_{\frac45} L_{-1} \Phi^2, \\
 \mu_{31} \left( A^{-1}_{-\frac15} \Phi^2 \right) &=&
  A^{-1}_{\frac45} A^{-1}_{-1} \Phi^{-2}, \\
 \mu_{32} \left( A^{-1}_{-\frac15} \Phi^2 \right) &=&
  A^{-1}_{\frac45} A^{-2}_{-\frac45} A^{-1}_{-\frac15} \Phi^{-2}, \\
 \mu_{33} \left( A^{-1}_{-\frac15} \Phi^2 \right) &=&
  A^{-1}_{\frac45} A^1_{-\frac45} A^{-1}_{-\frac15} \Phi^2, \\
 \mu_{34} \left( A^{-1}_{-\frac15} \Phi^2 \right) &=&
  A^{-1}_{\frac45} L_{-1} \Phi^2, \\
 \mu_{41} \Phi^2 &=& L_1 A^{-1}_{-1} \Phi^{-2}, \\
 \mu_{42} \Phi^2 &=& L_1 A^{-2}_{-\frac45} A^{-1}_{-\frac15} \Phi^{-2}, \\
 \mu_{43} \Phi^2 &=& L_1 A^1_{-\frac45} A^{-1}_{-\frac15} \Phi^2, \\
 \mu_{44} \Phi^2 &=& L_1 L_{-1} \Phi^2,
\eea
the degeneracy condition becomes
\beq
 \det \mu_{ij} = 0,
\label{deg3_5}
\eeq
where $\mu_{ij}$ is the $4 \times 4$ matrix defined by the above matrix
elements.

Six of these matrix elements are easily found from $\{T,\Psi\}\Phi$:
\bea
 \mu_{14} = \mu_{41} &=& \frac{8 h}{5} \\
 \mu_{24} &=& \frac{22 h}{5 \lambda_2^{1,1}} \\
 \mu_{34} &=& \frac75 \\
 \mu_{43} &=& \frac75 \mu_0 \\
 \mu_{44} &=& 2 \Delta.
\eea
(In the computation of $\mu_{24}$ we have also used Eq.~(\ref{D4}), and,
in the case of $\mu_{43}$, Eq.~(\ref{D13}).)
The last matrix element involving $L_n$ takes the form
\beq
 \mu_{42} \Phi^2 = \frac{11}{5} A^{-2}_{\frac15} A^1_{-\frac15} \Phi^{-2},
\eeq
once the $\{T,\Psi\}\Phi$ commutators have been applied.
The auxiliary matrix element appearing on the right-hand side is found 
from
$\{\Psi^1,\Psi^{-2}\}\Phi^{-2}$ with $n=m=0$, and using also
Eqs.~(\ref{D4}) and (\ref{D13}). The result is
\beq
 A^{-2}_{\frac15} A^1_{-\frac15} \Phi^{-2} =
 \frac{2 h \mu_0}{\lambda_2^{1,1}} \Phi^2, \label{D45}
\eeq
implying in particular
\beq
 \mu_{42} = \frac{22 h \mu_0}{5 \lambda_2^{1,1}}.
\eeq

It now remains to evaluate the nine matrix elements not involving $L_n$.
\begin{itemize}
\item \underline{$\mu_{11}$.} Using $\{\Psi^1,\Psi^{-1}\}\Phi^{-2}$ with
  $n=1$, $m=-1$ we readily obtain
  \beq
   \mu_{11} = \frac{h^2}{5} + \frac{7}{25} + \frac{16 \Delta}{5c}.
  \eeq
\item \underline{$\mu_{12}$.}
  {}From $\{\Psi^1,\Psi^{-2}\}(\Phi^{-1})_{-\frac15}$
  with $m=1$, $n=0$ we get the relation
  \beq
   \mu_{12}\Phi^{-2} = \frac25 A^1_0 \left( A^{-2}_{\frac15} 
A^1_{-\frac15}
   \Phi^{-2} \right) + \frac34 \lambda_{-1}^{1,-2} \left( A^{-1}_{\frac15}
   A^1_{-\frac15} \Phi^{-2} \right).
  \eeq
  Using further Eqs.~(\ref{D13}) and (\ref{D45}) we obtain
  \beq
   \mu_{12} = \mu_0 \left( \frac{4 h^2}{5 \lambda_2^{1,1}} +
   \frac34 \lambda_{-1}^{1,-2} \right).
  \eeq
\item \underline{$\mu_{13}$.} We use 
$\{\Psi^1,\Psi^1\}(\Phi^1)_{-\frac15}$
  with $m=0$, $n=2$ and, once again, Eqs.~(\ref{D13}) and (\ref{D45}).
  The result is
  \beq
   \mu_{13} = \frac{9 h \mu_0}{5}.
  \eeq
\item \underline{$\mu_{21}$.} Employing $\{\Psi^1,\Psi^{-2}\}\Phi^2$ with
  $m=1$, $n=-1$ as well as Eq.~(\ref{D4}) gives us
  \beq
   \mu_{21} = \frac{4 h^2}{5 \lambda_2^{1,1}} + \frac34 
\lambda_{-1}^{1,-2}.
  \eeq
\item \underline{$\mu_{23}$.} Using $\{\Psi^1,\Psi^2\}(\Phi^1)_{-\frac15}$
  with $n=2$, $m=0$, in conjunction with
  Eqs.~(\ref{D4}), (\ref{D5}) and (\ref{D13}), yields
  \beq
   \mu_{23} = \frac{2h}{\lambda_2^{1,1}} \left[
   \frac13 \left( \lambda_{-2}^{1,2} \right)^2 - \frac{2 \mu_0}{5} 
\right].
  \eeq
\item \underline{$\mu_{31}$.} {}From $\{\Psi^1,\Psi^1\}\Phi^2$ with
  $n=1$, $m=-1$ and Eq.~(\ref{D4}) we find
  \beq
   \mu_{31} = \frac{9h}{5}.
  \eeq
\item \underline{$\mu_{32}$.} By use of 
$\{\Psi^1,\Psi^2\}(\Phi^1)_{-\frac15}$
  with $m=2$, $n=0$ as well as Eqs.~(\ref{D5}) and (\ref{D45}) it is found
  that
  \beq
   \mu_{32} = \frac{2h}{\lambda_2^{1,1}} \left[
   \frac13 \left( \lambda_{-2}^{1,2} \right)^2 - \frac{2 \mu_0}{5} \right]
   \equiv \mu_{23}.
  \eeq
\item \underline{$\mu_{33}$.} {}From 
$\{\Psi^1,\Psi^{-1}\}(\Phi^1)_{-\frac15}$
  with $m=1$, $n=0$ and using Eq.~(\ref{D13}) we get
  \beq
   \mu_{33} = \frac{12}{125} - \frac{h^2}{5} + \frac{16}{25c}(6\Delta+1).
  \eeq
\end{itemize}

The final matrix element, $\mu_{22}$, is the most involved to calculate.
The reason is that the commutation relations $\{\Phi^2,\Phi^{-2}\}\Phi^1$
are not very predictive. We therefore rewrite
$A^{-2}_{-\frac45}A^1_{-\frac15}\Phi^{-2}$ by means of
$\{\Psi^1,\Psi^1\}(\Phi^1)_{-\frac15}$ with $n=1$, $m=0$. The result can
be stated as
\beq
 \mu_{\rm A} + \frac65 \mu_{\rm B} = \lambda_2^{1,1} \mu_{22}, \label{D56}
\eeq
with
\bea
 \mu_{\rm A} \left( A^1_{-\frac15} \Phi^{-2} \right) &=&
 \left(A^2_{\frac45} A^{-1}_0 \right) A^{-1}_{-\frac45} A^1_{-\frac15}
 \Phi^{-2} \label{D57} \\
 \mu_{\rm B} \left( A^1_{-\frac15} \Phi^{-2} \right) &=&
 \left(A^2_{\frac45} A^{-1}_{-1} \right) A^{-1}_{\frac15} A^1_{-\frac15}
 \Phi^{-2}.
\eea

The auxiliary matrix element $\mu_{\rm B}$ is easily calculated.
One first uses Eq.~(\ref{D13}) to obtain
\beq
 \mu_{\rm B} \left( A^1_{-\frac15} \Phi^{-2} \right) =
 \mu_0 A^2_{\frac45} A^{-1}_{-1} \Phi^{-2},
\eeq
and application of $\{\Psi^1,\Psi^{-2}\}\Phi^2$ with $m=1$, $n=-1$
in conjunction with Eq.~(\ref{D4}) then gives
\beq
 \mu_{\rm B} = \mu_0 \left( \frac{4h^2}{5 \lambda_2^{1,1}} +
 \frac34 \lambda_{-1}^{1,-2} \right). \label{D60}
\eeq

Eq.~(\ref{D57}) for $\mu_{\rm A}$ can be simplified by means of
$\{\Psi^{-2},\Psi^{1}\}(\Phi^{2})_{-1}$ with $m=1$, $n=0$:
\beq
 \mu_{\rm A} \left( A^1_{-\frac15} \Phi^{-2} \right) =
 \left[ A^{-1}_{\frac25} A^2_{\frac25} + \frac25 A^2_{-\frac15} A^{-1}_1
 \right] A^{-1}_{-\frac45} A^1_{-\frac15} \Phi^{-2}. \label{D61}
\eeq
This in turn calls for the evaluation of two auxiliary terms:
\begin{itemize}
 \item \underline{$A^{-1}_{\frac25} \left( A^2_{\frac25}
 A^{-1}_{-\frac45} A^1_{-\frac15} \Phi^{-2} \right)$.} The term between
 parentheses is found from $\{\Psi^1,\Psi^{-2}\}(\Phi^{1})_{-\frac15}$
 with $n=m=0$ as well as Eqs.~(\ref{D5}), (\ref{D13}) and (\ref{D15}):
 \beq
  A^2_{\frac25} A^{-1}_{-\frac45} A^1_{-\frac15} \Phi^{-2} =
  \frac12 \lambda_{-1}^{1,-2} A^1_{-\frac25} A^1_{-\frac15} \Phi^{-2} +
  \frac{2 \mu_0}{5} A^2_{-\frac35} \Phi^{-2} +
  \frac{h \lambda_{-2}^{1,2}}{3} A^{-2}_{-\frac35} \Phi^2.
 \eeq
 Acting on this by $A^{-1}_{\frac25}$ we see that the right-hand side
 produces a further three quantities which must be evaluated:
 \begin{enumerate}
   \item \underline{$\left( A^{-1}_{\frac25} A^1_{-\frac25} \right)
   A^1_{-\frac15} \Phi^{-2}$.}
   We use $\{\Psi^1,\Psi^{-1}\}(\Phi^{1})_{-\frac15}$ with $m=0$, $n=1$
   and Eq.~(\ref{D13}) to obtain:
   \beq
    \left( A^{-1}_{\frac25} A^1_{-\frac25} \right) A^1_{-\frac15} 
\Phi^{-2} =
    \left( \frac{1}{25} + h^2 + \frac{16}{25 c} \right) A^1_{-\frac15}
    \Phi^{-2}.
    \label{D63}
   \eeq
   \item \underline{$A^{-1}_{\frac25} A^2_{-\frac35} \Phi^{-2}$.}
   Using $\{\Psi^{1},\Psi^{-2}\}\Phi^{2}$ with $n=m=0$ together with
   Eq.~(\ref{D4}), we find that:
   \beq
    A^{-1}_{\frac25} A^2_{-\frac35} \Phi^{-2} =
    \left( \frac{2 h^2}{\lambda_2^{1,1}} + \frac14 \lambda_{-1}^{1,-2} 
\right)
    A^1_{-\frac15} \Phi^{-2}.
   \eeq
   \item \underline{$A^{-1}_{\frac25} A^{-2}_{-\frac35} \Phi^2$.}
   This is found from $\{\Psi^1,\Psi^2\}\Phi^{-2}$ with $n=1$, $m=0$,
   together with Eq.~(\ref{D4}):
   \beq
    A^{-1}_{\frac25} A^{-2}_{-\frac35} \Phi^2 =
    \frac{4 h \lambda_{-2}^{1,2}}{3 \lambda_2^{1,1}} A^1_{-\frac15} 
\Phi^{-2}.
    \label{D65}
   \eeq
 \end{enumerate}
 Collecting the pieces, Eqs.~(\ref{D63})--(\ref{D65}), we conclude that
 \bea
  \left( A^{-1}_{\frac25} A^2_{\frac25} A^{-1}_{-\frac45} \right)
  A^1_{-\frac15} \Phi^{-2} &=& \left\{ \frac12 \lambda_{-1}^{1,-2}
  \left( \frac{1}{25} + h^2 + \frac{16}{25c} \right) \right. \label{D66} 
\\
  &+& \left. \frac{2\mu_0}{5}
  \left( \frac{2h^2}{\lambda_2^{1,1}} + \frac14 \lambda_{-1}^{1,-2} 
\right) +
  \frac{4 h^2 \left( \lambda_{-2}^{1,2} \right)^2}{9 \lambda_2^{1,1}}
  \right\} A^1_{-\frac15} \Phi^{-2}. \nn
 \eea
 \item \underline{$A^{-1}_1 A^{-1}_{-\frac45} A^1_{-\frac15} \Phi^{-2}$.}
 We apply $\{\Psi^1,\Psi^1\}(\Phi^1)_{-\frac15}$ with $n=2$, $m=0$, and
 also Eqs.~(\ref{D13}) and (\ref{D45}), to obtain
 \beq
  \left( A^{-1}_1 A^{-1}_{-\frac45} \right) A^1_{-\frac15} \Phi^{-2} =
  \frac95 h \mu_0 \Phi^2. \label{D67}
 \eeq
\end{itemize}
To finish the calculation of $\mu_{\rm A}$, as given by Eq.~(\ref{D61}),
we assemble Eqs.~(\ref{D66}) and (\ref{D67}), using also Eq.~(\ref{D4}).
The result reads:
\beq
 \mu_{\rm A} = \lambda_{-1}^{1,-2} \left[ \frac{1}{125} + \frac25 h^2 +
 \frac{8}{25c}(\Delta+1) \right] + \frac{4 h^2 (\lambda_{-2}^{1,2})^2}
 {9 \lambda_2^{1,1}} + \frac{56 h^2 \mu_0}{25 \lambda_2^{1,1}}. 
\label{D68}
\eeq 

Finally, we combine Eqs.~(\ref{D56}), (\ref{D60}) and (\ref{D68}),
obtaining:
\beq
 \mu_{22} = \frac{\lambda_{-1}^{1,-2}}{\lambda_2^{1,1}} \left[
 -\frac{1}{10} - \frac{h^2}{2} + \frac{8}{25c} + \frac{16 \Delta}{5 c} 
\right]
 + \frac{16 h^2 \mu_0}{5 \left( \lambda_2^{1,1} \right)^2}
 + \frac{4 h^2}{9} \left( \frac{\lambda_{-2}^{1,2}}{\lambda_2^{1,1}} 
\right)^2.
\label{D69}
\eeq

We consider first the solution (\ref{h1}) with the condition (\ref{deg3_5});
in the following we label this case with the index $a$.
One obtains four possible values of the dimension $\Delta^{(i,a)}$ 
($i=1,\ldots,4$)  of the
doublet 2, given below as a function of $p$,  which solve the condition 
(\ref{deg3_5}):
\bea
\Delta^{(1,a)}&=&7/5 \label{1Asol}, \\
\Delta^{(2,a)}&=&\frac{1}{5}\frac{-25-4q+8q^{2}}{q(q+2)}, \label{2Asol}\\
\Delta^{(3,a)}&=&\frac{1}{5}\frac{20-q+2q^{2}}{q(q+2)}, \label{3Asol} \\
\Delta^{(4,a)}&=&\frac{1}{5}\frac{-16+7q}{q+2}. \label{4Asol} 
\eea

The solutions (\ref{1Asol}) and (\ref{2Asol}) cannot be obtained from
Eq.~(\ref{Kac1}) for integer values of the indices $(n_1,n_2)(n_1',n_2')$
and are therefore rejected
as non-physical solutions. 
For the remaining solutions, Eqs.~(\ref{3Asol}) and (\ref{4Asol}), it is
straightforward to verify that:
\bea
 \Delta^{(3,a)}&=& \Delta_{(1,3)(2,1)}, \\
 \Delta^{(4,a)}&=& \Delta_{(2,3)(1,1)}. 
\eea 

Analogous results are obtained by using the solution (\ref{h2}) with the
condition (\ref{deg3_5}); we label this as case $b$. The four solutions are:
\bea
\Delta^{(1,b)}&=&7/5, \label{1Bsol} \\
\Delta^{(2,b)}&=&\frac{1}{5}\frac{15+36q+8q^{2}}{q(q+2)}, \label{2Bsol}\\
\Delta^{(3,b)}&=&\frac{1}{5}\frac{30+9q+2q^{2}}{q(q+2)}, \label{3Bsol} \\
\Delta^{(4,b)}&=&\frac{1}{5}\frac{30+7q}{q}. \label{4Bsol} 
\eea

The solutions (\ref{1Bsol}) and (\ref{2Bsol}) are rejected as non-physical,
whilst the other two are physical:
\bea
\Delta^{(3,b)}&=& \Delta_{(2,1)(1,3)}, \\
\Delta^{(4,b)}&=& \Delta_{(1,1)(2,3)}. 
\eea

The above results constitute a strong verification of the validity of the
theory that we have built. First, it is in accordance with the positioning of
the doublet 2 operators. Second, it is easy to show that, according to
the reflections on the Kac table (see Section~\ref{sec4}), the modules of
the doublet 2 operators $\Phi_{(1,3)(2,1)}$, $\Phi_{(2,1)(1,3)}$,
$\Phi_{(2,3)(1,1)}$, and $\Phi_{(1,1)(2,3)}$ ought to be degenerate
at levels $3/5$ and $1$.

\end{document}